\documentclass[useAMS,usenatbib,onecolumn]{mn2e}
\usepackage{bm}
\usepackage{amsfonts}
\usepackage{amsmath}
\usepackage{graphicx}
\usepackage{subfigure}
\begin{document}
\title[Isopedically Magnetized Scale-Free Discs ]
{Scale-Free Thin Discs with an Isopedic Magnetic Field}
\author[Y. Wu \& Y.-Q. Lou]{Yue Wu$^1$
and Yu-Qing Lou$^{1,2,3}$\thanks{E-mail: louyq@mail.tsinghua.edu.cn;
wuyue00@mails.tsinghua.edu.cn }
\\$^1$Physics Department and Tsinghua Centre for Astrophysics
(THCA), Tsinghua University, Beijing 100084, China;
\\$^2$Department of Astronomy and Astrophysics, The University
of Chicago, 5640 South Ellis Avenue, Chicago, IL 60637, USA;
\\$^3$National Astronomical Observatories, Chinese Academy
of Sciences, A20, Datun Road, Beijing 100012, China.
\\}
\date{Accepted 2004... Received 2003...;
in original form 2003}\date{Accepted .
      Received ;
      in original form }
\pagerange{\pageref{firstpage}--\pageref{lastpage}} \pubyear{2006}

\maketitle

\label{firstpage}

\begin{abstract}
Global stationary configurations of both aligned and logarithmic
spiral magnetohydrodynamic (MHD) perturbations are constructed
analytically within an axisymmetric background of razor-thin
scale-free gas disc, which is embedded in an axisymmetric
gravitational potential of a dark matter halo and involves an
isopedic magnetic field almost vertically threaded through the
disc plane. The scale-free index $\beta$ of the disc rotation
speed $v_{\theta}\propto R^{-\beta}$ falls in the range of
$(-1/2\ ,\ 1/2)$ where $R$ is the cylindrical radius. With the
holding-back of a deep background dark matter halo potential, the 
isopedic magnetic field may be strong enough to allow for the 
magnetic tension force overtaking the disc self-gravity, which 
can significantly influence global stationary MHD perturbation 
configurations and stability properties of the scale-free disc 
system. Only for stationary logarithmic spiral MHD perturbations 
with a perturbation scale-free index $\beta_1=1/4$ or for aligned 
stationary MHD perturbations, can the MHD disc maintain a constant 
radial flux of angular momentum.
The variable radial flux of angular momentum in the radial 
direction corresponds to a complex dispersion relation. The 
marginal instabilities for axisymmetric MHD disturbances are 
also examined for a special case as an example. When the 
magnetic tension force overtakes the disc self-gravity, the 
scale-free disc can be completely stable against axisymmetric 
MHD disturbances of all wavelengths.
We predict the possible existence of an isopedically magnetized
gas disc system in rotation primarily confined by a massive
dark matter halo potential.
\end{abstract}


\section{Introduction}

As an important intermediate stage of many astrophysical processes, 
the dynamics of magnetized disc systems are frequently invoked to 
model different phenomena on various spatial and temporal scales, 
such as spiral or barred-spiral galaxies, magnetized accretion discs 
surrounding supermassive black holes (SMBHs) anchored at galactic 
centres, accreting environs involving binary stars and protoplanetary 
systems etc. It is therefore of considerable interest to study 
magnetohydrodynamic (MHD) processes in discs for both theoretical and 
applied purposes. The pioneer development of the classic density wave 
theory for large-scale perturbation structures in differentially 
rotating discs (Lin \& Shu 1964, 1966) has opened up a physical 
scenario to understand the basic dynamics of spiral galaxies (Lin 1987; 
Binney \& Tremaine 1987; Bertin \& Lin 1996). Linear perturbation or 
nonlinear theories are widely used as powerful tools of analysis to 
explore the structure and stability of disc systems. In many processes, 
initially small perturbations may lead to more violent consequences 
(e.g., gravitational collapse of a central core) which are crucial for 
the dynamical evolution of astrophysical systems. In various contexts, 
MHD perturbation configurations are valuable indicators of transitions 
to instabilities and of further dynamical structure and evolution.

For simplicity and clarity, the scale-free disc system as an idealized
model is a frequent target of theoretical investigation. The concept 
of a `scale-free' or a `self-similar' disc means all relevant physical
quantities in the disc system vary as powers of cylindrical radius
$R$ such that the disc system carries no characteristic length scale
(e.g., the linear speed of disc rotation $v_{\theta}\propto R^{-\beta}$
and the equilibrium disc surface mass density $\Sigma_0\propto 
R^{-2\beta-1}$ etc. with $\beta$ being a constant exponent). Given 
possible situations,
one may introduce `boundary conditions' to describe a modified
scale-free disc by cutting a central hole in a disc (e.g., Zang 1976;
Evans \& Read 1998a, b) or by limiting the radial extent of a disc.
Observationally, many disc galaxies show more or less flat rotation
curves, indicating the presence of massive dark matter halos according
to the Newtonian gravity theory. Theoretically, discs with flat rotation
curves (i.e., the azimuthal velocity $v_\theta=\mbox{const}$ or $\beta=0$)
belong to a specific subclass of scale-free discs, referred to as singular
isothermal discs (SIDs). Since the seminal analysis of Mestel (1963), the
SID model has attracted considerable attention in various astrophysical
contexts (e.g., Zang 1976; Toomre 1977; Lemos, Kalnajs \& Lynden-Bell
1991; Shu et al. 2000; Lou 2002; Lou \& Fan 2002; Lou \& Shen 2003; Shen
\& Lou 2004a,b; Lou \& Zou 2004, 2006; Shen, Liu \& Lou; Lou \& Wu 2005;
Lou \& Bai 2006). In most normal spiral galaxies, disc rotation curves
tend to be flat over extended radii, giving a strong evidence for the
existence of massive dark matter halos (e.g., Rubin et al. 1982; Kent
1986, 1987, 1988).
Given the inferred dominant masses, we recognize that a dark matter
halo plays extremely important roles in the dynamical evolution of
a disc galaxy (Miller et al. 1970; Ostriker \& Peebles 1973; Hohl
1971; Miller 1978; Binney \& Tremaine 1987).

Besides the normal mode perturbation analysis in scale-free discs
(e.g., Zang 1976; Binney \& Tremaine 1987; Fan \& Lou 1997; Evans
\& Read 1998a,b; Goodman \& Evans 1999; Shu et al. 2000), the
zero-frequency neutral perturbation modes or stationary
perturbation configurations are emphasized as the marginal
instability in scale-free discs (e.g., Lemos et al. 1991; Syer \&
Tremaine 1996; Shu et al. 2000; Lou \& Shen 2003; Shen \& Lou
2003; Shen \& Lou 2004a; Shen et al. 2005; Lou \& Zou 2004, 2006;
Lou \& Wu, 2005; Lou \& Bai 2006). Axisymmetric instabilities are
thought to set in through transitions across such neutral modes
(Safronov 1960; Toomre 1964; Lynden-Bell \& Ostriker 1967; Lemos
et al. 1991; Shu et al. 2000). Parallel to zero-frequency modes
for axisymmetric disturbances, Shu et al. (2000) tentatively
proposed that logarithmic spiral modes of stationary perturbation
configurations also signal onsets of non-axisymmetric
instabilities; the resulting criterion appears to be compatible
with the criterion of Goodman \& Evans (1999) for instabilities in
their normal mode approach. Recently, Shen \& Lou (2004a) extended
the analysis of Shu et al. (2000) in a gravitationally coupled
composite scale-free disc system of gaseous and stellar discs
using a two-fluid formalism. In solving nonlinear MHD equations, 
it is a challenge to achieve analytical results unless some 
simplified assumptions are introduced, such as the stationarity 
($\omega=0$) condition etc. 
Most results are analytic in this paper and it is highly desirable 
that numerical simulations should be developed as touchstones for 
the analytical results and further venture into new regimes where 
analytical methods cannot reach (Syer \& Tremaine 1996).

Magnetic field receives considerable attention in
astrophysics,
and contributes to structures, dynamics and diagnostics on various 
scales of different astrophysical systems (Sofue et al. 1986; Beck 
et al. 1996; Balbus \& Hawley 1998; Kaburaki 2000, 2001; Balbus 
2003; Vall\'ee 2004; Hu \& Lou 2004; Yu \& Lou 2006; Yu, Lou, Bian 
\& Wu 2006). Prompted by numerical simulations of magnetized cloud 
core formation through the ambipolar diffusion (e.g., Nakano 1979; 
Lizano \& Shu 1989), Shu \& Li (1997) introduced the so-called 
`isopedic' condition requiring that the $z-$component flux ($B_z$ 
in cylindrical coordinates) of an open magnetic field, which 
threads through the disc almost vertically, is proportional to the 
disc surface mass density ($\Sigma$) or $B_z/\Sigma=\mbox{const}$. 
With this simple and justifiable assumption (Lou \& Wu 2005), 
effects of such an isopedic magnetic field can be simply subsumed 
into relevant terms in MHD equations (Shu \& Li 1997); this 
simplification is powerful in analytical analysis in comparison with 
the usual awesome MHD equations describing the roles of magnetic 
field. Lou \& Wu (2005) discussed the global MHD perturbation 
configurations in a composite SID system of coupled gas and stellar 
components with an isopedic (vertical) magnetic field. Lou \& Zou 
(2004, 2006) investigated the similar kind of composite MHD SID 
problems except that their magnetic field is coplanar rather than 
isopedic. Lou \& Bai (2006) analyzed MHD perturbation structures in 
a composite system of two coupled scale-free discs with a coplanar
magnetic field.

%

The purpose of this paper is to couple an isopedic magnetic field 
with a single scale-free gas disc embedded in an axisymmetric dark 
matter halo potential and to construct physically allowed global
stationary MHD perturbation configurations in such a disc as well
as to investigate the instability of such a disc system. Parallel 
to the hydrodynamic treatment of Syer \& Tremaine (1996), our 
analysis leads to further MHD extensions and generalizations. To 
be specific, six components of forces are involved in the disc
equilibrium of our analysis, namely, the radial centrifugal force,
the thermal gas pressure force in the disc, the self-gravity of
the disc, the holding-back gravity force of an axisymmetric dark
matter halo and the tension and pressure forces resulting from 
an isopedic magnetic field within the disc plane. While highly 
idealized, the unique combination of these essential elements 
of a disc galaxy does provide a new perspective and has already 
shown some novel characteristics. For example, the introduction 
of an isopedic magnetic field geometry in a scale-free disc does 
bring several new and interesting features in reference to prior 
related disc model analyses (Syer \& Tremaine 1996; Shu et al. 
2002; Shen \& Lou 2004b; Lou \& Zou 2004, 2006; Shen, Liu \& Lou 
2005; Lou \& Wu 2005; Lou \& Bai 2006).

This paper is organized as follows. In Section 2, we describe the 
basic MHD model of a scale-free razor-thin disc. In Section 3, 
isopedic magnetic field are discussed under the scale-free condition. 
In Section 4, we construct the stationary background axisymmetric 
equilibrium in a rotational MHD balance. In Section 5, a generalized 
MHD dispersion relationship is derived for MHD perturbations. We 
investigate the global stationary aligned MHD cases in Section 6 and 
stationary unaligned MHD (logarithmic spiral) cases in Section 7. We 
summarize the results and provide discussions in Section 8. Specific 
details of derivations are contained in Appendices.


\section{A Class of Scale-Free Razor-Thin Discs}

Formally, a razor-thin disc has a three-dimensional
mass density mathematically described by
\begin{equation}\label{razorthin}
\rho(R,\ \theta,\ z,\ t)=\Sigma(R,\ \theta,\ t)\delta(z)\ ,
\end{equation}
where cylindrical coordinates $(R, \theta, z)$ are adopted,
$\Sigma(R, \theta, t)$ is the two-dimensional surface mass density
and $\delta(z)$ is the Dirac $\delta$-function with the vertical
coordinate $z$ as the argument. Except for the three-dimensional
gravitational potential, any physical quantity $q$ is a function of 
$R$, $\theta$ and $t$ as $q(R, \theta,\ t)$ for a razor-thin disc.

In a razor-thin disc, the vertically integrated barotropic
equation of state has a `two-dimensional' form of
\begin{equation}\label{barotropic}
\Pi=k\Sigma^{\Gamma}\ ,
\end{equation}
where $\Pi(R, \theta, t)$ represents the two-dimensional gas
pressure, constant coefficient $k\geq0$ and the barotropic 
index $\Gamma>0$. We here take the general `three-dimensional' 
polytropic equation
of state $p=\tilde k\rho^{\gamma}$ for a comparison
\begin{eqnarray*}
\qquad\qquad
\Pi(R,\ \theta,\ t)\equiv\int_{\Delta z}p(R,\ \theta,\
z,\ t)\mathrm{d}z \ \qquad \leftrightarrow \ \qquad \Sigma(R,\
\theta,\ t)\equiv\int_{\Delta z}\rho(R,\ \theta,\ z,\ t)\mathrm{d}z\ ,
\end{eqnarray*}
where $\Pi$ is determined by vertically integrating the thermal gas 
pressure $p$. Given certain assumptions such as $\Pi\propto\Sigma 
p_c/\rho_c$ with $p_c$ and $\rho_c$ being the central pressure and 
density, a relationship between the two indices $\Gamma$ and $\gamma$ 
can be established as $\Gamma=3-2/\gamma$ (see equation 3.5 of Lemos 
et al. 1991).

The sound speed $v_s$ in a barotropic disc can be defined as
\begin{equation}\label{soundspeed}
v_s^2\equiv\frac{\mathrm{d} \Pi}{\mathrm{d}
\Sigma}=k\Gamma\Sigma^{\Gamma-1}=\Gamma\Pi/\Sigma
\end{equation}
and the enthalpy $H$ is expressed by
\begin{equation}\label{enthalpy}
H=\int\frac{\mathrm{d} \Pi}{\Sigma}=\frac{k\Gamma
\Sigma^{\Gamma-1}}{(\Gamma-1)}\ .
\end{equation}
The gravitational potential $\Phi(R,\ \theta,\ t)$
in the disc plane and the disc surface mass density
is related by the Poisson integral
\begin{equation}\label{Poisson}
\Phi(R,\ \theta,\ t)=\int\int\frac{-G\Sigma(r,\ \phi,\ t)r
\mathrm{d}r\mathrm{d}\phi}
{[r^2+R^2-2Rr \cos(\theta-\phi)]^{1/2}}\ ,
\end{equation}
where $G$ is the gravitational constant.

Among various model problems in disc dynamics, the scale-free disc
model is often picked up by theorists for its relative simplicity
and is explored as a powerful vehicle for a global analytical
analysis. Scale-free discs carry no characteristic spatial scales.
For a mathematical approach to scale-free discs, we adopt the
definition given by Lynden-Bell \& Lemos (1999). A flat disc is
scale-free if any physical quantity $q(R,\ \theta)$ of a disc
configuration has the following relationship
\begin{equation}\label{scalefreecondition}
q\left(\zeta R,\ \theta+h(\zeta)\right)
=A_1(\zeta)q(R,\ \theta)\ ,
\end{equation}
where $\zeta$ is a parameter, and $h(\zeta)$
and $A_1(\zeta)$ are two arbitrary functions.
By requirement (\ref{scalefreecondition}),
$q$ has a general solution form of
\begin{equation}\label{qform}
q=R^{\alpha}F(\theta+\mu\ln R)\ ,
\end{equation}
where $\alpha$ is complex, $\mu$ is real, and 
$F$ is an arbitrary function. The proof can 
be found in Lynden-Bell \& Lemos (1999).

Fluid equations provide a valid dynamical description for a gas disc
component, while distribution function equations are more appropriate
for a collisionless stellar disc component (e.g., Binney \& Tremaine
1987). After an introduction of the velocity dispersion of the stellar
disc component regarded as an effective `sound speed', fluid equations
can be approximately applied to stellar disc component as well (Lin \&
Shu 1964, 1966; Lin 1987; Binney \& Tremaine 1987; Bertin \& Lin 1996;
Lou \& Shen 2003; Shen \& Lou 2004a; Lou \& Zou 2004, 2006; Lou \& Wu
2005; Lou \& Bai 2006).

In treating a razor-thin scale-free disc system, we adopt
two-dimensional fluid equations. The three basic fluid equations
of describing razor-thin disc dynamics can be written as
\begin{equation}\label{continuity}
\frac{\partial\Sigma}{\partial t}+
\frac{1}{R}\frac{\partial}{\partial R}(R\Sigma v_R)
+\frac{1}{R}\frac{\partial}{\partial\theta}(\Sigma v_\theta)=0\ ,
\end{equation}
\begin{equation}\label{radialeuler}
\frac{\partial v_R}{\partial t}+ v_R\frac{\partial v_R}
{\partial R}+\frac{v_\theta}{R}\frac{\partial v_R}{\partial\theta}
-\frac{v_\theta^2}{R}=-\frac{1}{\Sigma}\frac{\partial\Pi}
{\partial R} -\frac{\partial (\Phi+\bar{\Phi})}{\partial R}
=-\frac{\partial (H+\Phi+\bar{\Phi})}{\partial R}\ ,
\end{equation}
\begin{equation}\label{azimuthaleuler}
\frac{\partial v_\theta}{\partial t}+ v_R\frac{\partial
v_\theta}{\partial R} +\frac{v_\theta}{R}\frac{\partial
v_\theta}{\partial \theta} +\frac{v_\theta v_R}{R}
=-\frac{1}{\Sigma R}\frac{\partial\Pi}{\partial\theta}
-\frac{\partial (\Phi+\bar{\Phi})}{R\partial\theta}
=-\frac{\partial (H+\Phi+\bar{\Phi})}{R\partial \theta}\ ,
\end{equation}
where $v_R$ is the radial bulk flow velocity and $v_\theta$ is 
the azimuthal bulk flow velocity and $\bar{\Phi}$ represents 
the axisymmetric background halo potential presumed to be 
unperturbed, i.e., $\bar{P}$ parameter remains unchanged in 
equation (\ref{backgroundpowerlaw}). The last equalities in 
equations (\ref{radialeuler}) and (\ref{azimuthaleuler})
corresponds to the barotropic condition.

In order to meet the scale-free requirement (Syer \& Tremaine 1996;
Goodman \& Evans 1999; Shen \& Lou 2004a; Shen, Liu \& Lou 2005),
we presume relevant physical quantities in the following forms of
\begin{eqnarray}\nonumber
\qquad\qquad \Sigma=R^{-2\beta-1}S(\varphi)\ , \quad\qquad
v_R=R^{-\beta}a(\varphi)\ , \quad\qquad
v_\theta=R^{-\beta}b(\varphi)\ ,
\\ \label{backgroundpowerlaw}
\qquad\qquad \Phi=-R^{-2\beta}P(\varphi)\ , \quad\qquad
\bar\Phi=-R^{-2\beta}\bar{P}\ , \quad\qquad
H=R^{-2\beta}Q(\varphi)\ ,
\end{eqnarray}
where we define $\varphi(R,\ \theta)\equiv\theta+\mu\ln R$ 
and $\beta$ is a scale-free index. Both $\mu$ and $\beta$ 
are constant real coefficients here.

To satisfy the scale-free condition in equation (\ref{enthalpy}),
we should require
\begin{equation}\label{Gamma}
(-2\beta-1)(\Gamma-1)=-2\beta\quad\Rightarrow\quad
\Gamma=(1+4\beta)/(1+2\beta)\ .
\end{equation}
The constraint $\Gamma>0$ for a warm disc ($k>0$) implies either
$\beta>-1/4$ or $\beta<-1/2$ and thus ensures $v_s^2>0$. For a 
cold disc ($k=0$), such a constraint is unnecessary.

Other constraints on $\beta$ come from the disc mass distribution.
First, $\Sigma$ is assumed to decrease with increasing $R$ and 
this implies $\beta>-1/2$ by the first relation in equation 
(\ref{backgroundpowerlaw}). Secondly, the central mass should 
be finite. This simply means
\begin{eqnarray*}
\qquad\qquad\lim_{r\rightarrow0}\int_0^r\int_0^{2\pi}\Sigma(R,
\theta)R \mathrm{d}R \mathrm{d}\theta=
\lim_{r\rightarrow0}\int_0^r\int_0^{2\pi}R^{-2\beta}S(\theta+\mu\ln
R) \mathrm{d}R \mathrm{d}\theta
=\int_0^{2\pi}S(\theta)\mathrm{d}\theta
\lim_{r\rightarrow0}\left.\frac{R^{1-2\beta}}{1-2\beta}\right|_0^r
<+\infty\ .
\end{eqnarray*}
We then require $1-2\beta>0$, indicating $\beta<1/2$.
In short, we therefore have $-1/2<\beta<1/2$ based
on the requirement of the disc mass distribution.

Moreover, the allowed range of $\beta$ is constrained by the
convergence of the Poisson integral relating $\Sigma$ and
$\Phi$. According to the analytical results of Qian (1992),
for $\Sigma(R,\theta)=R^{-2\beta-1+\mathrm{i}\nu}
\exp({\mathrm{i}m\theta})=R^{-2\beta-1}\exp[{\mathrm{i}
(m\theta+\nu\ln R)}]$ with $\nu$ being a real constant
parameter, we have correspondingly
\begin{equation}\label{potential}
\Phi(R, \theta)=-GY_m\left(\beta-\mathrm{i}\nu/2\right)
R^{-2\beta+\mathrm{i}\nu}\exp({\mathrm{i}m\theta})\ ,
\end{equation}
where $Y_m(x)$ is defined in terms
of the standard $\Gamma$ function as
\begin{equation}\label{Ym}
Y_m(x) \equiv\frac{\pi\displaystyle\Gamma({m}/{2}-x+{1}/{2})
\Gamma({m}/{2}+x)}{\displaystyle\Gamma({m}/{2}-x+1)
\Gamma({m}/{2}+x+{1}/{2})}\ .
\end{equation}
The above expression remains valid only for $-m/2<\beta<(m+1)/2$;
outside this range of $\beta$, the force from gas materials at
either $r\rightarrow\infty$ or $r\rightarrow0$ diverges (Qian 1992). 
When $m\geq1$, potential expression (\ref{potential}) remains valid 
for $\beta\in (-1/2,\ 1/2)$. For the $m=0$ case, although the 
potential $\Phi$ in equation (\ref{potential}) converges only for 
$\beta\in (0,\ 1/2)$, we can extend the valid range to $\beta\in 
(-1/2, 1/2)$ if we only make use of $\nabla\Phi$ to compute the 
force, which suffices for our purpose (Syer \& Tremaine 1996).

We frequently refer to the Kalnajs function (Kalnajs
1971; Shu et al. 2000) which is explicitly defined by
\begin{equation}\label{Kalnajs}
{\mathcal N}_m(\nu)\equiv
K(\nu,m)=\frac{1}{2}\frac{\Gamma[(m+1/2+\mathrm{i}\nu)/2]
\Gamma[(m+1/2-\mathrm{i}\nu)/2]}
{\Gamma[(m+3/2+\mathrm{i}\nu)/2]\Gamma[(m+3/2-\mathrm{i}\nu)/2]}
=\frac{Y_m\left({1}/{4}-\mathrm{i}\nu/2\right)}{2\pi}\ .
\end{equation}
For $m\neq0$ and setting $\mu\equiv\nu/m$, we write
$\exp[{\mathrm{i}(m\theta+\nu\ln R)}]=\exp({\mathrm{i}m\varphi})$.
For $m=0$, expression $\exp({\mathrm{i}\nu\ln R})$ should replace
expression $\exp({\mathrm{i}m\varphi})$ in treating axisymmetric 
perturbations (Syer \& Tremaine 1996). We write perturbations in 
the form of $\exp[{\mathrm{i}(m\theta+\nu\ln R)}]$ instead of 
$\exp({\mathrm{i}m\varphi})$ throughout our analysis unless 
otherwise stated.

We note that from the scale-free condition (Lynden-Bell \&
Lemos 1999), there is no a priori reason that the scale-free index
$\alpha$ (see equation \ref{qform}) should be necessarily real.
However, the imaginary part $\Im(\alpha)$ will make such term as
$\exp[{\mathrm{i}\Im(\alpha)\ln R}]$ oscillating around 0. For
example, the surface mass density $\Sigma=R^\alpha$ could be
negative when $\Im(\alpha)\neq0$. From the physical perspective,
only a perturbation term may have such non-zero imaginary part
of the scale-free index $\alpha$.

Finally, we conclude that the gravitational potential of a
scale-free disc is given by expression (\ref{potential}) and
the constraint for scale-free index $\beta$ is $\beta\in(-1/2,
\ 1/2)$ for cold discs (i.e., $k=0$) and $\beta\in(-1/4,\ 1/2)$
for warm discs (i.e., $k>0$).

\section{An Isopedic Magnetic Field across a Scale-Free Thin Disc}

The generation, amplification, and activities of magnetic field are
widely believed to be caused by nonlinear MHD dynamo processes in an
electrically conducting gas medium (e.g., Parker 1979; Moffatt 2000; Sofue
et al. 1996; Beck et al. 1996). In contexts of star formation, numerical
simulations (e.g., Nakano 1979; Lizano \& Shu 1989; Basu \& Mouschovias
1994) lend support to the notion that the mass-to-magnetic flux ratio
$\Lambda$ may be constant in both space and time. In Shu \& Li (1997),
$\Lambda$ parameter was assumed to be a constant and two important
theorems were derived based on this assumption. From the ideal nonlinear
MHD equations, Lou \& Wu (2005) demonstrated in a straightforward manner
that a constant $\Lambda$ is a natural consequence of the frozen-in
condition for the magnetic flux.

As in Section 2, the cylindrical coordinate system $(R, \theta,
z)$ is adopted to describe the dynamics in an infinitely
conducting razor-thin disc coincident with the $z=0$ plane. This
razor-thin disc is threaded across by an `isopedic' magnetic field
that exists for $|z|>0$ in vacuum and is almost vertical to the
disc plane at $z=0$. As a natural consequence of the standard
ideal nonlinear MHD equations (Lou \& Wu 2005), we demonstrates
the magnetic field to be isopedic (Shu \& Li 1997; Shu et al.
2000), with a spatially constant dimensionless ratio $\lambda$ 
of the mass per unit area $\Sigma (R,\ \theta,\ t)$ to the 
magnetic flux per unit area $B_z(R,\ \theta,\ t)$ (perpendicular 
component of the magnetic field evaluated at $z=0$), namely
\begin{equation}\label{lambda}
\lambda=2\pi G^{1/2}{\Sigma}/{B_z}
=2\pi G^{1/2}\Lambda=\mathrm{constant}\ ,
\end{equation}
where the factor $2\pi G^{1/2}$ makes $\lambda$ parameter dimensionless.

Such kind of frozen-in isopedic magnetic field adds no extra
complications to the hydrodynamic treatment of stationary
perturbations in scale-free razor-thin discs for a global analysis.
From the two MHD theorems established by Shu \& Li (1997), we
introduce the following transformation
for an isopedically magnetized razor-thin scale-free disc
\begin{equation}\label{transformation}
\Phi\rightarrow \epsilon \Phi\ ,
\qquad\qquad
\Pi\rightarrow\Theta\Pi\ ,
\end{equation}
where parameter $\epsilon$ is the dilution factor for the
disc self-gravity caused by the magnetic tension force,
\begin{equation}\label{epsilon}
\epsilon\equiv 1-{1}/{\lambda^2}\ ,
\end{equation}
and $\Theta$ is the enhancement factor for the thermal
gas pressure caused by the magnetic pressure force,
\begin{equation}\label{Theta}
\Theta=1+{(1+\eta^2)}/{(\lambda^2+\eta^2)}\ .
\end{equation}
In expression (\ref{Theta}), parameter $\eta$ is defined
as the ratio of the horizontal gravity $g_\parallel$
to the vertical gravity $2\pi G\Sigma$ (in absolute
values) just above or below the disc plane,
\begin{equation}\label{eta}
\eta\equiv {|g_\parallel|}/{(2\pi G \Sigma)}\ .
\end{equation}
For an unmagnetized razor-thin scale-free gas disc, we simply have
$\lambda\rightarrow\infty$ and set $\epsilon=1$ and $\Theta=1$.
If $\Theta$ is independent of $R$ and $\theta$ or if such
dependence is nonessential and ignorable for simplicity, the
replacement of $\Pi\rightarrow\Theta\Pi$ is equivalent to
$H\rightarrow\Theta H$ by definition (\ref{enthalpy}). For an
isolated scale-free disc sustained by its own self-gravity,
the disc equilibrium would require $|\lambda|\geq 1$ and thus
$\epsilon>0$. However, the existence of a massive background 
dark matter halo potential would allow $|\lambda|<1$ for a 
disc equilibrium and hence $\epsilon<0$. This latter 
possibility leads to potentially interesting consequences.

For a scale-free disc, it is possible to determine $\eta$ value more
accurately. Considering an axisymmetric scale-free disc equilibrium,
we take $\beta\in(-1/2,\ 1/2)$ and $\Sigma=cR^{-2\beta-1}$ with $c$
being a constant coefficient. By the analysis of Qian (1992), we 
have the corresponding gravitational potential in the form of
\begin{eqnarray*}
\Phi=-cGR^{-2\beta}
\frac{\pi\Gamma(1/2-\beta)\Gamma(\beta)}
{\Gamma(1-\beta)\Gamma(1/2+\beta)}
=-cGR^{-2\beta}Y_0(\beta)\ .
\end{eqnarray*}
The horizontal gravitational force $g_\parallel$
tangential to the disc plane is given by
\begin{eqnarray*}
g_\parallel=-\nabla_\parallel\Phi
=-\left(\hat{\bf{e}}_R\frac{\partial}{\partial
R}+\frac{\hat{\bf{e}}_\theta}{R}\frac{\partial}{\partial
\theta}\right)\Phi
=-2cGR^{-2\beta-1}\beta Y_0(\beta)\hat{\bf{e}}_R\ .
\end{eqnarray*}
We therefore obtain
\begin{equation}\label{eta-axisymmetric}
\eta\equiv\frac{|g_\parallel|}{2\pi G
\Sigma}=\frac{2cGR^{-2\beta-1}\beta Y_0(\beta)}{2\pi cG
R^{-2\beta-1}}=\frac{\beta Y_0(\beta)}{\pi}\ ,
\end{equation}
leading to the following conclusions: $\eta\in (0,\ +\infty)$
when $\beta\in (-1/2,\ 1/2)$ and $\eta$ grows as $\beta$ grows.
Note that $\eta=1$ when $\beta=0$ corresponds to a singular
isothermal disc (SID; e.g., Shu et al. 2000; Lou 2002; Lou \&
Shen 2003; Shen \& Lou 2004; Lou \& Zou 2004, 2006).

Effects of an isopedic magnetic field on a razor-thin scale-free
disc are summarized below. The horizontal self-gravity force
$g_\parallel$ and the two-dimensional thermal gas pressure
$\Pi$ of a scale-free disc will be effectively modified as
\begin{equation}\label{modified}
g_\parallel\rightarrow \epsilon g_\parallel\ ,
\qquad\qquad
\Pi\rightarrow\Theta\Pi\ ,
\end{equation}
respectively. Or equivalently, we can 
express such MHD modifications as
\begin{equation}\label{furthermodified}
\Phi\rightarrow \epsilon\Phi\ , \qquad\qquad H\rightarrow\Theta H\
.
\end{equation}

\section{The Stationary Axisymmetric Rotational MHD Equilibrium}

We first specify the stationary axisymmetric rotational MHD
equilibrium for a razor-thin scale-free disc with an isopedic
magnetic field. This requires that every physical quantity 
does not depend on time $t$ and azimuthal angle $\theta$. 
In the following, we use subscript `0' to denote physical 
variables in the stationary axisymmetric equilibium. For a 
rotating disc, we also require $v_{R0}=0$. Using these 
constraints in equations 
$(\ref{continuity})-(\ref{azimuthaleuler})$, we see that 
equations (\ref{continuity}) and (\ref{azimuthaleuler}) 
are satisfied automatically. The radial force balance
(\ref{radialeuler}) appears in the form of
\begin{equation}\label{equili}
\frac{v_{\theta 0}^2}{R} =\frac{\mathrm{d} (\Theta
H_0+\epsilon\Phi_0+\bar{\Phi}_0)}{\mathrm{d} R}\ .
\end{equation}
Naturally, condition (\ref{equili}) represents that the
centrifugal force, thermal pressure force, magnetic tension 
and pressure forces, the gravities from the disc and the 
background dark matter halo achieve a force equilibrium in 
the radial direction. The self-gravity potential $\Phi_0$ 
and the enthalpy $H_0$ are respectively replaced by $\epsilon 
\Phi_0$ and $\Theta H_0$ such that the effects of an isopedic 
magnetic field are included [see expression (\ref{furthermodified})].


We first take $\Sigma_0=c_0 R^{-2\beta-1}$ and $v_{\theta 0}=b_0
R^{-\beta}$, where $c_0$ and $b_0$ are two positive constants.
Then the enthalpy becomes
\begin{equation}\label{H0}
H_0=\frac{k\Gamma}{(\Gamma-1)}\Sigma_0^{\Gamma-1}
=\frac{(1+4\beta)}{2\beta}
kc_0^{{2\beta}/{(1+2\beta)}}R^{-2\beta}\ ,
\end{equation}
and the corresponding self-gravity potential appears as
\begin{equation}\label{Phi0}
\Phi_0=-Gc_0Y_0(\beta)R^{-2\beta}\ .
\end{equation}
Since the gravitational potential of the background dark matter
halo is assumed to be axisymmetric ($\bar{P}=\mathrm{const}$)
and unperturbed ($\bar\Phi=\bar\Phi_0$) in the presence of disc
perturbations, we define the ratio of the background dark matter
halo potential to the disc self-gravity potential as
\begin{equation}\label{f}
f\equiv {\bar\Phi}/{\Phi_0}={\bar\Phi_0}/{\Phi_0}\ .
\end{equation}
Here, $f$ is a measure for the dark matter halo effect. The larger
the $f$ value, the more massive a dark matter halo is.

The MHD radial force balance as described by equation
(\ref{equili}) then becomes
\begin{equation}\label{equili2}
b_0^2R^{-2\beta-1}=\frac{\mathrm{d}(\Theta
H_0+\epsilon\Phi_0+\bar{\Phi}_0)}{\mathrm{d} R}=
-2\beta\left[\frac{(1+4\beta)}{2\beta}\Theta
kc_0^{{2\beta}/{(1+2\beta)}}-(\epsilon+f)Gc_0Y_0(\beta)\right]R^{-2\beta-1}\
.
\end{equation}
The background rotational MHD equilibrium simply requires the
following relationship
\begin{equation}\label{equili3}
b_0^2=2\beta Gc_0Y_0(\beta)(\epsilon+f)-(1+4\beta)\Theta
kc_0^{2\beta/(1+2\beta)}\ .
\end{equation}
According to definition (\ref{soundspeed}), the sound speed
$v_{s0}$ in the axisymmetric  background disc is explicitly given
by
\begin{equation}\label{soundspeedwithisopedic}
v_{s0}^2= k\Gamma\Sigma_0^{\Gamma-1}=\frac{(1+4\beta)}{(1+2\beta)}
kc_0^{2\beta/(1+2\beta)}R^{-2\beta}\ ,
\end{equation}
varying with $R$. A parameter $w$ is defined as the square of the
ratio of the sound speed $v_{s 0}$ to the background rotation
speed $v_{\theta 0}$, which is the inverse square of the
rotational Mach number
\begin{equation}\label{w}
w\equiv\frac{v_{s 0}^2}{v_{\theta 0}^2}
=\frac{(1+4\beta)}{(1+2\beta)}
\frac{kc_0^{2\beta/(1+2\beta)}}{b_0^2}\ .
\end{equation}
Physically, $w$ is a measure for the disc temperature. A larger 
$w$ value means that the disc temperature is higher, corresponding
to a more random thermal motion in reference to the regular disc
rotation.

As a normalization, we set the radial self-gravity force in the
axisymmetric disc be $-1$ at $R=1$. Since the radial self-gravity
force in the disc plane $g_\parallel=-2\beta
Gc_0Y_0(\beta)R^{-2\beta-1}\hat{\bf{e}}_R$, we then require
\begin{equation}\label{force1}
2\beta Gc_0Y_0(\beta)=1\ .
\end{equation}
Finally, the axisymmetric MHD disc equilibrium equation
(\ref{equili3}) can be simply written as
\begin{equation}\label{equili4}
b_0^2=\epsilon+f-b_0^2(1+2\beta)\Theta w\ \qquad \mbox{ or }
\qquad b_0^2=\frac{(\epsilon+f)}{[1+(1+2\beta)\Theta w]}\ .
\end{equation}
By the very definitions, both $b_0^2$ and $w$ should be positive.
For a physical background dark matter halo potential, it is clear
that $f\geq0$ for an attractive background halo potential. We now
demonstrate the physical requirement $f+\epsilon\geq0$. When the
disc is gravitationally isolated ($f=0$), we should have
$\epsilon\geq0$, equivalent to $\lambda\geq1$ (see equation
\ref{epsilon}). The physical interpretation is that the isopedic
magnetic field should not be too strong to overtake the
self-gravity in an isolated scale-free disc, otherwise the disc
would be torn apart by the magnetic tension force. In the presence
of a background dark matter halo potential, the tension force of
the isopedic magnetic field is allowed to be stronger than the
disc self-gravity to the limit of $f+\epsilon>0$. The similar 
kind of effect can be seen for a composite disc system of two
gravitationally coupled discs with an isopedic magnetic field, 
in which one is a stellar disc and the other is a magnetized 
gas disc. The stellar disc acts as the background halo 
potential, whose gravity counteracts the isopedic magnetic 
tension force in the gas disc (Lou \& Wu 2005).

\section{Dispersion Relations for MHD Disc Perturbations}

We now introduce MHD perturbations in the axisymmetric 
equilibrium of a scale-free disc with subscript `$_1$' 
revealing the association with the perturbation of a 
relevant physical variable. More specifically, we write
\begin{eqnarray}\label{perturb}
\Sigma=\Sigma_0+\Sigma_1(R,\theta, t)\ ,\ v_R=v_{R1}(R,\theta, t)\
,\ v_\theta=v_{\theta 0}+v_{\theta 1}(R,\theta, t)\ , \
\Phi=\Phi_0+\Phi_1(R,\theta, t)\ ,\ \bar{\Phi}=\bar{\Phi}_0\ ,\
H=H_0+H_1(R,\theta, t)\ ,
\end{eqnarray}
respectively. Meanwhile, the magnetic flux should be also
perturbed following the surface density perturbation. By
the frozen-in condition, the ratio of the surface mass
density to the magnetic flux remains unchanged at all times.
That is, $\lambda$ remains constant and thus $\epsilon$
should be unperturbed (see equation \ref{epsilon}).
For disc perturbations, changes in $\eta$ would lead to
variations in $\Theta$; following Shu et al. (2000), we
presume that such induced variations of $\Theta$ are
nonessential and are thus ignored.

Substituting expression (\ref{perturb}) into the three basic MHD
equations $(\ref{continuity})-(\ref{azimuthaleuler})$, we readily
come to the linearized MHD perturbation equations in the forms of
\begin{equation}\label{linear1}
\frac{\partial \Sigma_1}{\partial t}+
\frac{1}{R}\frac{\partial}{\partial R} (R\Sigma_0v_{R1})
+\frac{\partial}{R\partial\theta}
(\Sigma_0v_{\theta1}+\Sigma_1v_{\theta0})=0\ ,
\end{equation}
\begin{equation}\label{linear2}
\frac{\partial v_{R1}}{\partial t} +\frac{v_{\theta
0}}{R}\frac{\partial v_{R1}}{\partial \theta} -\frac{2v_{\theta
0}v_{\theta 1}}{R} =-\frac{\partial (\Theta
H_1+\epsilon\Phi_1)}{\partial R}\ ,
\end{equation}
\begin{equation}\label{linear3}
\frac{\partial v_{\theta 1}}{\partial t}+ v_{R1}\frac{\partial
v_{\theta 0}}{\partial R} +\frac{v_{\theta 0}}{R} \frac{\partial
v_{\theta 1}}{\partial \theta} +\frac{v_{\theta 0}v_{R1}}{R}
=-\frac{\partial (\Theta H_1+\epsilon\Phi_1)}{R\partial \theta}\ .
\end{equation}
As both $\Phi_1$ and $H_1$ can be expressed in terms of $\Sigma_1$
together with the background equilibrium condition, we first take
\begin{equation}
\label{sigma1} \Sigma_1(R, \theta, t)=
c_1R^{-2\beta_1-1}\exp[{\mathrm{i}(\omega t+m\theta+\nu\ln R)}]
=c_1R^{-2\beta'-1}\exp[{\mathrm{i}(\omega t+m\theta)}]\ ,
\end{equation}
where $c_1$ specifies a small amplitude coefficient and
\begin{equation}\label{beta'}
\beta'\equiv\beta_1-i\nu/2\ .
\end{equation}
In expression (\ref{sigma1}) , $\exp({\mathrm{i}m\theta})$ and
$\exp({\mathrm{i}\nu\ln R})$ represent the azimuthal and radial
perturbations, respectively. As there is no a priori reason to
force the scale-free index of the surface density perturbation 
to be the same as that of the background equilibrium, we allow 
$\beta_1$ and $\beta$ to be different, representing constant 
scale-free indices of the perturbation and background surface 
mass densities respectively.

We then take both radial and azimuthal bulk flow velocity
perturbations $v_{R1}$ and $v_{\theta 1}$ in the form of
\begin{equation}\label{vr1vtheta1}
{\mathcal A}(R)\exp[{\mathrm{i}(\omega t+m\theta)}]\ ,
\end{equation}
where ${\mathcal A}(R)$ is a function of $R$ to be
specified by equations (\ref{vr1}) and (\ref{vtheta1})
for $v_{R1}$ and $v_{\theta 1}$, respectively.


With equation (\ref{sigma1}) and the Poisson integral, we obtain
(Qian 1992)
\begin{equation}\label{Phi1}
\Phi_1
=-Gc_1Y_m(\beta')R^{-2\beta'}\exp[{i(\omega t+m\theta)}]\ ,
\end{equation}
where $\beta'$ and $Y_m(u)$ are defined by equations 
(\ref{beta'}) and (\ref{Ym}). Properties of $Y_m(u)$ 
are summarized in Appendix A.

For $H_1$ being a perturbation of 
enthalpy $H$, we readily obtain 
\begin{equation}\label{H1}
H_1
=k\Gamma\Sigma_0^{\Gamma-2}\Sigma_1
=k\frac{(1+4\beta)}{(1+2\beta)}c_0^{-1/(1+2\beta)}c_1
R^{-2\beta'}\exp[{i(\omega t+m\theta)}]
=\frac{b_0^2wc_1}{c_0}R^{-2\beta'}\exp[{i(\omega t+m\theta)}]\ .
\end{equation}
As $\beta_1$ is only constrained by the surface mass distribution,
we require $\beta_1\in(-1/2,\ 1/2)$ for either warm or cold
discs.

Here, the angular speed $\Omega_0$ and the epicyclic frequency
$\kappa_0$ of the axisymmetric background are given explicitly by
\begin{equation}\label{angularspeed}
\Omega_0={v_{\theta 0}}/{R}=b_0R^{-\beta-1}\ , \qquad\qquad
\kappa_0^2=R{\mathrm{d}\Omega_0^2}/{\mathrm{d}R}+4\Omega_0^2\
=2(1-\beta)\Omega_0^2=2b_0^2(1-\beta)R^{-2\beta-2}\ .
\end{equation}
After substituting expressions (\ref{sigma1}) and
(\ref{vr1vtheta1}) into equations $(\ref{linear1})-
(\ref{linear3})$, we readily obtain
\begin{equation}\label{linear4}
i(\omega+m\Omega_0)\Sigma_1+ \frac{1}{R}\frac{\partial}{\partial
R}(R\Sigma_0v_{R1}) +\frac{imv_{\theta1}\Sigma_0}{R} =0\ ,
\end{equation}
\begin{equation}\label{linear5}
i(\omega+m\Omega_0) v_{R1} -2\Omega_0 v_{\theta 1}
=-\frac{\partial (\Theta H_1+\epsilon\Phi_1)}{\partial R}\ ,
\end{equation}
\begin{equation}\label{linear6}
i(\omega+m\Omega_0) v_{\theta 1}+
\frac{\kappa_0^2}{2\Omega_0}v_{R1}=-\frac{im(\Theta
H_1+\epsilon\Phi_1)}{R}\ .
\end{equation}
Recombining equations (\ref{linear5}) and (\ref{linear6}), 
we obtain
\begin{equation}\label{vr1express}
v_{R1}=\frac{i}{[(\omega+m\Omega_0)^2-\kappa_0^2]}
\left[\frac{2m\Omega_0}{R}+(\omega+m\Omega_0)\frac{\partial}{\partial
R} \right](\Theta H_1+\epsilon\Phi_1)\ ,
\end{equation}
and
\begin{equation}\label{vtheta1express}
v_{\theta1}=-\frac{1}{[(\omega+m\Omega_0)^2-\kappa_0^2]}
\left[(\omega+m\Omega_0)\frac{m}{R}+\frac{\kappa_0^2}{2\Omega_0}
\frac{\partial}{\partial R}\right](\Theta H_1+\epsilon\Phi_1)\ .
\end{equation}
%
%
Substituting expressions (\ref{sigma1}), (\ref{Phi1}), (\ref{H1}),
(\ref{vr1express}) and (\ref{vtheta1express}) into equation
(\ref{linear4}), we immediately arrive at
\begin{eqnarray}\nonumber
\qquad (\omega+m\Omega_0)R^{-2\beta'}+\left[w\Theta b_0^2
-\epsilon c_0 GY_m(\beta')\right] \frac{\mathrm{d}}{\mathrm{d} R}
\left\{\frac{R^{-2\beta}}{[(\omega+m\Omega_0)^2-\kappa_0^2]}
\left[\frac{2m\Omega_0}{R}+(\omega+m\Omega_0)\frac{\mathrm{d}}{\mathrm{d}
R} \right]\right\}R^{-2\beta'}
\\ \label{dispersionall}
-\left[w\Theta b_0^2 -\epsilon c_0 GY_m(\beta')\right]
\frac{mR^{-2\beta-1}}{[(\omega+m\Omega_0)^2-\kappa_0^2]}
\left[(\omega+m\Omega_0)\frac{m}{R}+\frac{\kappa_0^2}{2\Omega_0}
\frac{\mathrm{d}}{\mathrm{d} R}\right]R^{-2\beta'} =0\ ,
\end{eqnarray}
where $\Omega_0$ and $\kappa_0$ are defined in equation
(\ref{angularspeed}). As the angular frequency $\omega$ is a
constant, we derive a simpler form of equation
(\ref{dispersionall}) by writing $\hat\omega\equiv\omega
R^{\beta+1}$, namely
\begin{eqnarray}
\begin{split}
\frac{\hat\omega+mb_0}{w\Theta b_0^2 -\epsilon c_0 GY_m(\beta')}=
\frac{mb_0\left[m^2+2(\beta+\beta'-2\beta'^2)\right]
+\hat\omega\left[m^2-2\beta'(2\beta'-1)\right]}
{(\hat\omega+mb_0)^2-2(1-\beta)b_0^2} \label{dispersionfinal}
\qquad\qquad\\ \qquad\qquad
+\frac{4(1+\beta)\left[mb_0-\beta'(\hat\omega+mb_0)\right]
\hat\omega(\hat\omega+mb_0)}
{\left[(\hat\omega+m b_0)^2-2(1-\beta)b_0^2\right]^2}\ .
\end{split}
\end{eqnarray}
Equation (\ref{dispersionfinal}) can be viewed as the {\it local}
dispersion relation for time-dependent MHD perturbations in a
scale-free razor-thin disc because of the explicit $R-$dependence
of $\hat \omega$. Apparently, it is a complex quintic equation in 
terms of $\hat\omega$. According to the Abel Impossibility 
Theorem\footnote{A general polynomial equation of order higher 
than the fourth degree cannot be solved algebraically in terms of 
finite additions, multiplications and root extractions (e.g., 
Hungerfort 1997).}, we cannot solve for $\hat\omega$ analytically 
from equation (\ref{dispersionfinal}) generally. Nevertheless, 
this does not prevent straightforward numerical solutions.

The classical Wentzel-Kramers-Brillouin-Jeffreys (WKBJ) relationship
in the so-called tight-winding regime can be readily recovered with
certain simplifications and approximations. In the parameter regime
of $\Theta w\ll1$,
$\nu\gg m$ and $\omega\rightarrow0$, dispersion relation
(\ref{dispersionfinal}) naturally reduces to
\begin{equation}\label{WKBJ1}
(\hat\omega+mb_0)^2=2(1-\beta)b_0^2+w\Theta b_0^2
\nu^2-2\pi|\nu|\epsilon c_0 G\ .
\end{equation}
By defining the radial wavenumber $k$ as $\nu/R$, the WKBJ
dispersion relation (\ref{WKBJ1}) can be cast into the form of
\begin{equation}\label{WKBJ}
(\omega+m\Omega_0)^2=\kappa_0^2+k^2\Theta v_{s 0}^2-2\pi\epsilon
\Sigma_0 G|k|\ ,
\end{equation}
(see Lou \& Fan 1998a for fast MHD density waves) which is
essentially the same as equation (39) of Shu et al. (2000).
This consistent correspondence to the classical WKBJ
dispersion relation is a necessary check.
Equation (\ref{dispersionfinal}) can be viewed as a more
comprehensive local description for MHD perturbations in 
a scale-free disc, in comparison to the standard WKBJ
approximation. Detailed derivation procedures and formulae
can be found in both Appendices B and A.

By equation (\ref{epsilon}), $\epsilon$ is a function of $\lambda$
and by equations (\ref{Theta}) and (\ref{eta-axisymmetric}),
$\Theta$ is a function of both $\lambda$ and $\beta$. From equation
(\ref{force1}) that normalizes the tangential gravity force in the
disc plane, we determine $c_0$ by a specified $\beta$ value. According
to equation (\ref{equili4}), $b_0$ parameter can be expressed in
terms of $f$, $w$, $\beta$ and $\lambda$. We thus formally write
out the solution of dispersion relation (\ref{dispersionfinal}) as
$\hat\omega={\mathcal F}(m,\ \nu,\ \beta_1;\ \beta,\ f,\ w,\
\lambda)$, yielding $\omega=R^{-\beta-1}{\mathcal F}(m,\ \nu,\
\beta_1;\ \beta,\ f,\ w,\ \lambda)$, where parameters $m$ (an
integer for the azimuthal wavenumber), $\nu$ (for the radial
wavenumber) and $\beta_1$ (the scale-free index for perturbations)
characterize MHD perturbations, while parameters $\beta$ (the
scale-free index for the background), $f$ (the indicator for the
background dark matter halo), $w$ (the indicator for the disc
temperature) and $\lambda$ (the indicator for the magnetic field
strength) specify the axisymmetric background rotational MHD
equilibrium. As dispersion relation (\ref{dispersionfinal}) is a
complex polynomial,
MHD perturbations in a disc are stable only if $\omega$ or
$\mathcal F$ is real.

In order to derive corresponding forms of $v_{R1}$ and $v_{\theta1}$
from equation (\ref{vr1vtheta1}), we directly substitute expressions
(\ref{sigma1}), (\ref{Phi1}) and (\ref{H1}) into equations
(\ref{vr1express}) and (\ref{vtheta1express}) to explicitly obtain
\begin{eqnarray}\label{vr1}
\qquad v_{R1}&=&\frac{c_1}{c_0}\left[w\Theta b_0^2 -\epsilon c_0
GY_m(\beta')\right]\frac{2\mathrm{i}[mb_0(1-\beta')-\beta'\hat\omega]}
{[(mb_0+\hat\omega)^2-2(1-\beta)b_0^2]}R^{-2\beta'+\beta}
\exp[{\mathrm{i}(\omega t+m\theta)}]\ ,
\\ \label{vtheta1}
\qquad v_{\theta1}&=&-\frac{c_1}{c_0}\left[w\Theta b_0^2-\epsilon
c_0GY_m(\beta')\right]\frac{[m^2-2\beta'(1-\beta)]b_0+m\hat\omega}
{[(mb_0+\hat\omega)^2-2(1-\beta)b_0^2]}
R^{-2\beta'+\beta}\exp[{\mathrm{i}(\omega t+m\theta)}]\ .
\end{eqnarray}
Because $\hat\omega$ in expressions (\ref{vr1}) and (\ref{vtheta1})
contains the $R-$dependence of $\hat\omega=\omega R^{\beta+1}$,
both velocity perturbation components $v_{R1}$ and $v_{\theta1}$
no longer take a simple power-law form in $R$ besides the wave
factor $\exp[{\mathrm{i}(\omega t+m\theta+\nu\ln R)}]$.

A special subcase of equation (\ref{dispersionfinal}) is the
axisymmetric case with $m=0$, from which we immediately have
\begin{equation}\label{axisymmetric}
\hat\omega\left\{\frac{1}{w\Theta b_0^2
-\epsilon c_0GY_0(\beta')}
+\frac{2\beta'(2\beta'+2\beta+1)\hat\omega^2
-4\beta'(2\beta'-1)(1-\beta)b_0^2}
{\left[\hat\omega^2-2(1-\beta)b_0^2\right]^2}\right\}=0\ .
\end{equation}
Equation (\ref{axisymmetric}) can be solved analytically and obviously
$\omega=0$ is one of the solutions. In discussing the axisymmetric
case, we therefore adopt $\omega\rightarrow 0$ instead of $\omega=0$,
for the stationary condition in order to avoid the singularity caused
by $\omega=0$.

Having admitted our limited capability to study time-dependent
local dispersion relation (\ref{dispersionfinal}) analytically,
we now explore the global stationary perturbation configurations,
which is simpler yet important (Shu et al. 2000; Lou 2002; Shen
\& Lou 2004a; Shen, Liu \& Lou 2005; Lou \& Bai 2006). Stationary
MHD perturbation configurations or zero-frequency neutral MHD
perturbation modes correspond to $\omega=0$ in dispersion
relation (\ref{dispersionfinal}). For clarity, we emphatically
distinguish the axisymmetric and non-axisymmetric cases at this 
point.

For non-axisymmetric cases with $m\neq0$, we immediately have
\begin{equation}\label{nonaxisymmetric-stationary}
\frac{b_0^2}{w\Theta b_0^2 -\epsilon c_0 GY_m(\beta')}=
\frac{m^2+2(\beta+\beta'-2\beta'^2)} {m^2-2(1-\beta)}\ .
\end{equation}
With normalization (\ref{force1}), we readily cast equation
(\ref{nonaxisymmetric-stationary}) into the following form of
\begin{equation}\label{syertremaine-nonaxisymmetric}
\frac{\epsilon Y_m(\beta')}{2\beta Y_0(\beta)}= b_0^2\left[\Theta
w-\frac{m^2-2(1-\beta)}{m^2+2(\beta+\beta'-2\beta'^2)}\right] \ .
\end{equation}
Because of the symmetry in $m$ for $Y_m(\beta')=Y_{-m}(\beta')$ as
shown in Appendix A,
equation (\ref{nonaxisymmetric-stationary})
is valid for both $m>0$ and $m<0$
where $m$ is an integer. From now on, we focus 
on $m\geq0$ without loss of generality.

The axisymmetric case of $m=0$ is somewhat special. We easily see
the axisymmetric dispersion relation (\ref{axisymmetric}) is
satisfied for $\omega=0$, because zero is a valid solution for
$\omega$ when $m=0$. Alternatively, we let $\omega$ approach zero
($\omega\rightarrow 0$) but do not vanish ($\omega\neq0$). We adopt
this limiting procedure to deal with the axisymmetric case (Lou \&
Shen 2003; Shen \& Lou 2004b; Lou \& Zou 2004, 2006; Lou \& Wu 2005;
Lou \& Bai 2006). Using this axisymmetric limiting procedure, we
determine the global stationary axisymmetric MHD dispersion
relation in the form of
\begin{equation}\label{axisymmetric-stationary}
\frac{b_0^2}{[w\Theta b_0^2 -\epsilon c_0 GY_0(\beta')]}
=\frac{\beta'(2\beta'-1)}{(1-\beta)}\ .
\end{equation}
Using normalization (\ref{force1}), dispersion relation
(\ref{axisymmetric-stationary}) can be rewritten as
\begin{equation}\label{syertremaine-axisymmetric}
\frac{\epsilon Y_0(\beta')}{2\beta Y_0(\beta)}= b_0^2\left[\Theta
w-\frac{(\beta-1)}{\beta'(1-2\beta')}\right]\ .
\end{equation}

Equations (\ref{syertremaine-nonaxisymmetric}) and
(\ref{syertremaine-axisymmetric}) appear similar to the
nonaxisymmetric self-consistent relation (39) and the axisymmetric
self-consistent relation (57) of Syer \& Tremaine (1996)
respectively except for the two extra parameters $\epsilon$ and
$\Theta$ representing the effect of an isopedic magnetic field and
for a differently defined $\beta'$ parameter. In the absence of an
isopedic magnetic field with $\epsilon=1$ and $\Theta=1$ and with
the perturbation scale-free index $\beta_1$ being the same as the
background scale-free index $\beta$, equations
(\ref{syertremaine-nonaxisymmetric}) and
(\ref{syertremaine-axisymmetric}) respectively reduce to equation
(39) and (57) of Syer \& Tremaine (1996) precisely.


In the analysis of Syer \& Tremaine (1996), the global aligned and
logarithmic spiral perturbation configurations are integrated together
by the parameter $\nu$, i.e., $\nu=0$ for aligned cases and $\nu\neq 0$
for logarithmic spiral cases. However in their work, the axisymmetric
and nonaxisymmetric cases are handled in different approaches. Here as
a unification in our analysis, equation (\ref{dispersionfinal}) is
equally valid for axisymmetric ($m=0$) and nonaxisymmetric ($m\neq 0$)
perturbations, as well as for aligned ($\nu=0$) and logarithmic spiral
perturbations ($\nu\neq0$).
Therefore, equation (\ref{dispersionfinal}) becomes a very general and
rigorous dispersion relation for global stationary MHD perturbation
configurations and for axisymmetric stability of a MHD scale-free disc.

Under the stationarity condition, the two velocity components
$v_{R1}$ and $v_{\theta1}$ can be written as
\begin{eqnarray}\label{vr1stationary}
\qquad v_{R1}&=&\frac{c_1\left[w\Theta b_0^2 -\epsilon c_0
GY_m(\beta')\right]}{c_0b_0}\frac{2\mathrm{i}m(1-\beta')}
{[m^2-2(1-\beta)]}R^{-2\beta'+\beta}\exp({\mathrm{i}m\theta})\ ,
\\ \label{vtheta1stationary}
\qquad v_{\theta1}&=&-\frac{c_1\left[w\Theta b_0^2 -\epsilon c_0
GY_m(\beta')\right]}{c_0b_0}\frac{[m^2-2\beta'(1-\beta)]}
{[m^2-2(1-\beta)]} R^{-2\beta'+\beta}\exp({\mathrm{i}m\theta})\ .
\end{eqnarray}
When $m=0$, we have $v_{R1}=0$ according
to expression (\ref{vr1stationary})
indicating no radial velocity perturbation.

By convention, the real parts of expressions (\ref{vr1stationary})
and (\ref{vtheta1stationary}) represent the physical solutions for
$v_{R1}$ and $v_{\theta 1}$. In addition to the wave perturbation
factor $\exp[{\mathrm{i}(\omega t+m\theta+\nu\ln R)}]$, amplitudes
of $v_{R1}$ and $v_{\theta 1}$ carry the power-law of $R$ in the
form of $R^{-2\beta_1+\beta}$. For $\beta_1=\beta$, perturbation
velocity components $v_{R1}$ and $v_{\theta 1}$ share the same
power law of $R$ as $R^{-\beta}$, similar to the background power
law in $R$ such as expression (\ref{backgroundpowerlaw}).

In general, it would be natural to take $\beta_1=\beta$, which
can be explained by a direct proportionality between a MHD
perturbation and the background equilibrium. However, for
logarithmic spiral perturbations, $\beta_1=1/4$ can insure a
constant radial flux of angular momentum (Goldreich
\& Tremaine 1979; Syer \& Tremaine 1996; Fan \& Lou 1997; Shen
\& Lou 2004a). For aligned cases, there is no angular momentum
flux inwards or outwards. For logarithmic spiral cases, the
total radial angular momentum flux (including the gravity torque
$\Lambda_\mathrm{G}$, advective transport $\Lambda_\mathrm{A}$
and magnetic torque $\Lambda_\mathrm{B}$) is given by the sum
of $\Lambda_\mathrm{total}=\Lambda_\mathrm{G}+\Lambda_\mathrm{A}
+\Lambda_\mathrm{B}\propto R^{1-4\beta_1}$. Therefore, only for
$\beta_1=1/4$ can the radial angular momentum flux be independent
of $R$. Detailed procedures can be found in Appendix G.


Although dispersion relation (\ref{dispersionfinal}) is derived 
for MHD perturbations without the WKBJ approximation, it should be 
properly viewed as a local dispersion relation for $\omega\neq 0$, 
otherwise $\omega$ cannot remain constant but varies with $R$. As 
such, equation
(\ref{dispersionfinal}) is complex in general. Only under two special
circumstances can dispersion relation (\ref{dispersionfinal}) become
real. One is when $\nu=0$, while the other is when $\omega=0$ and
$\beta_1=1/4$. The first case of $\ \nu=0$ means no radial wave
oscillations and thus no radial flux of angular momentum,
while the second case of $\omega=0$ and $\beta_1=1/4$ assures a
constant radial flux of angular momentum (see Appendix G).
Furthermore, dispersion relation (\ref{dispersionfinal}) should be
regarded as a global stationary dispersion relation when $\omega=0$.
It might be suggestive of a link between whether the dispersion
relation (\ref{dispersionfinal}) is real or complex and whether the
radial angular momentum flux is constant or not, as no angular
momentum flux ($\nu=0$) can be seen as a special case of a constant
radial flux of angular momentum in $R$. It seems that a non-constant
radial flux of angular momentum makes dispersion relation
(\ref{dispersionfinal}) complex. It is a common choice to assume a 
constant angular momentum flux in $R$
(Goldreich \& Tremaine 1979; Lemos et al. 1991; Shu et al. 2000;
Lou \& Shen 2003; Shen \& Lou 2004a; Lou \& Wu 2005).

By imposing the stationary condition of $\omega\rightarrow0$,
dispersion relation (\ref{dispersionfinal}) represents a constraint
on the background parameters ($f$, $\bar w$, $\beta$, $\lambda$) and
the perturbation parameters ($m$, $\nu$, $\beta_1$) simultaneously.
While a real form of equation (\ref{dispersionfinal}) gives one
constraint, a complex form would then give two.
The additional constraint is due to the imaginary part of
equation (Syer \& Tremaine 1996), which gives a more strict
condition on possible parameters (see Figs. 2a and 2b below).

In the following analysis, we mainly focus on global stationary
MHD perturbation configurations. Since the $\Theta$ parameter
appears always together with $w$ in the dispersion relations, we
introduce a new variable combination $\bar w\equiv\Theta w$ for
simplicity. A natural interpretation for this can be related to
the very definition of $w$ for $v_{s0}^2/v_{\theta0}^2$. In other
words, $\bar w$ can be defined as
$(\sqrt{\Theta}v_{s0})^2/v_{\theta0}^2$ with $\sqrt{\Theta}v_{s0}$
representing the magnetosonic speed (Lou \& Wu 2005) and seen as a
measure of disc temperature and magnetic pressure together. We
emphasize here that by equation (\ref{Theta}), the value of
$\Theta$ parameter can be very large as $\eta\rightarrow0$ and
$\lambda\rightarrow0$. In disc galaxies, effects of magnetic
pressure and tension can be limited in a certain manner under
specific situations. From now on, we shall deal with $\bar
w=\Theta w$ throughout the remaining analysis.

Obviously, we should require $\bar w\geq0$ on the ground of
physics. As a repulsive background dark matter halo potential
is unphysical, we also demand $f\geq0$. According to equation
(\ref{equili4}), we need to impose $f+\epsilon\geq0$ to maintain
a background rotational MHD equilibrium of axisymmetry. In
constructing global stationary MHD perturbation configurations,
we keep in mind these three constraints, namely, $\bar w\geq0$,
$f\geq0\ $ and $f+\epsilon\geq0$.

For the perturbation scale-free index $\beta_1$, we have
$\beta_1\in(-1/2,\ 1/2)$, while for the background scale-free
index $\beta$, we have $\beta\in(-1/2,\ 1/2)$ for cold discs and
$\beta\in(-1/4,\ 1/2)$ for warm discs. But for the convenience of
analysis, we still use $\beta\in(-1/2,\ 1/2)$ for general
situations hereafter. One should keep in mind that when
$\beta\in(-1/2,\ -1/4]$, the variable $\bar w$ can only
take on zero value.

\section{Global Stationary Aligned Cases with $\nu=0$}

For aligned global MHD perturbation configurations, the isodensity
contours of perturbations are aligned in azimuth at various radii.
Physically, such aligned cases should be interpreted as purely
azimuthal propagation of MHD density waves (Lou 2002). Without the
constraint on the angular momentum flux transport and as an example
of illustration, we may take $\beta_1=\beta$ indicating that MHD
perturbations carry the same scale-free index as the background one.

\subsection{Axisymmetric Disturbances with $m=0$}

For axisymmetric configurations with $m=0$, expressions
(\ref{sigma1}), (\ref{vr1stationary}) and
(\ref{vtheta1stationary}) give the following results
\begin{eqnarray}
\Sigma_1=c_1R^{-2\beta-1}\ ,\qquad\qquad v_{R1}=0\ , \qquad\qquad
v_{\theta1}=-\frac{ c_1\left(2\beta w\Theta b_0^2
-\epsilon\right)}{2c_0b_0} R^{-\beta}\ ,
\end{eqnarray}
which are trivial in the sense of a perturbation rescaling for the
axisymmetric background rotational MHD equilibrium (Shu et al. 2000;
Lou 2002; Lou \& Shen 2003; Lou \& Zou 2004, 2006; Lou \& Wu 2005).

\subsection{Nonaxisymmetric Disturbances with $m>0$}

When $m>0$, equation (\ref{syertremaine-nonaxisymmetric})
can be readily rearranged into the form of
\begin{equation}\label{aligned}
\frac{\epsilon Y_m(\beta)}{2\beta Y_0(\beta)}= b_0^2\left[
\bar w-\frac{m^2-2(1-\beta)}{m^2+4\beta(1-\beta)}\right]=
\frac{(\epsilon+f)}{[1+(1+2\beta)\bar w]}\left[\bar w
-\frac{m^2-2(1-\beta)}{m^2+4\beta(1-\beta)}\right]\ ,
\end{equation}
where $\bar w=\Theta w$ and $b_0^2$ is determined by equation
(\ref{equili4}). For the range of $\beta\in(-1/2,\ 1/2)$, 
numerical computations show that $2\beta Y_0(\beta)$ remains 
always positive after taking limit at certain points, such as 
$\beta=0$. Furthermore, with the asymptotic form of $Y_m(\beta)
\rightarrow 2\pi/m$ as $m\rightarrow\infty$, it is guaranteed 
that $Y_m(\beta)>0$ for $m\geq1$ based on carefully designed 
numerical tests.

For simplicity of expressions, we introduce a handy parameter
\begin{equation}\label{Ca}
C_a\equiv\frac{[m^2-2(1-\beta)]}{[m^2+4\beta(1-\beta)]}\ .
\end{equation}


When the scale-free index $\beta$, the azimuthal wavenumber 
$m$ and the gravity dilution factor $\epsilon$ are specified, 
equation (\ref{aligned}) establishes a unique relation between 
the inverse square of the magnetosonic Mach number (denoted by 
$\bar w$ parameter) and the background gravitational potential 
ratio (represented by $f$ parameter) that sustains a global
nonaxisymmetric stationary MHD perturbation pattern.

\begin{figure}
\begin{center}
\subfigure[]{
\label{fig1a} 
\includegraphics[angle=0,scale=0.49]{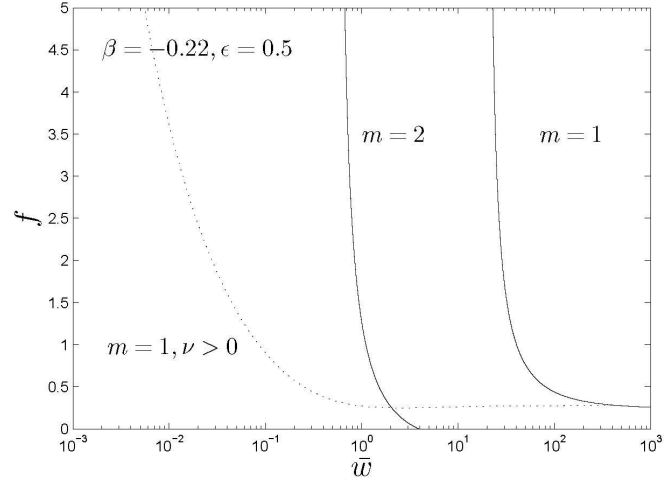}}
\hfill \subfigure[]{
\label{fig1b} 
\includegraphics[angle=0,scale=0.49]{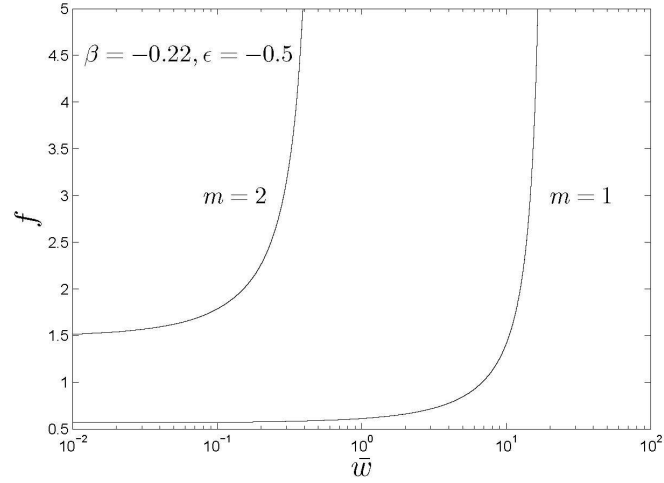}}
\caption{
(a) The physically allowed combinations of $\bar w$ (a measure for
the disc temperature and magnetic pressure together) and $f$ (a
measure for the background dark matter halo potential) for global 
stationary MHD perturbation configurations with the scale-free index
$\beta=-0.22$ and $\epsilon=0.5$ (i.e., the disc self-gravity
overtakes the magnetic tension force); the two solid curves
separately stand for $m=1$ and $m=2$ aligned cases, while the dotted 
curve stands for $m=1$ unaligned case and the curves of $m\geq2$ 
unaligned cases are unphysical. Note that different
points on the dotted curve have different $\nu$ values
and the bifurcation point
is at (461.2,\ 0.2764) in the ($\bar w, f$) diagram. (b) The same
as in panel (a), except for $\epsilon=-0.5$ (i.e., the magnetic
tension force overtakes the disc self-gravity) and thus the
requirement $f>0.5$. The unaligned logarithmic spiral cases all
become unphysical.}
\label{fig1} 
\end{center}
\end{figure}

\begin{figure}
\begin{center}
\subfigure[]{ \label{fig2a}
\includegraphics[angle=0,scale=0.49]{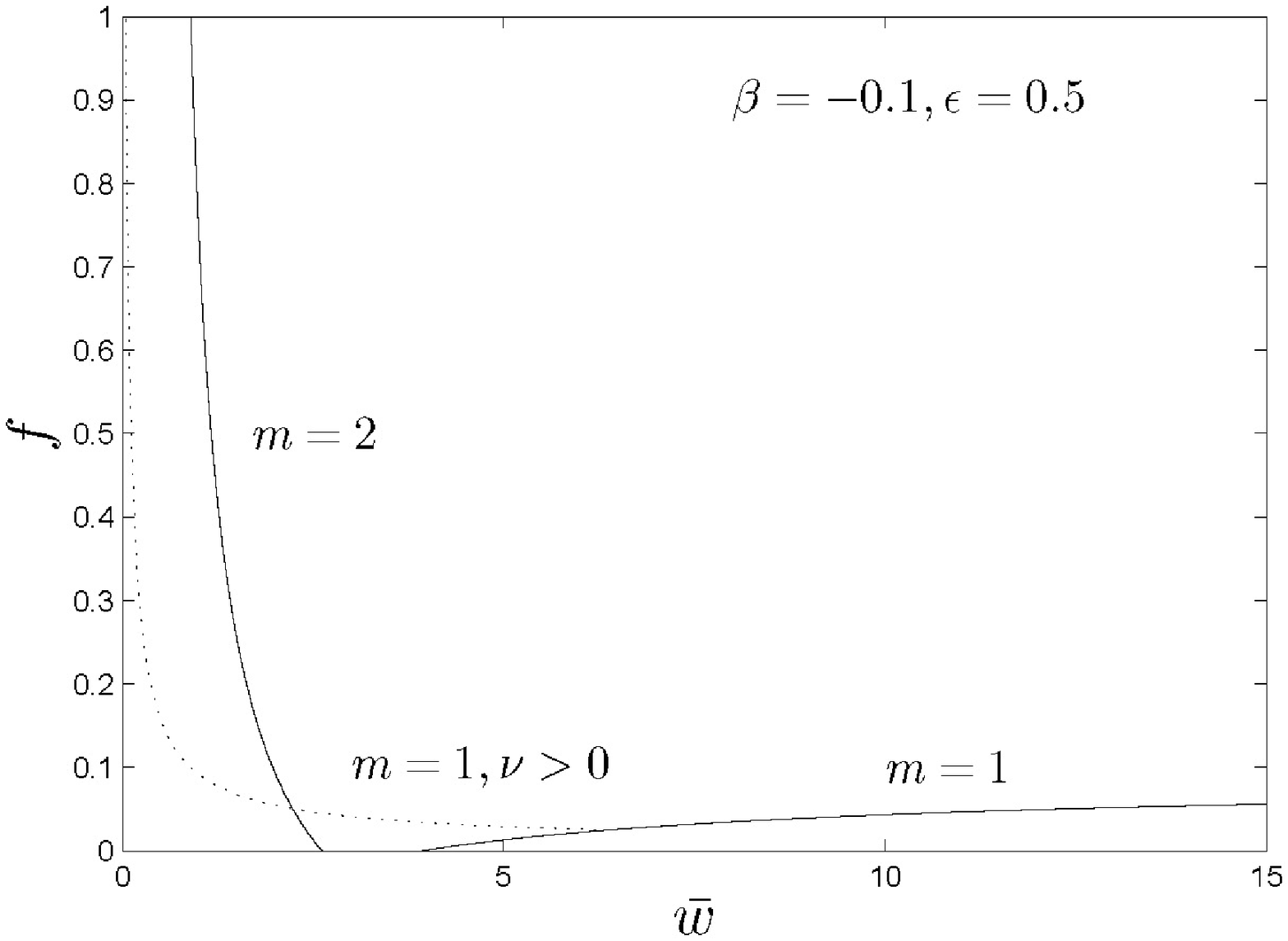}}
\hfill \subfigure[]{ \label{fig2b}
\includegraphics[angle=0,scale=0.49]{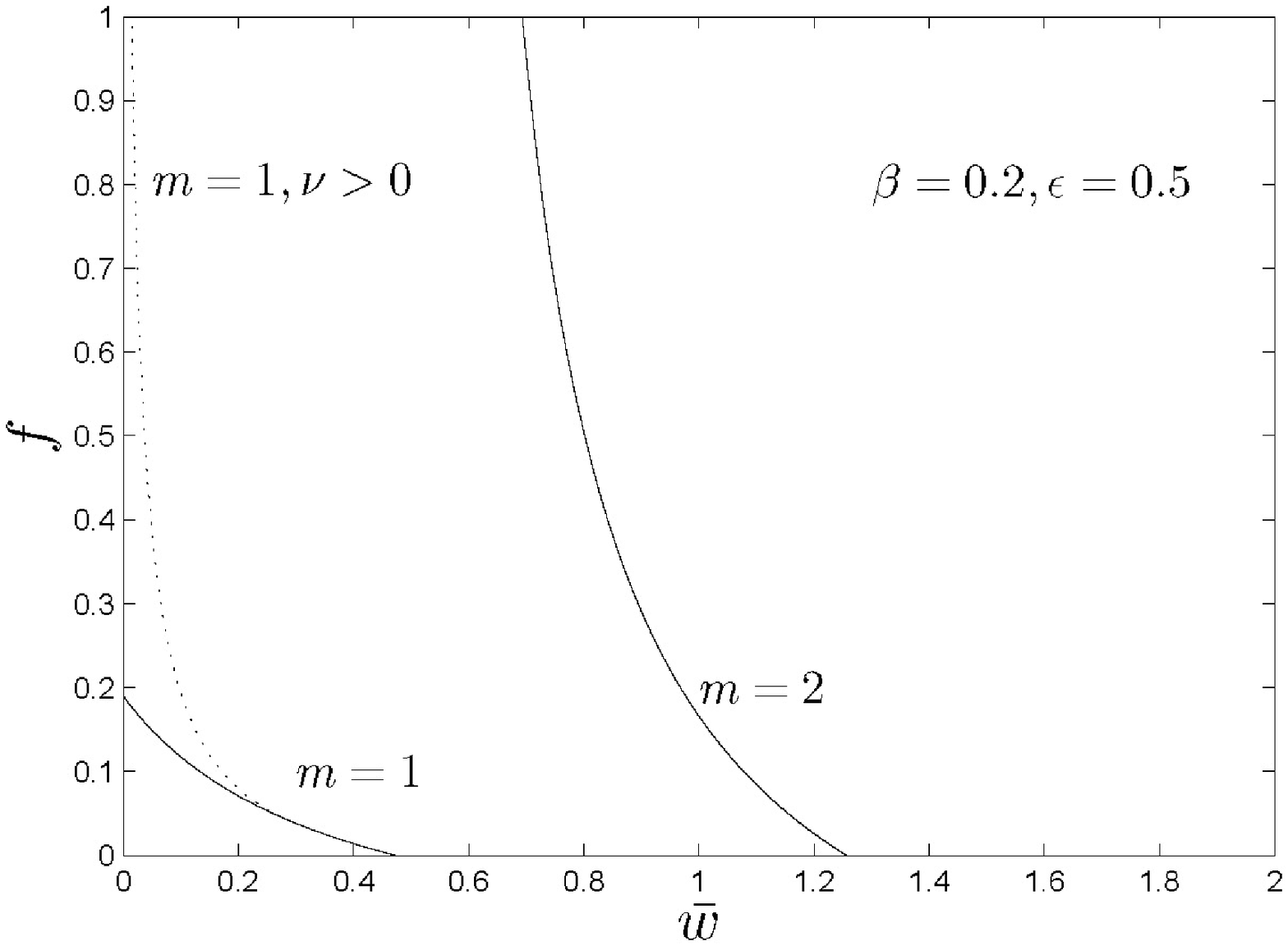}}
\hfill \subfigure[]{ \label{fig2c}
\includegraphics[angle=0,scale=0.49]{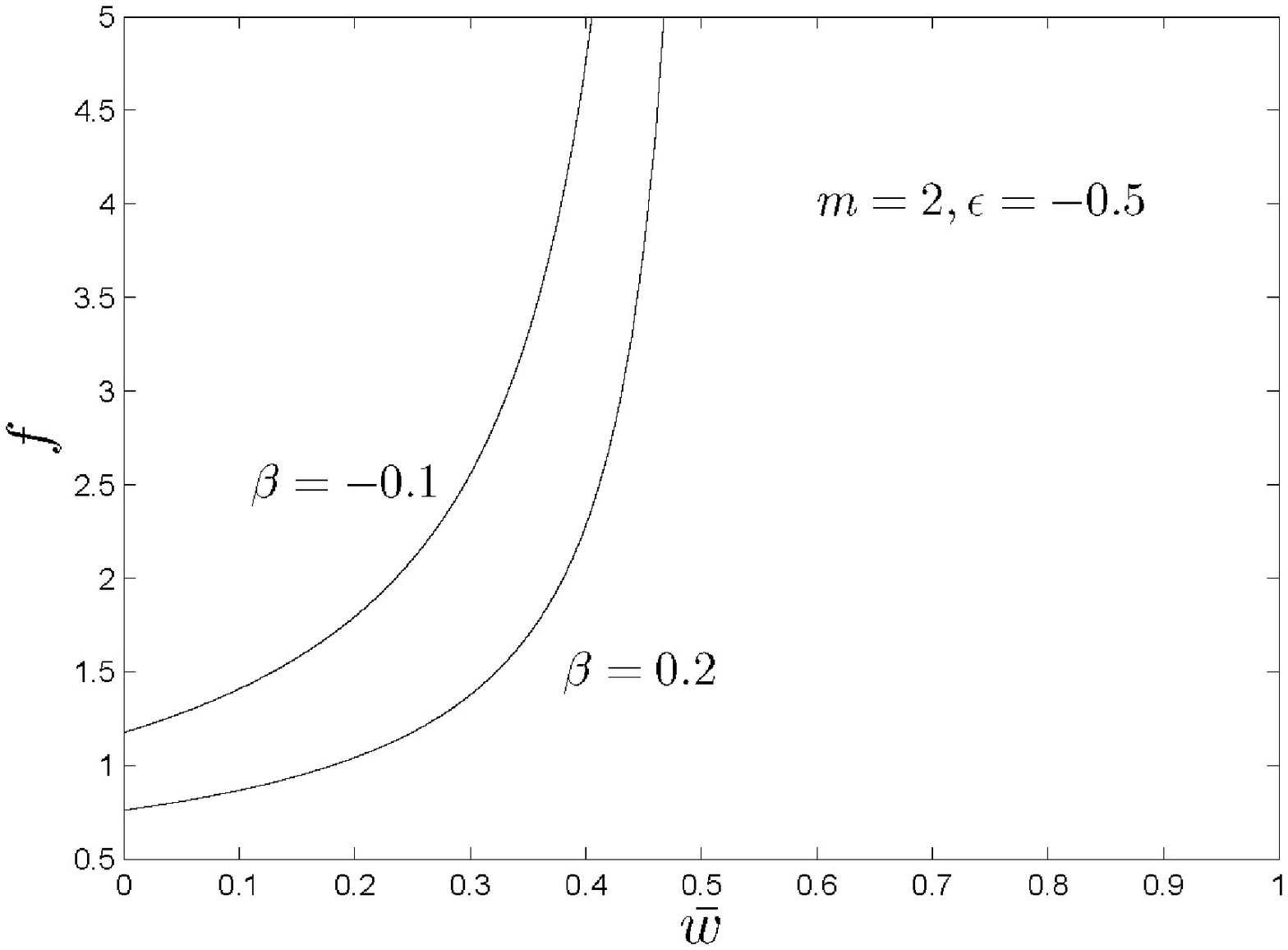}}
\caption{
(a) The physically allowed combinations of $\bar w$ (a measure of
the disc temperature and magnetic pressure together) and $f$ (a
measure of the background dark matter halo potential) for global
stationary MHD perturbation configurations when the scale-free
index $\beta=-0.1$
and $\epsilon=0.5$
(i.e., the horizontal disc self-gravity overtakes the magnetic
tension force); the two solid curves separately stand for $m=1$
and $m=2$ aligned cases, while the dotted curve stands for $m=1$
unaligned case and the curves of $m\geq2$ unaligned cases are all
unphysical. The bifurcation point
for $m=1$ case from the aligned to unaligned case is at (6.383,\ 0.02475).
(b) The same as in panel (a), except for $\beta=0.2$. For the $m=1$ case,
the bifurcation point is at (0.2943,\ 0.03955). (c) For $\epsilon=-0.5$
(i.e., the magnetic tension force overtakes the horizontal disc
self-gravity), the physically allowed combinations of $\bar w$ and $f$
for $m=2$ and $m=1$ stationary MHD perturbation configurations with
$\beta=-0.1$ and $\beta=0.2$. The aligned and unaligned cases of $m=1$
and the unaligned case of $m=2$ are all unphysical. Two solid curves
stand for $m=2$ aligned case with $\beta=-0.1$ and $\beta=0.2$,
respectively.}
\label{fig2}
\end{center}
\end{figure}

A more straightforward relation between $f$ and $\bar w$ can be
simply expressed as
\begin{equation}\label{f-w}
f=\epsilon\left(\frac{C_1}{\bar w-C_a}-C_2\right)\ ,
\end{equation}
where $C_1$ and $C_2$ are defined explicitly by
\begin{eqnarray}\label{C1}
C_1&\equiv&\frac{[1+(1+2\beta)C_a]Y_m(\beta)}{2\beta Y_0(\beta)}\ ,
\\
\label{C2} C_2&\equiv&1-\frac{(1+2\beta)Y_m(\beta)}{2\beta
Y_0(\beta)}\ .
\end{eqnarray}

When $\epsilon=0$ corresponding to the situation of the magnetic
tension exactly cancelled by the horizontal disc self-gravity,
equation (\ref{nonaxisymmetric-stationary}) and background
equilibrium relation (\ref{equili4}) then give
\begin{equation}\label{epsilon0}
\bar w=C_a\ , \qquad\qquad f=b_0^2[1+(1+2\beta)C_a]\ .
\end{equation}
In order to ensure $C_a\geq0$, we need  $m\geq2$ or
$\beta\in(-1/2,\ -0.2071)$. Thus for $\epsilon=0$, aligned MHD
perturbation configurations of $m\geq2$ can exist, while the $m=1$
mode can only exist when $\beta\in(-1/2,\ -0.2071)$. As the
horizontal disc self-gravity is cancelled by the magnetic tension
force, only the dark matter halo potential resists the centrifugal
force, the thermal pressure force and the magnetic pressure force
to maintain the disc equilibrium. Note that when $\epsilon=0$, we
have a fixed $\bar w$ and $b_0^2$ is determined by $f$ as compared
to the expression of $b_0^2$ using $w$ and $f$ in equation
(\ref{equili4}). This situation is strikingly similar to those of
low-mass discs ($f\rightarrow\infty$) referred to by Syer \&
Tremaine (1996). Such a similarity can be easily understood
because when $\epsilon=0$, the horizontal disc self-gravity is
completely cancelled by the magnetic tension force and the
effective $f$ (defined as the ratio of the force from the
background halo to the force from the disc) apparently approaches
infinity. In a certain sense, $\epsilon=0$ represents a critical
point separating two qualitatively different cases. For
$0<\epsilon\leq1$, the magnetic tension force is weaker than the
horizontal self-gravity force of the disc, while for $\epsilon<0$,
the magnetic tension force becomes stronger than the horizontal
self-gravity of the disc.
One would expect systems dominated by magnetic Lorentz 
force or gravity to show different physical properties.

When $0<\epsilon\leq1$, the entire analysis here parallels the
corresponding part of aligned discs in Syer \& Tremaine (1996),
because we can simply take $f/\epsilon$ and $\Theta w$ to
replace their $f$ and the $w$ variables, respectively. In
other words, the two factors $1/\epsilon$ and $\Theta$ can be
regarded as linear rescaling factors of $f$ and $w$ variables
when the magnetic tension force is weaker than the horizontal
disc self-gravity force.

When $\epsilon<0$, the magnetic tension force becomes stronger than the
horizontal self-gravity of the disc. From the background rotational
MHD equilibrium condition (\ref{equili4}), the gravitational
potential of a dark matter halo must exist and be strong enough
in order to create an MHD disc equilibrium in the first place.
As already noted earlier, physical constraints on the disc system
are $\bar w\geq0$ and $f\geq\max\{0,\ -\epsilon\}$. With these
requirements, we can infer the physically allowed part of a
hyperbolic curve of equation (\ref{f-w}) in the ($\bar w,\ f$)
diagram and determine the relevant range of $\bar w$.

For the aligned cases, we reach the following conclusions drawn
from the analysis detailed in Appendix A. For $m\geq2$ MHD
perturbation modes (bar-like), we have $\bar w\in (C_a\ ,\ 
C_a+C_1/C_2]$ when $0<\epsilon\leq1$ (see Figs. 1a,\ 2a,\ 2b) 
and $\bar w\in[0\ ,\ C_a)$ when $\epsilon<0$ (see Figs. 1b,\ 2c). 
For $m=1$ MHD perturbation modes with $0<\epsilon\leq1$, we have
$\bar w\in (C_a\ ,\ +\infty)$ when $\beta\in(-1/2\ ,\ -0.2071)$ 
(see Fig. 1a), $\bar w\in [C_1/C_2+C_a\ ,\ +\infty)$ when 
$\beta\in(-0.2071\ ,\ 0)$ (see Fig. 2a), and $\bar w\in 
[0\ ,\ C_1/C_2+C_a]$ when $\beta\in(0\ ,\ 1/2)$ (see Fig. 2b). 
For $m=1$ MHD perturbation modes with $\epsilon<0$, only when 
$\beta\in(-1/2\ ,\ -0.2071)$ and $\bar w\in[0\ ,\ C_a)$ can 
global stationary MHD perturbation patterns possibly exist 
(see Fig. 1b).

We now discuss several special cases below.

The case of $\beta=0$ corresponds to a flat rotation curve of
constant $v_{\theta 0}$ and a singular isothermal disc (SID) 
under scale-free conditions (Shu et al. 2000; Lou 2002; Lou 
\& Shen 2003; Shen \& Lou 2004a; Lou \& Zou 2004, 2006; Lou 
\& Wu 2005; Lou \& Bai 2006). Using the asymptotic expression
$\Gamma(x)\sim x^{-1}$ in the limit of $x\rightarrow 0$,
we write equation (\ref{f-w}) as
\begin{equation}\label{beta0}
f=-\frac{\epsilon(m-1)(m\bar w-m-2)}{(m^2\bar
w-m^2+2)}=\frac{\epsilon(m-1)}{m} \left[\frac{2(m+1)}{m^2\bar
w-(m^2-2)}-1\right]
\end{equation}
where $\beta=0$ has been set. One can readily see that $f=0$ for
an arbitrary $\bar w$ when $m=1$ and $\beta=0$. In other words,
an isopedically magnetized isolated disc with a flat rotation curve
supports an $m=1$ MHD perturbation mode for an arbitrary Mach
number (Syer \& Tremaine 1996; Shu et al. 2000; Lou 2002; Lou \& 
Shen 2003; Lou \& Zou 2004, 2006; Lou \& Wu 2005). For $m\geq2$
MHD perturbation modes, we have $\bar w\in (1-2/m^2\ ,\ 1+2/m]$ when
$0<\epsilon\leq1$ and $\bar w\in[\ 0\ ,\ 1-2/m^2)$ when $\epsilon<0$,
respectively.

Another singular case appears when $\beta=(1-\sqrt{2})/2=-0.2071$
and $m=1$. According to equation (\ref{nonaxisymmetric-stationary}),
we have $b_0=0$ corresponding to a nonrotating disc. In other words,
for a disc with a scale-free index $\beta=-0.2071$, it has to be
nonrotational in order to support a stationary $m=1$ MHD 
perturbation mode.

\section{Global Stationary Unaligned Logarithmic Spirals with $\nu\neq0$}

For unaligned logarithmic spiral cases, the isodensity contours of 
MHD perturbations entail a systematic phase shift in azimuth as 
$R$ increases, which gives a logarithmic spiral pattern. Unaligned 
stationary logarithmic spirals involve both azimuthal and radial 
propagations of MHD density waves. For a constant radial flux of 
angular momentum, we require $\beta_1=\Re(\beta')=1/4$. To be 
general, we take $\beta_1$ and $\beta$ to be two different parameters.

\subsection{Nonaxisymmetric Disturbances with $m\neq0$}

For $m\neq0$, the corresponding relation
(\ref{syertremaine-nonaxisymmetric}) can be written as
\begin{eqnarray}\label{unaligned}
\epsilon B=b_0^2(\bar w-C) \ ,
\end{eqnarray}
where
\begin{eqnarray}\label{BandC}
B\equiv\frac{Y_m(\beta')}{2\beta Y_0(\beta)}=B(m,\ \nu,\ \beta_1;
\ \beta)\ , \qquad\qquad
C\equiv\frac{m^2-2(1-\beta)}{[m^2+2(\beta+\beta'-2\beta'^2)]}=C(m,\
\nu,\ \beta_1;\ \beta)\ .
\end{eqnarray}
Here, parameters $b_0^2$, $\bar w$ and $\epsilon$ are all real,
while functions $B$ and $C$ may be complex in general. We then have
\begin{eqnarray}\label{ReIm}
\epsilon\Re(B)=b_0^2[\bar w-\Re(C)]\ , \qquad\qquad
\epsilon\Im(B)=-b_0^2\Im(C)\ .
\end{eqnarray}
We can also write the real and imaginary parts of $C$, $\Re(C)$
and $\Im(C)$, explicitly as
\begin{eqnarray}\label{ReC}
\Re(C)&=&\frac{[m^2-2(1-\beta)](m^2+2\beta+2\beta_1-4\beta_1^2+\nu^2)}
{(m^2+2\beta+2\beta_1-4\beta_1^2+\nu^2)^2+\nu^2(1-4\beta_1)^2}\
,\\
\label{ImC} \Im(C)&=&\frac{\nu(1-4\beta_1)[m^2-2(1-\beta)]}
{(m^2+2\beta+2\beta_1-4\beta_1^2+\nu^2)^2+\nu^2(1-4\beta_1)^2}\ .
\end{eqnarray}
Based on the analysis of Appendix A and equations (\ref{ReC}) and
(\ref{ImC}), we notice the following relations that $C(m,\ -\nu,\
\beta_1;\ \beta)=C^\ast(m,\ \nu,\ \beta_1;\ \beta)$, $B(m,\ -\nu,\
\beta_1;\ \beta)=B^\ast(m,\ \nu,\ \beta_1;\ \beta)$, $B(m,\ \nu,\
\beta_1;\ \beta)=B(-m,\ \nu,\ \beta_1;\ \beta)$ and $C(m,\ \nu,\
\beta_1;\ \beta)=C(-m,\ \nu,\ \beta_1;\ \beta)$, where the asterisk
$^\ast$ indicates the complex conjugate operation. Therefore,
equation (\ref{ReIm}) remains valid for $\pm\nu$ and $\pm m$. This
conclusion can be seen as a manifestation of the anti-spiral theorem
(Lynden-Bell \& Ostriker 1967), which states that trailing spiral 
and leading spiral share the same solution forms under stationary 
and time-reversible conditions. Therefore, there is no loss of 
generality to consider $m\geq0$ and $\nu>0$.

When $\beta\in(-1/2,\ 1/2)$ and $m\geq1$, it is clear that
$m^2-2(1-\beta)\neq0$. According to equation (\ref{ImC}), 
we thus have $\Im(C)=0$ only if $\nu=0$ or $\beta_1=1/4$. 
Obviously, $\nu=0$ corresponds to the aligned case as has 
been discussed already. The requirement $\beta_1=1/4$ 
implies a constant radial flux of angular momentum 
associated with logarithmic spiral MHD perturbations.

When $\epsilon=0$, we need $\beta_1=1/4$ and $m\geq2$ to satisfy
equation (\ref{ReIm}).

\subsubsection{Discs with $\beta_1=1/4$}

When $\beta_1=1/4$, equation (\ref{unaligned})
becomes real and appears as
\begin{equation}\label{beta1/4}
\frac{\epsilon\pi}{2\beta Y_0(\beta)}
\left|\frac{\Gamma(m/2+1/4+\mathrm{i}\nu/2)}
{\Gamma(m/2+3/4+\mathrm{i}\nu/2)}\right|^2
=\frac{(f+\epsilon)}{[1+(1+2\beta)\bar w]}\left[\bar
w-\frac{m^2-2(1-\beta)}{(m^2+2\beta+1/4+\nu^2)}\right]\ ,
\end{equation}
which can be further transformed into
\begin{equation}\label{beta1/4-1}
f=\epsilon\left(\frac{D_1}{\bar w-D_0}-D_2\right)
\qquad \mbox{or}
\qquad (f+\epsilon D_2)(\bar w-D_0)=\epsilon D_1\ ,
\end{equation}
where the three coefficients $D_0$, $D_1$
and $D_2$ are explicitly defined by
\begin{eqnarray*}
D_0&\equiv&\frac{[m^2-2(1-\beta)]}{(m^2+2\beta+1/4+\nu^2)}\ ,
\\
D_1&\equiv&\frac{\pi}{2\beta Y_0(\beta)}
\left|\frac{\Gamma(m/2+1/4+\mathrm{i}\nu/2)}
{\Gamma(m/2+3/4+\mathrm{i}\nu/2)}\right|^2
\frac{[2m^2(1+\beta)+4\beta^2-7/4+\nu^2]}
{(m^2+2\beta+1/4+\nu^2)}\ ,
\\
D_2&\equiv&1-\frac{\pi(1+2\beta)}{2\beta Y_0(\beta)}
\left|\frac{\Gamma(m/2+1/4+\mathrm{i}\nu/2)}
{\Gamma(m/2+3/4+\mathrm{i}\nu/2)}\right|^2\ .
\end{eqnarray*}

\begin{figure}
\begin{center}
\subfigure[]{ \label{fig3a}
\includegraphics[angle=0,scale=0.49]{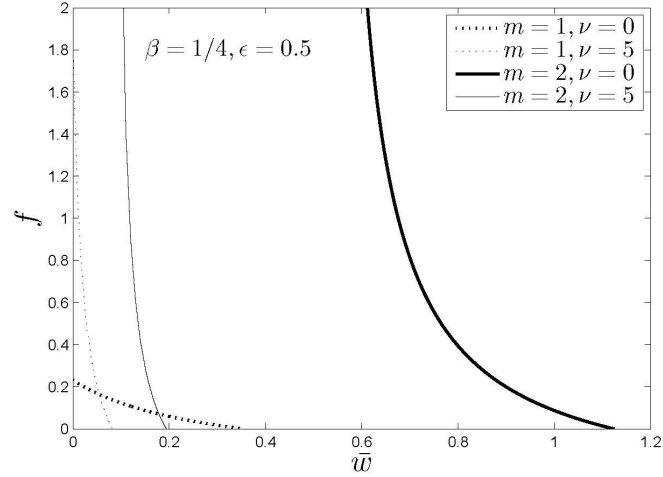}}
\hfill \subfigure[]{ \label{fig3b}
\includegraphics[angle=0,scale=0.49]{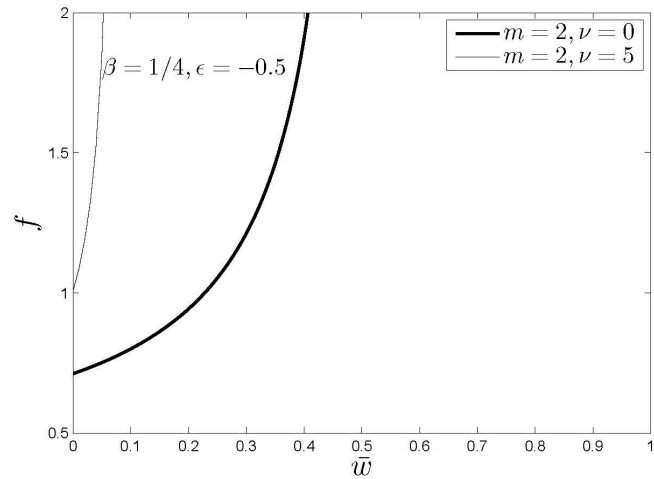}}
\caption{(a) Given the background scale-free index $\beta=1/4$
and $\epsilon=0.5$ (i.e., the horizontal disc self-gravity
overtakes the magnetic tension force), the physically allowed
combinations of $\bar w$ (a measure of the disc temperature and
magnetic pressure together) and $f$ (a measure of the background
dark matter halo potential) for $m=1$ and $m=2$ global stationary
MHD perturbation configurations of the perturbation scale-free
index $\beta_1=1/4$ disc with $\nu=0$ (aligned) and $\nu=5$
(unaligned), respectively. Two solid curves stand for $m=2$ MHD
perturbation mode and two dotted curves stand for $m=1$ MHD
perturbation mode. (b) The same as panel (a), except for
$\epsilon=-0.5$ (i.e., the magnetic tension force overtakes the
horizontal disc self-gravity). Note that a $m=1$ MHD perturbation
mode does not exist for $\epsilon<0$.
}
\label{fig3}
\end{center}
\end{figure}
A relationship is thus established by equation (\ref{beta1/4-1})
between $\bar w$ and $f$. By specifying parameters $m$, $\nu$ 
and $\beta$, this relation gives rise to a hyperbolic curve in 
the $(\bar w, f)$ diagram. From Appendix B, we thus draw the 
following conclusions. For $m\geq2$ MHD perturbation modes, we 
have $\bar w\in(D_0\ ,\ D_0+D_1/D_2]$ when $0<\epsilon\leq1$ 
(see Fig. 3a) and $\bar w\in[0\ ,\ D_0)$ when $\epsilon<0$ 
(see Fig. 3b); for $m=1$ MHD perturbation modes, we have 
$\bar w\in[0\ ,\ D_0+D_1/D_2]$ when $0<\epsilon\leq1$ (see 
Fig. 3a) and these modes do not exist when $\epsilon<0$.

When $\epsilon=0$, we have a fixed $\bar w$ as $\bar w=D_0$
and thus $m=1$ MHD perturbation mode cannot exist for a
negative $D_0$. A relationship between $f$ and $b_0^2$
can be established in the form of
\begin{equation}\label{epsilon0spiral1}
f=b_0^2\left[1+(1+2\beta)D_0\right]\ ,
\end{equation}
which is similar to equation (\ref{epsilon0})
for the aligned cases.

\subsubsection{Stationary Logarithmic Spiral
Patterns in Discs with $\beta_1\neq1/4$ }

When $\beta_1\neq1/4$, we have $\Im(C)\neq0$. Equation
(\ref{unaligned}) is then complex, giving rise to two
real equations. From equation (\ref{ReIm}), we solve
for $\bar w$ and $f$ to obtain
\begin{eqnarray}\label{w-unaligned}
\bar w&=&\Re(C)-\frac{\Re(B)\Im(C)}{\Im(B)}\ ,\\
\label{f-unaligned} f&=&-\epsilon\left\{\left[1+(1+2\beta)
\left(\Re(C)-\frac{\Re(B)\Im(C)}{\Im(B)}\right)\right]
\frac{\Im(B)}{\Im(C)}+1\right\}\ .
\end{eqnarray}
From extensive numerical computations, we gain the following useful
information. First, $\Re(B)>0$ and decreases with increasing $\nu$.
Secondly, $\Im(B)>0$ for $-1/2<\beta_1<1/4$ and $\Im(B)<0$ for
$1/4<\beta_1<1/2$, respectively.

We emphasize here that the $\bar w$ and $f$ solutions
(\ref{w-unaligned}) and (\ref{f-unaligned}) give a specific point in
the $(\bar w, f)$ diagram, which differs from the aligned case with
$\nu=0$ or $\beta_1=1/4$ discs whose real dispersion relation gives a
hyperbolic curve in the $(\bar w, f)$ diagram. Under the conditions
$\nu\neq0$ and $\beta_1\neq1/4$ (non-constant radial flux of angular
momentum, see Appendix G), dispersion relation (\ref{unaligned}) is
generally complex, corresponding to two constraints (i.e., real and
imaginary parts of equation \ref{unaligned}) on the combination of
$\bar w$ and $f$ and reducing a solution curve to a solution point
in the $(\bar w, f)$ diagram (see Figs. 1a,\ 2a,\ 2b).

Finally, we reach the following conclusions numerically: $\bar
w<0$ when $m\geq2$; for $m=1$ MHD perturbation modes, $\bar w$
remains always positive, while $f>0$ only for $\epsilon>0$.
Therefore, global stationary spiral MHD perturbation
configurations with $\beta_1\neq1/4$ can only exist 
for $m=1$ MHD perturbation modes and when $\epsilon>0$ 
(see Figs. 1a,\ 2a,\ 2b).

\subsubsection{Bifurcation Points from Aligned Cases}

The sequence of stationary logarithmic spirals with $\nu\in(0,\
+\infty)$ bifurcates from the aligned discs with $\nu=0$ at the
critical point where $\nu\rightarrow0$. When $\nu$ becomes small,
$\Im(B)$ and $\Im(C)$ may be replaced by $-\nu(\partial B/\partial
\beta')/2$ and $-\nu(\partial C/\partial \beta')/2$ respectively,
where the two first derivatives are evaluated in the limit of
$\nu\rightarrow0\ \ (\beta'=\beta_1)$.

Therefore at the bifurcation point (indicated by a subscript
$_{\mbox{bif}}$ of a physical variable), equations (\ref{ReIm}) become
\begin{eqnarray}
\epsilon B=b_0^2(\bar w-C)\ , \qquad\qquad \epsilon\frac{\partial
B}{\partial\beta'}=-b_0^2\frac{\partial C}{\partial \beta'}\ ,
\end{eqnarray}
where all quantities are evaluated at $\nu=0$ ($\beta'=\beta_1$).
Eliminating $b_0^2$ between the above two expressions, we obtain
\begin{eqnarray*}
\bar w_{\mbox{bif}}=-\left(\frac{\partial C}{\partial
\beta'}\right)\left(\frac{\partial \ln B}{\partial
\beta'}\right)^{-1}+C\ ,
\end{eqnarray*}
where the derivatives are evaluated at $\nu=0$ or
$\beta'=\beta_1$.

We then obtain
\begin{eqnarray}\label{w-bif}
\bar w_{\mbox{bif}}=\frac{2(1-4\beta_1)[m^2+2(\beta-1)]}
{[(m^2+2\beta+2\beta_1-4\beta_1^2)^2\Psi_m(\beta_1)]}
+\frac{[m^2+2(\beta-1)]}{(m^2+2\beta+2\beta_1-4\beta_1^2)}\ ,
\end{eqnarray}
where
\begin{eqnarray*}
\Psi_m(z)\equiv\psi\left(\frac{m}{2}+z\right)+\psi\left(\frac{m}{2}-z+1\right)
-\psi\left(\frac{m}{2}-z+\frac{1}{2}\right)
-\psi\left(\frac{m}{2}+z+\frac{1}{2}\right)\ ,
\end{eqnarray*}
and $\psi(z)$ is the digamma function (e.g., Shen \& Lou 2004).
Once $\bar w$ is determined, the expression of $f$ follows from
equation (\ref{f-w}), namely
\begin{eqnarray}\label{f-bif}
f_{\mbox{bif}}=\epsilon\left\{\frac{Y_m(\beta_1)}{2\beta
Y_0(\beta)}\left[(1+2\beta)+\frac{(m^2+2\beta+2\beta_1-4\beta_1^2)
[m^2(1+\beta)-(1-\beta_1)+2(\beta^2-\beta_1^2)]}
{(1-4\beta_1)[m^2+2(\beta-1)]}\Psi_m(\beta_1)\right]-1\right\}\ .
\end{eqnarray}
Together, this pair of $\bar w_{\mbox{bif}}$ (equation
\ref{w-bif}) and $f_{\mbox{bif}}$ (equation \ref{f-bif}) gives a
specific point in the $(\bar w, f)$ diagram which signifies the
bifurcation place between the aligned and unaligned sequences (see
Figs. 1a,\ 2a,\ 2b). Numerical explorations have demonstrated
that when $m\geq2$, $\bar w_{\mbox{bif}}<0$, while for $m=1$, both
$\bar w_{\mbox{bif}}$ and $f_{\mbox{bif}}$ are positive. Therefore
only for $m=1$ MHD perturbation modes, the bifurcation point is 
physically meaningful.

\subsection{Marginal Instability of Axisymmetric Disturbances ($m=0$)}

When $m=0$, dispersion relation (\ref{syertremaine-axisymmetric})
can be written as
\begin{eqnarray}\label{axis1}
\frac{\epsilon Y_0(\beta')}{2\beta Y_0(\beta)}
=b_0^2\left[ \bar
w-\frac{(\beta-1)}{\beta_1(1-2\beta_1)+\nu^2/2-\mathrm{i}
(1-4\beta_1)\nu/2}\right]\ .
\end{eqnarray}
We adopt $\beta_1=1/4$ to make equation (\ref{axis1}) real, namely
\begin{eqnarray}\label{axis2}
\frac{\epsilon \pi{\mathcal N}_0(\nu)}{\beta Y_0(\beta)}
=b_0^2\left[\bar w+\frac{8(1-\beta)}{(1+4\nu^2)}\right]\ ,
\end{eqnarray}
where ${\mathcal N}_0(\nu)$ is the Kalnajs function defined 
by equation (\ref{Kalnajs}). From equation (\ref{axis2}),
we then obtain $\bar w$ as
\begin{eqnarray}\label{axis3}
\bar w\equiv\frac{1}{D^2}=\frac{1}{(1+2\beta)} \left[\mathcal{
N}_0(\nu)-\frac{8
Y_0(\beta)\beta(1-\beta)(1+f/\epsilon)}{\pi(1+4\nu^2)}\right]
\bigg/\left[\frac{\beta
Y_0(\beta)(1+f/\epsilon)}{\pi(1+2\beta)}-{\mathcal N}_0(\nu)\right]\ ,
\end{eqnarray}
where $D^2$ is defined as $1/\bar w=1/(\Theta w)$, which is the same
as the $D^2$ introduced by Shu et al. (2000). When $\epsilon<0$, we
require $\epsilon+f>0$ and thus $1+f/\epsilon<0$. Therefore when
$\epsilon<0$ with $1+f/\epsilon<0$, $\bar w$ remains always negative
according to equation (\ref{axis3}). When $\epsilon=0$ in equation
(\ref{axis2}), there is no global stationary configuration, as in
the situation of $\epsilon<0$.

\begin{figure}
\begin{center}
\includegraphics[angle=0,scale=0.49]{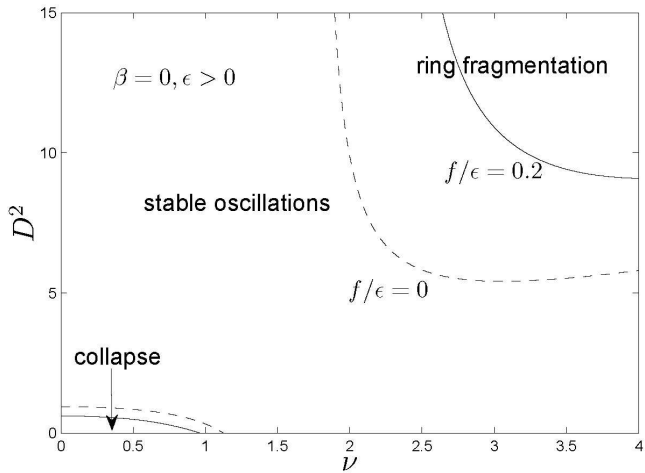}
\caption{The axisymmetric ($m=0$) marginal instability curves of
$D^2$ $(1/\bar w)$ (the reciprocal of the disc temperature and
magnetic pressure together) versus $\nu$ (the dimensionless radial
wavenumber) with different ratio $f/\epsilon$ equal to 0 (dashed
curves) and 0.2 (solid curves), respectively, for specified values
of the perturbation scale-free index $\beta_1=1/4$ and the
background scale-free index $\beta=0$. We take $\epsilon>0$ (i.e.,
the disc self-gravity overtakes the magnetic tension force). For
$\epsilon<0$ (i.e., the magnetic tension force dominates the disc
self-gravity), the instability does not exist. Note that both the
Jeans collapse and ring fragmentation unstable regions shrink with
increasing $f/\epsilon$.}
\end{center}
\end{figure}

When equation (\ref{axis3}) is displayed in the $(\nu,\ D^2)$
diagram, the curves correspond to the marginal instability
($\omega=0$). Here $\nu$ represents the radial wavenumber of
MHD perturbations, and a smaller $\nu$ signifies a larger
perturbation scale.
Also, $D^2$ is the square of the rotational Mach number, and a
larger $D^2$ thus signifies a faster rotation. Using equation
(\ref{WKBJ}), we can analyze stability properties of the $(\nu,\
D^2)$ diagram in different parameter regimes separated by the
marginal instability curves (Shu et al. 2000; Lou 2002; Lou \&
Shen 2003; Shen \& Lou 2004b; Lou \& Wu 2005; Lou \& Zou 2006).
Here $\omega^2>0$ means stability and $\omega^2<0$ means
instability, respectively. According to equation (\ref{WKBJ}),
we see that when $\epsilon<0$, $\omega^2$ remains always positive
so that the disc is completely linearly stable to axisymmetric
perturbations at all wavelengths. As a consequence, the marginal
stability curve in Figure 4 does not exist when $\epsilon<0$;
in other words, both Jeans collapse and ring fragmentation
instabilities are suppressed (i.e., lower-left curves become
negative and upper-right curves move to infinity).

For $\epsilon>0$, we discuss only a special case of $\beta=0$ as
an example of illustration, which corresponds to the SID case
(Shu et al. 2000; Lou 2002; Lou \& Shu 2003; Lou \& Wu 2005). 
A more general treatment (Shen \& Lou 2004a, b) will be given
in a forthcoming paper especially aiming at the axisymmetric
instability.

Knowing the asymptotic limiting expression $\beta
Y_0(\beta)\rightarrow\pi$ as $\beta\rightarrow 0$, we obtain
$1/\bar w$ as
\begin{eqnarray}\label{axisSID}
\frac{1}{\bar w}\equiv D^2=\frac{(1+f/\epsilon)-{\mathcal
N}_0(\nu)}{{\mathcal N}_0(\nu)-8(1+f/\epsilon)/(1+4\nu^2)}\ .
\end{eqnarray}
From Appendix F, the solution curves of equation (\ref{axisSID})
separate the $(\nu,\ D^2)$ diagram into three or two regimes. The
lower-left corner and the upper-right corner are the unstable
regions with $\omega^2<0$. The lower-left corner corresponding to
MHD perturbations of large radial spatial scales represents the
region of Jeans collapse instability modified by the disc rotation
and the isopedic magnetic field, while the upper-right corner
corresponding to perturbations of smaller radial spatial scales
represents the region of the MHD ring fragmentation instability
(Lou \& Fan 1998a; Shu et al. 2000; Lou 2002). Sandwiched between
these two unstable regimes is the stable region of $\omega^2>0$
(Shen \& Lou 2004a; see Fig. 4 for direct visual impressions).

We can also demonstrate that as $f$ increases or $\epsilon$
decreases, both the MHD Jeans collapse and ring fragmentation
instability regions tend to shrink. In other words, an enhanced
background dark matter halo potential or a stronger isopedic
magnetic field tend to squash instabilities both in small ( MHD
ring fragmentation) and large (MHD Jeans collapse) spatial scales.
In particular, when $1+f/\epsilon>{\mathcal N}_0(0)=4.337$, the
MHD Jeans collapse instability region disappears. In contrast, the
MHD ring fragmentation region always exists. For example, a cold
disc ($\bar w=0$ or $D^2\rightarrow+\infty$) will always subject
to the ring fragmentation instability. Our numerical explorations
are supportive of some conclusions of numerical simulation results
of Miller (1978), namely, a certain level of velocity dispersion
is necessarily required for the disc stability, and furthermore, a
massive dark matter halo cannot stabilize a `cold' disc galaxy
completely.

In short, we come to the conclusion that a massive background dark
matter halo (i.e., a high $f$ ratio) and a strong isopedic
magnetic field (i.e., a small $\epsilon$) tend to make the
magnetized gas disc more stable against axisymmetric MHD
disturbances.

\section{Conclusions, Summary and Discussion}

Using the standard linearization procedure for MHD perturbations
in a rotating disc, we have in a more general manner constructed
global stationary MHD perturbation configurations (i.e., small
deviations from the axisymmetric background) for both aligned and
unaligned logarithmic spiral patterns in a razor-thin scale-free
disc system embedded in an axisymmetric dark matter halo potential
and with an isopedic magnetic field almost vertically threaded 
through the gas disc plane. The stability analysis of axisymmetric 
MHD perturbations is performed in a disc of flat rotation curve 
with $\beta=0$ (i.e., SID) as an example of illustration. The most 
important and interesting contribution of our paper is that we 
integrate several essential aspects, such as isopedic magnetic 
field and dark matter halo, into the disc dynamics and adopt an 
analytic approach for the stability analysis. A generalized 
relationship is achieved and some interesting properties directly 
derived from the relationship have been revealed.

As global stationary MHD perturbation configurations may represent
transitions between stabilities to instabilities and vice versa (Shu
et al. 2000), we can use the obtained global MHD perturbation
configurations to explore initial conditions for possible
instabilities and the dynamical evolution of a razor-thin scale-free
disc system. Thus, a more realistic magnetized disc system to model
actual spiral galaxies may provide sensible physical information.
Since the systematic galactic observations by Rubin et al. (1982)
and Kent (1986, 1987, 1988),
the massive dark matter halo becomes an indispensable component for
models of disc galaxies, which are referred to as a partial disc
system (e.g., Binney \& Tremaine 1987; Syer \& Tremaine 1996; Shu et al.
2000; Lou \& Shen 2003; Shen \& Lou 2004a; Lou \& Wu 2005; Lou \& Bai
2006). Meanwhile, the ubiquitous presence of galactic magnetic fields
requires a serious and systematic consideration of the magnetized gas
disc in a typical disc spiral galaxy in order to trace the dynamic
evolution in a more realistic manner. In our idealized MHD model,
the more massive dark matter halo is presumed to be axisymmetric and
compatible with the scale-free conditions, while the magnetic field is
taken to be of an isopedic geometry. Our MHD disc model does include
several essential elements of disc galaxies and have already shown some
interesting and intricate phenomena. In short, self-gravity and thermal
pressure of a rotating scale-free disc, the axisymmetric gravitational
potential of a dark matter halo and the isopedic magnetic field across
the disc are synthesized together in our current model framework.

In the model analysis, we consider the self-gravity component in the 
disc without the usual WKBJ or tight-winding approximation and allow 
the background disc scale-free index $\beta$ and the perturbation 
scale-free index $\beta_1$ to be different in general. In other 
words, the long-range gravitational effect has been taken into 
account in full. Therefore equation (\ref{dispersionfinal}) is fairly 
comprehensive and rigorous linear dispersion relationship, from 
which much information can be gained, such as the WKBJ relationship 
in Appendix B, global stationary MHD dispersion relations 
(\ref{nonaxisymmetric-stationary}) and (\ref{axisymmetric-stationary}). 
By choosing different parameters, equation (\ref{dispersionfinal}) is 
sufficiently general to include aligned ($\nu=0$) and unaligned 
($\nu\neq0$) cases, as well as axisymmetric ($m=0$) and 
nonaxisymmetric ($m\neq0$) cases.

The MHD discs are specified in a range for the background
scale-free index $\beta\in(-1/2,\ 1/2)$ [n.b., $\beta\in(-1/4,\
1/2)$ for warm discs only], the inverse-square magnetosonic Mach
number (disc temperature and magnetic pressure together) $\bar
w>0$, the ratio of the gravity force of the dark matter halo to
the unperturbed disc gravity force $f>0$ and the gravity dilution
(due to the magnetic tension force) factor $\epsilon\leq1$.
Meanwhile, MHD perturbations are characterized by the perturbation
scale-free index $\beta_1\in(-1/2,\ 1/2)$, azimuthal wavenumber
$m$ (integer) and radial wavenumber $\nu$. With all these
parameters in physical ranges, we proceed to construct global
stationary MHD perturbation configurations in a specific
background disc (characterized by $\beta$, $\bar w$, $f$ and
$\epsilon$ parameters) under a specific perturbation
(characterized by $\beta_1$, $m$ and $\nu$). In comparison with
the WKBJ dispersion relation (Appendix B), the marginal
instability of axisymmetric MHD disturbances is examined for discs
with flat rotation curves ($\beta=0$).

Our conclusions are summarized below.

(i) {\it Aligned Cases with $\nu=0$}.

For global stationary aligned MHD perturbations, we take $\beta_1=\beta$,
although the disc system does allow for $\beta_1\neq\beta$ in general.
The axisymmetric ($m=0$) aligned MHD perturbation represents purely a
slight rescaling of the axisymmetric background (Lou \& Shen 2003; Shen
\& Lou 2004a; Lou \& Zou 2004, 2006; Lou \& Wu 2005). For both
$\epsilon>0$ and $\epsilon<0$ regimes, the global $m\geq2$ stationary
MHD perturbation configurations exist when $\bar w$ varies within a
certain limited region (see Figs. 1, 2 and C1). For neutral $m=1$
MHD perturbation modes with $\epsilon>0$, there are three types of
relations between $\bar w$ and $f$ when $\beta$ parameter falls
within the three intervals $(-1/2,\ -0.2071)$, $(-0.2071,\ 0)$ and 
$(0,\ 1/2)$, respectively (see Figs. 1a,\ 2a,\ 2b and Fig. C2). When 
$\epsilon<0$, neutral $m=1$ MHD perturbation modes can only exist for 
$\beta\in(-1/2,\ -0.2071)$ (see Figs. 1b and C2). Having specified 
values of $m$, $\beta$ and $\epsilon$ parameters, the global stationary 
MHD perturbation configurations can be displayed in the ($\bar w,\ f$) 
diagram. One point in the ($\bar w,\ f$) diagram then corresponds to 
one specific global stationary MHD perturbation configuration, and all 
of these points together constitute continuously a hyperbolic curve in 
the ($\bar w,\ f$) diagram. A detailed description and figures can be 
found in Appendix C.

(ii) {\it Unaligned Cases with $\nu\neq0$}.

To maintain a constant radial flux of angular momentum associated with
MHD perturbations (e.g., Goldreich \& Tremaine 1979; Shen et al. 2005),
the perturbation scale-free index $\beta_1$ should be set to $1/4$.
Furthermore, the $\beta_1=1/4$ stationary MHD perturbation configurations,
in which the angular momentum flux is constant (see Appendix G), restrict
parameter $\bar w$ to a certain limited region for $m\geq1$ neutral MHD
perturbations. In particular, the $m=1$ neutral MHD perturbation mode does
not exist for $\epsilon<0$. For given values of $m$, $\beta$, $\nu$ and
$\epsilon$ parameters, global stationary MHD perturbation configurations
correspond to a hyperbolic curve shown in the ($\bar w,\ f$) diagram (see
Fig. 3); this is similar to the aligned MHD perturbation configurations.
Details of analysis can be found in Appendix D. When $\beta_1\neq1/4$
(i.e., non-constant radial flux of angular momentum), the dispersion 
relation becomes complex and its imaginary part gives an additional 
constraint. Thus for given values of $m$, $\beta$, $\beta_1$, $\nu$ 
and $\epsilon$ parameters, the continuous point set of global 
stationary MHD perturbation configurations shrink from an extended 
one-dimensional curve to a single point in the ($\bar w,\ f$) 
diagram (see Figs. 1a,\ 2a,\ 2b). Numerical explorations show 
that only $m=1$ lopsided MHD perturbation modes exist.

(iii) {\it Marginal Instabilities for Axisymmetric
Disturbances with $m=0,\ \beta_1=1/4$ and $\nu\neq0$}.

Given a special value of $\beta=0$, we have shown that the increase
of a dark matter halo mass (larger $f$ parameter) or of magnetic field
flux (smaller $\epsilon>0$ parameter) can squeeze both the Jeans
collapse and ring fragmentation instability regions. In the regime
of $\epsilon<0$, the MHD disc becomes stable for axisymmetric MHD
perturbations of all radial wavelengths, according to the WKBJ analysis.

From the model results summarized above, we can see the existence of an
isopedic magnetic field can greatly influence the global stationary MHD
perturbation configurations in a scale-free disc system, as well as the
axisymmetric stability properties. When $0<\epsilon<1$, an isopedic
magnetic field gives rise to an effective modification factor on the
gravitational constant $G$, from $G$ to $\epsilon G$ (Shu et al. 2000),
and an effective enhancement factor on the sound speed $v_s$ (thermal 
pressure), from $v_s$ to $\sqrt\Theta v_s$. In the absence of the dark 
matter halo, $\epsilon$ has to be positive in order to maintain a disc 
equilibrium. In other words, the magnetic field tension force cannot 
dominate the self-gravity component in the disc plane. However, a 
massive dark matter halo exerts an extra gravity force to hold the 
magnetized disc together,
allowing the magnetic tension force to overtake the self-gravity
component in the disc plane, corresponding to a situation of
$\epsilon<0$. In the parameter regime $\epsilon<0$, we can no
longer regard the isopedic magnetic field as a modification but
must face an entirely new physical situation involving several
interesting features in a strongly magnetized disc system. An
interesting example is to derive the MHD WKBJ relation in equation
(\ref{WKBJ}) for the instability of axisymmetric disturbances
($m=0$). When $\epsilon<0$, we always have $\omega^2>0$,
indicating the disc system being stable for axisymmetric MHD
disturbances of all wavelengths.

Starting from the case of $\epsilon<0$, we may extend the same idea 
to more generalized situations. As a system has a strong magnetic 
field attached to plasma or gas, the magnetic field would have torn 
the system apart in the absence of an external massive dark matter 
halo potential. Two possible physical examples come into mind: the 
hot magnetized gas trapped in the centre of a galaxy cluster (e.g., 
Makino 1997; Dolag et al. 2001; Hu \& Lou 2004) and the accretion 
disc around a SMBH (e.g., Kudoh et al. 1996; Kaburaki 2000) where 
strong magnetic field may exist. The former can be perceived in
reference to a quasi-spherical system in which a hot gas is mainly 
confined by the dark matter potential and the hot gas itself is 
strongly magnetized. Our current model offers another possibility
in that a more or less flat gas disc system in rotation can be 
primarily confined by a dark matter potential and the gas disc
itself is isopedically magnetized with a considerable strength. 
By observing the existence of unusually strong magnetic field 
in a localized region, we may predict some `unseen' dark matter 
components of the system. For instance, if a source region has 
strong cyclotron or synchrotron emissions with much reduced 
optical emissions, the scenario outlined above may provide a 
possible explanation.

Fujita \& Kato (2005) proposed that Weibel instability may be
responsible for generating strong magnetic fields in galaxies 
and clusters of galaxies at redshift as high as $z\sim 10$; in 
such a scenario, the magnetic field can play a crucial role in 
forming galaxies and galaxy clusters. Our model results, 
especially those for a strong magnetic field (i.e., 
$\epsilon<0$), offer several interesting clues for the dynamics 
of galaxy or galaxy cluster formation. With extensive 
observations for magnetic fields in clusters of galaxies 
(Taylor et al. 2006), a more sensible MHD model may be 
constructed and tested.


The final point of interest involves the radial flux of angular momentum.
If the radial flux of angular momentum is constant, only the central
part will collapse and the outer portion of a disc is stable. If the
radial flux of angular momentum increases with increasing $R$, the
disc mass at all radii will lose their angular momentum by interacting
with stationary MHD density waves, and hence gradually spiral inward
and drift towards the centre. In this situation, not only the central
part, but also the entire disc drifts inward. For a single rotating disc
embedded in an axisymmetric dark matter halo, the angular momentum of the
disc system is conserved. In the presence of an external nonaxisymmetric
potential, such as a companion or satellite galaxy, this conservation no
longer holds. Implications of this issue will be investigated further.

\section*{Acknowledgment}
We thank Y. Shen for useful discussions on scale-free discs. This
research has been supported in part by the ASCI Center for
Astrophysical Thermonuclear Flashes at the University of Chicago,
by the Special Funds
for Major State Basic Science Research Projects of China, by
Tsinghua Center for Astrophysics, by the Collaborative Research
Fund from the National Natural Science Foundation of China (NSFC)
for Young Outstanding Overseas Chinese Scholars (NSFC 10028306) at
National Astronomical Observatories of China, Chinese Academy of
Sciences, by the NSFC grants 10373009 and 10533020 at Tsinghua
University, and by the SRFDP 20050003088
and by the Yangtze Endowment from the Ministry of Education through
Tsinghua University. The hospitality and support of the Mullard
Space Science Laboratory at University College London, U.K. and of
Centre de Physique des Particules de Marseille (CPPM/IN2P3/CNRS)
+Universit\'e de la M\'editerran\'ee Aix-Marseille II, France are
also gratefully acknowledged. Affiliated institutions of Y-QL
share this contribution.

\begin{appendix}
\section{Several Properties of Function $Y_m(\beta')$}

The function $Y_m(\beta')$ can be explicitly expressed as
\begin{eqnarray}
Y_m(\beta')=\frac{\pi\Gamma(z_1)\Gamma(z_2)}
{\Gamma(z_1+1/2)\Gamma(z_2+1/2)}\
\end{eqnarray}
in terms of Gamma function $\Gamma(\cdots)$, where
\begin{eqnarray} z_1=m/2-\beta'+1/2\ , \qquad
z_2=m/2+\beta'\ , \qquad \beta'=\beta_1-{\mathrm{i}\nu}/{2}\ .
\end{eqnarray}

Since $\Gamma(z^\ast)=\Gamma(z)^\ast$ with the asterisk $\ast$
denoting the complex conjugate operation, it then follows that
\begin{eqnarray}
Y_m(\beta'^\ast)
=\frac{\pi\Gamma(z_1^\ast)\Gamma(z_2^\ast)}
{\Gamma(z_1^\ast+1/2)\Gamma(z_2^\ast+1/2)}
=\frac{\pi\Gamma(z_1)^\ast\Gamma(z_2)^\ast}
{\Gamma(z_1+1/2)^\ast\Gamma(z_2+1/2)^\ast}=Y_m(\beta')^\ast\ .
\end{eqnarray}
Using the recursion relation $\Gamma(z)\Gamma(1-z)
=\pi/\sin(\pi z)$, we obtain
\begin{eqnarray}\nonumber
\begin{split}
Y_{-m}(\beta')=\frac{\pi\Gamma(1/2-z_2)\Gamma(1/2-z_1)}
{\Gamma(1-z_2)\Gamma(1-z_1)}=Y_m(\beta')
\frac{\sin[\pi(1-z_2)]\sin[\pi(1-z_1)]}
{\sin[\pi(1/2-z_2)]\sin[\pi(1/2-z_1)]}
\qquad\qquad\\ \qquad\qquad
=Y_m(\beta')\frac{\cos[\pi (z_1-z_2)]-\cos[\pi (z_1+z_2)]}
{\cos[\pi (z_1-z_2)]+\cos[\pi (z_1+z_2)]}=Y_m(\beta')\ .
\end{split}
\end{eqnarray}
That is, $Y_m(\beta')$ function is symmetric
or even with respect to the subscript $m$.

According to the Stirling formula, the asymptotic
series of Gamma function $\Gamma(z)$ is given by
\begin{eqnarray}\label{A4}
\Gamma(z)=(2\pi)^{1/2}e^{-z}z^{z-1/2}\left(1+
\frac{1}{12z}+\frac{1}{288z^2}+\cdots\right)\ .
\end{eqnarray}
In terms of asymptotic series expansion
(\ref{A4}), $\Gamma(z_1)$, $\Gamma(z_2)$, $\Gamma(z_1+1/2)$ and
$\Gamma(z_2+1/2)$ are given by
\begin{eqnarray*}
\Gamma(z_1)&=&(2\pi)^{1/2}e^{-z_1}z_1^{z_1-1/2}
\left[1-\frac{\mathrm{i}}{6\nu}+O(\nu^{-2})\right]\ ,
\\
\Gamma(z_2)&=&(2\pi)^{1/2}e^{-z_2}z_2^{z_2-1/2}
\left[1+\frac{\mathrm{i}}{6\nu}+O(\nu^{-2})\right]\ ,
\\
\Gamma(z_1+1/2)&=&(2\pi)^{1/2}e^{-(z_1+1/2)}(z_1+1/2)^{z_1}
\left[1-\frac{\mathrm{i}}{6\nu}+O(\nu^{-2})\right]\ ,
\\
\Gamma(z_2+1/2)&=&(2\pi)^{1/2}e^{-(z_2+1/2)}(z_2+1/2)^{z_2}
\left[1+\frac{\mathrm{i}}{6\nu}+O(\nu^{-2})\right]\ .
\end{eqnarray*}
To the leading order, we then have $Y_m(\beta')$ in the form of
\begin{eqnarray}
Y_m(\beta')=e\pi z_1^{z_1-1/2}z_2^{z_2-1/2}(z_1+1/2)^{-z_1}
(z_2+1/2)^{-z_2}[1+O(\nu^{-2})]\ .
\end{eqnarray}
A formula frequently used in the following analysis is given below
\begin{eqnarray}\label{A6}
z^{z'}=\exp\left[z'(\ln|z|+\mathrm{i}\arg z)\right]\ ,
\end{eqnarray}
where $z$ and $z'$ are two arbitrary complex numbers 
and $\arg z$ is the principal value of the argument $z$.

In addition, we have
\begin{eqnarray*}
\ln\left|z_i+1/2\right|
&=&\ln\left\{\sqrt{[\Re(z_i+1/2)]^2+[\Im(z_i+1/2)]^2}\right\}
=\ln(|\nu|/2)+{2[\Re(z_i+1/2)]^2}/{\nu^2}+O(\nu^{-4})\ ,
\quad (\hbox{for } i=1,2)
\\
\ln\left|z_i\right|
&=&\ln\left\{\sqrt{[\Re(z_i)]^2+[\Im(z_i)]^2}\right\}
=\ln(|\nu|/2)+{2[\Re(z_i)]^2}/{\nu^2}+O(\nu^{-4})\ ,
\qquad\qquad (\hbox{for } i=1,2)
\end{eqnarray*}
and
\begin{eqnarray*}
\arg z_1&=&
\ \hbox{ }\ \mbox{sgn}(\nu)\pi/2- {2\Re(z_1)}/{\nu}+O(\nu^{-3})\ ,
\\
\arg z_2&=&
-\mbox{sgn}(\nu)\pi/2 +{2\Re(z_2)}/{\nu}+O(\nu^{-3})\ ,
\\
\arg\left(z_1+1/2\right)&=&
\ \hbox{ }\ \mbox{sgn}(\nu)\pi/2- {2\Re(z_1+1/2)}/{\nu}+O(\nu^{-3})\ ,
\\
\arg\left(z_2+1/2\right)&=&
-\mbox{sgn}(\nu)\pi/2 +{2\Re(z_2+1/2)}/{\nu}+O(\nu^{-3})\ ,
\end{eqnarray*}
where sgn stands for the signum function. Hence, 
we arrive at the following two relations
\begin{eqnarray*}
L_1=(z_1-1/2)\ln|z_1|+(z_2-1/2)\ln|z_2|-z_1\arg\ln|z_1+1/2|-z_2\ln|z_2+1/2|
=-\ln(|\nu|/2)-{\mathrm{i(1-4\beta_1)}}/{2\nu}+O(\nu^{-2})\ ,
\end{eqnarray*}
and \begin{eqnarray*} L_2=(z_1-1/2)\arg z_1+(z_2-1/2)\arg z_2
-z_1\arg\left(z_1+1/2\right) -z_2\arg\left(z_2+1/2\right)
=\mathrm{i}+{(1-4\beta_1)}/{\nu}+O(\nu^{-2})\ .
\end{eqnarray*}

From formula (\ref{A6}), it is easy to see
\begin{eqnarray*}
\exp(L_1+\mathrm{i}L_2)=z_1^{z_1-1/2}z_2^{z_2-1/2}(z_1+1/2)^{-z_1}
(z_2+1/2)^{-z_2}\ .
\end{eqnarray*}

Finally, we derive
\begin{eqnarray}
Y_m(\beta')=\pi \exp\left[ -\ln
\frac{|\nu|}{2}+\frac{\mathrm{i(1-4\beta_1)}}{2\nu}
+O(\nu^{-2})\right][1+O(\nu^{-2})]
\label{Approx} =\frac{2\pi}{|\nu|}\left[1
+\frac{\mathrm{i(1-4\beta_1)}}{2\nu} \right]+O(\nu^{-3})\ .
\end{eqnarray}

\section{The WKBJ or Tight-Winding Approximation}

According to equation (\ref{dispersionfinal}), we have
\begin{eqnarray}\nonumber
\frac{(\hat\omega+mb_0)^2-2(1-\beta)b_0^2}{w\Theta b_0^2
-\epsilon c_0GY_m(\beta')}&=&\nu^2+m^2+2\beta_1-4\beta_1^2
+\mathrm{i}\nu\left[4\beta_1-1+\frac{2(1+\beta)
\hat\omega(\hat\omega+mb_0)}
{(\hat\omega+m b_0)^2-2(1-\beta)b_0^2}\right]
\\ \nonumber& &\qquad\qquad
+\frac{2\beta mb_0} {\hat\omega+mb_0}
+\frac{4(1+\beta)\left[mb_0-\beta_1(\hat\omega+mb_0)\right]\hat\omega}
{(\hat\omega+m b_0)^2-2(1-\beta)b_0^2}\ ,
\end{eqnarray}
which can be further cast into the form of
\begin{eqnarray}\nonumber
\frac{R^2[(\omega+m\Omega_0)^2-\kappa_0^2]}
{\Theta v_{s0}^2-\epsilon R\Sigma_0GY_m(\beta')}
&=&\nu^2+m^2+2\beta_1-4\beta_1^2+\mathrm{i}\nu\left[
4\beta_1-1+\frac{2(1+\beta)\omega(\omega+m\Omega_0)}
{(\omega+m\Omega_0)^2-\kappa_0^2}\right]
\\ \label{WKBJ-appendix}& & \qquad\qquad
+\frac{2\beta m\Omega_0} {\omega+m\Omega_0}
+\frac{4(1+\beta)\left[m\Omega_0
-\beta_1(\omega+m\Omega_0)\right]\omega}
{(\omega+m\Omega_0)^2-\kappa_0^2}\ .
\end{eqnarray}
First, it is clear that equation (\ref{WKBJ-appendix}) is singular
when $\omega+m\Omega=0$ and $(\omega+m\Omega_0)^2-\kappa_0^2=0$. 
These correspond to the corotation resonance and the inner and 
outer Lindblad resonances 
(Goldreich \& Tremaine 1979). In our analysis, such singular 
points do not appear due to the stationary condition $\omega=0$.

In the regime of $\nu\gg\mbox{max}(m,\ \omega,\ 1)$, equation
(\ref{WKBJ-appendix}) can be significantly simplified to
\begin{eqnarray}\label{WKBJ-appendix1}
\frac{(\omega+m\Omega_0)^2-\kappa_0^2}{\Theta v_{s0}^2
-\epsilon R\Sigma_0 GY_m(\beta')}
=\left(\frac{\nu}{R}\right)^2[1+O(\nu^{-1})]\ .
\end{eqnarray}
From Appendix A, we know $Y_m(\beta')\approx 2\pi/|\nu|$ 
in the limit of $\nu\gg \mbox{max}(m,\ 1)$. By defining the 
radial wavenumber $k$ as $\nu/R$, we can rearrange equation
(\ref{WKBJ-appendix1}) into the following form
\begin{eqnarray}\label{WKBJ-appendix2}
(\omega+m\Omega_0)^2=\kappa_0^2+k^2\Theta v_{s0}^2 
-2\pi\epsilon G|k|\Sigma_0\ ,
\end{eqnarray}
which shares the classic WKBJ form (Lou \& Fan 1998a).
According to equation (\ref{WKBJ-appendix1}), dispersion relation
(\ref{WKBJ-appendix2}) has an error of the order of $O(\nu^{-1})$. From
equation (\ref{WKBJ-appendix}), we note that errors can be reduced to
$O(\nu^{-2})$ when $\beta_1=1/4$ and $\omega\rightarrow0$; for these
special cases, a more accurate dispersion relation can be obtained
(Shu et al. 2000).

\section[]{Aligned Cases}

For aligned cases, we begin with the expression
\begin{equation}\label{a1}
f=\epsilon\left(\frac{C_1}{\bar w-C_a}-C_2\right)\ ,
\end{equation}
where the three coefficients $C_a$, $C_1$,
and $C_2$ are explicitly defined by
\begin{eqnarray}
C_a&\equiv&\frac{m^2-2(1-\beta)}{m^2+4\beta(1-\beta)}\ , \\
C_1&\equiv&\frac{[1+(1+2\beta)C_a]Y_m(\beta)}{2\beta Y_0(\beta)}\ , \\
C_2&\equiv&1-\frac{(1+2\beta)Y_m(\beta)}{2\beta Y_0(\beta)}\ .
\end{eqnarray}
Clearly, we can write equation (\ref{a1}) in the form of
$(f+\epsilon C_2)(\bar w-C_a)=\epsilon C_1$. When plotted on
a two-dimensional diagram of $(\bar w, f)$, this relation
takes a hyperbolic shape.

The three constraints $\bar w\geq0$, $f\geq0$ and
$f+\epsilon\geq0$ can also be effectively written
as $\bar w\geq0$ and $f\geq\max\{0,-\epsilon\}$.

In short, equation (\ref{a1}) depicts a hyperbolic curve in the
$(\bar w,\ f)$ diagram for specified $m$, $\beta$ and $\epsilon$
parameters. The two constraints $\bar w\geq0$ and
$f\geq\max\{0,-\epsilon\}$ enclose a region of physical interest.
In other words, only the enclosed parts of the hyperbolic curve
are physically meaningful.

For different combinations of $m$, $\beta$ and $\epsilon$, we
will discuss whether the physically meaningful solutions of the
hyperbolic curves exist and the possible ranges of the curves,
which are expressed as ranges of $\bar w$. From the hyperbolic
curves, the ranges of $f$ can be readily calculated for the
given ranges of $\bar w$.
Note that $m$ is a positive integer, $\beta\in(-1/2,\ 1/2)$
and $\epsilon\leq1$. The special case of $\epsilon=0$ is
investigated in the main text.

\begin{figure}
\begin{center}
\subfigure[]{ \label{fig5a}
\includegraphics[angle=0,scale=0.49]{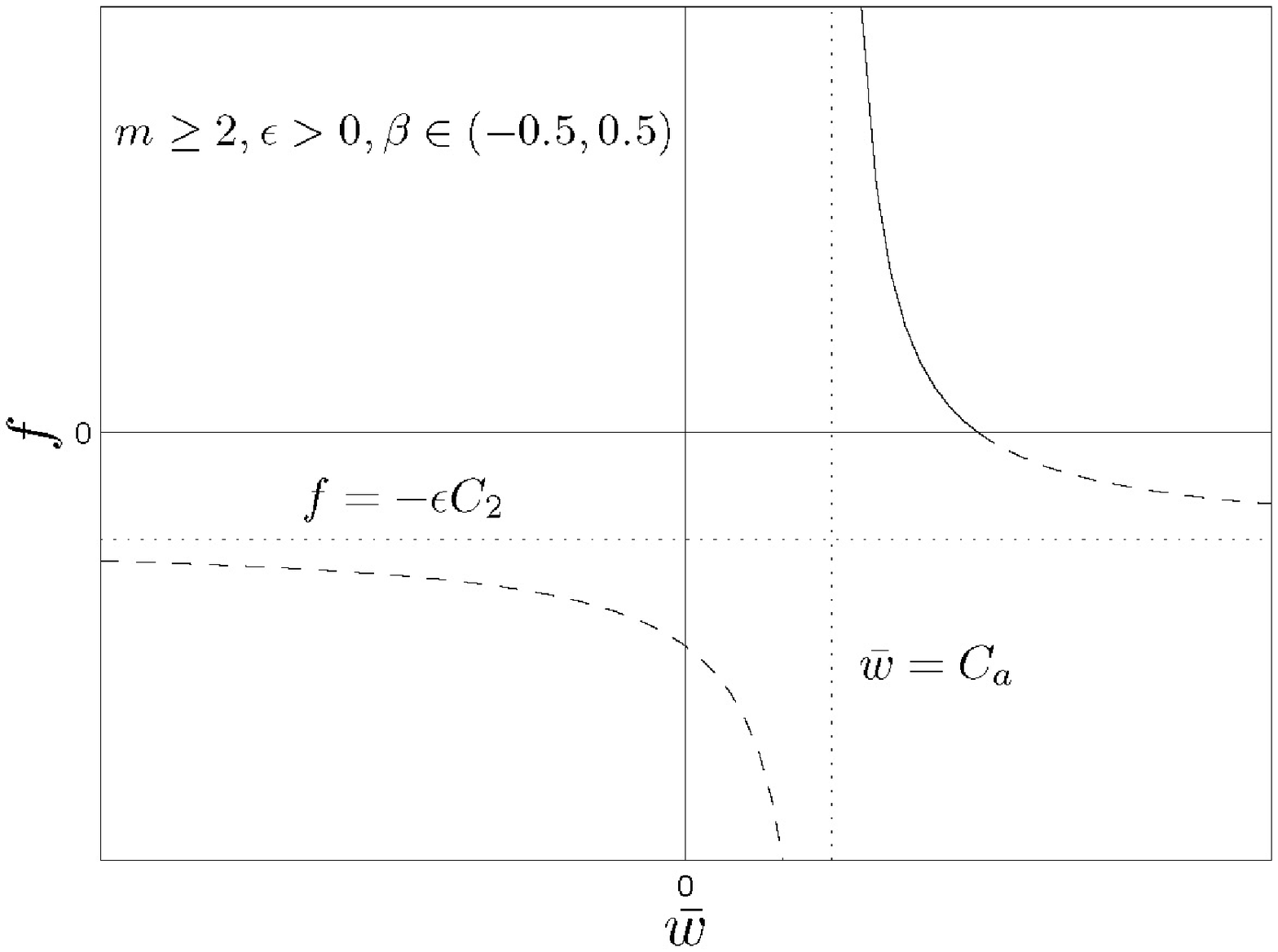}}
\hfill \subfigure[]{ \label{fig5b}
\includegraphics[angle=0,scale=0.49]{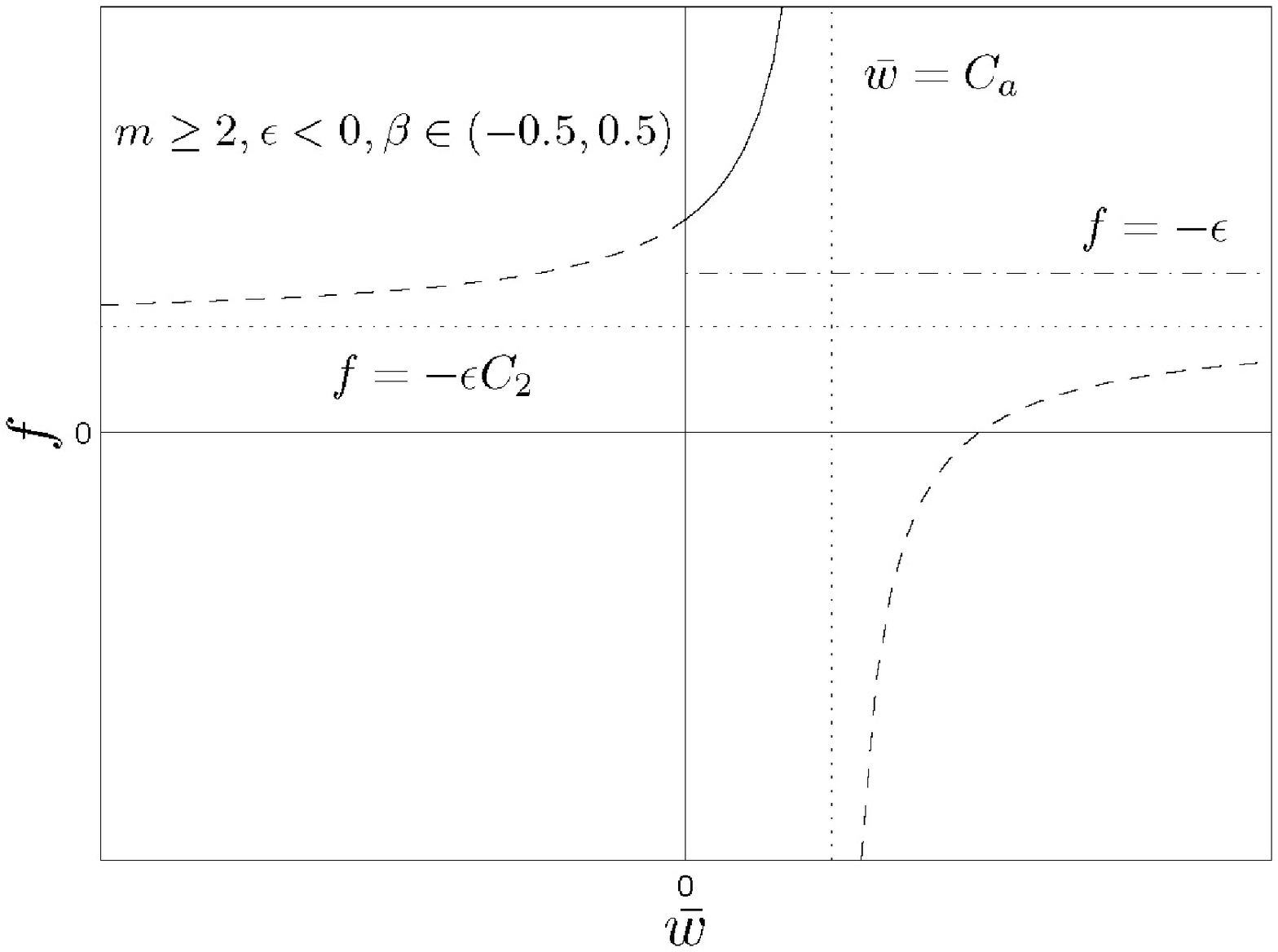}}
\caption{
Combinations of $\bar w$ (a measure for the disc temperature and
magnetic pressure together) and $f$ (a measure for the background
dark matter halo potential strength) for $m\geq2$ global
stationary MHD perturbation configurations when $\beta\in(-1/2,\
1/2)$. Panel (a) shows the case of $\epsilon>0$ (i.e., the disc
self-gravity overtakes the magnetic tension force) and panel (b)
is the case of $\epsilon<0$ (i.e., the magnetic tension force
overtakes the disc self-gravity). The physically meaningful parts
are plotted in solid curve and the unphysical parts are plotted in
dashed lines. In panel (b), the dash-dotted line indicates the
constraint on $f$ when $\epsilon<0$, due to $f+\epsilon\geq0$.
Several asymptotes are given in dotted lines, $f=-\epsilon C_2$
and $\bar w=C_a$.
}
\label{fig5}
\end{center}
\end{figure}

\begin{figure}
\begin{center}
\subfigure[]{ \label{fig6a}
\includegraphics[angle=0,scale=0.45]{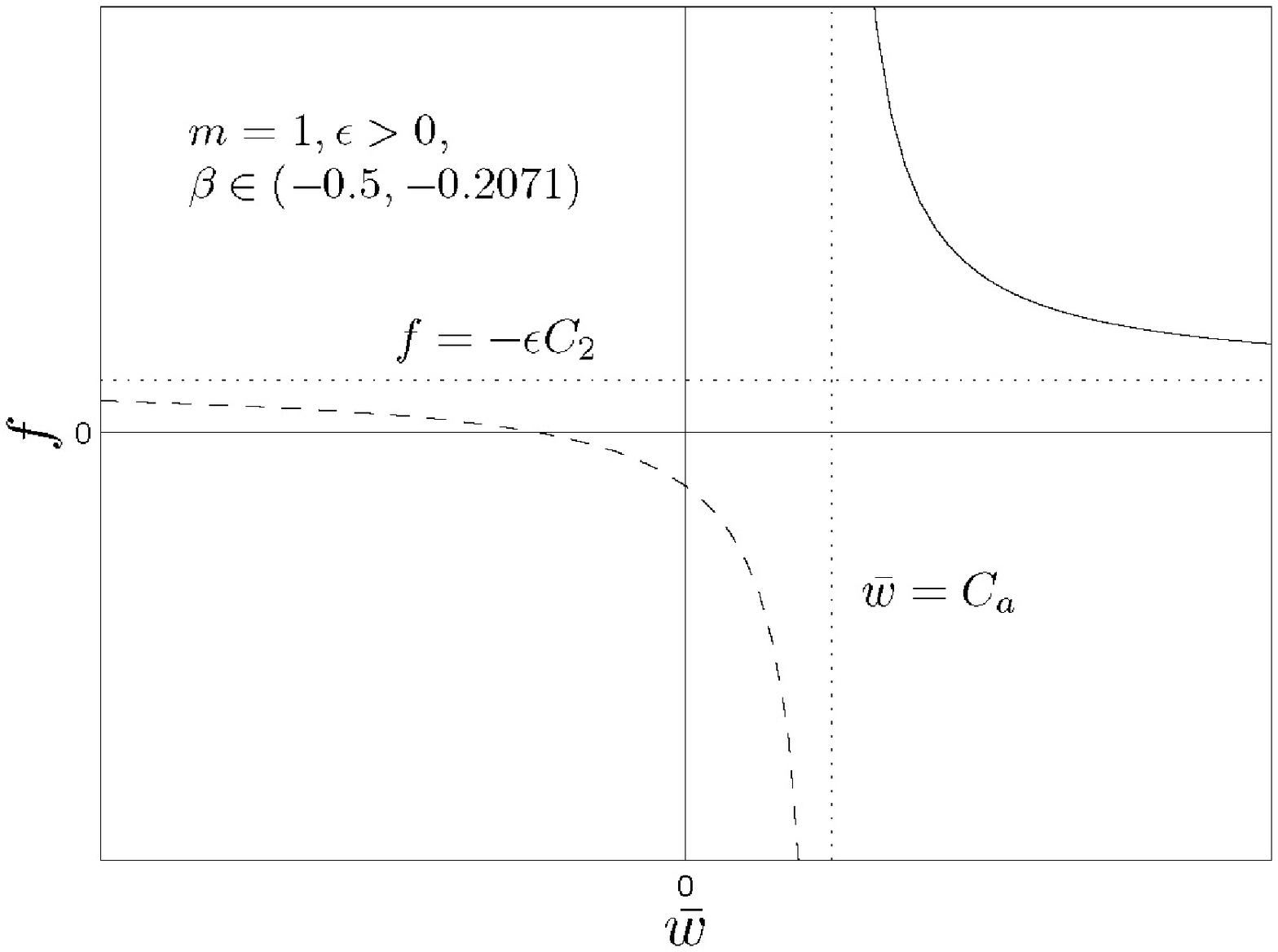}}
\hfill \subfigure[]{ \label{fig6b}
\includegraphics[angle=0,scale=0.45]{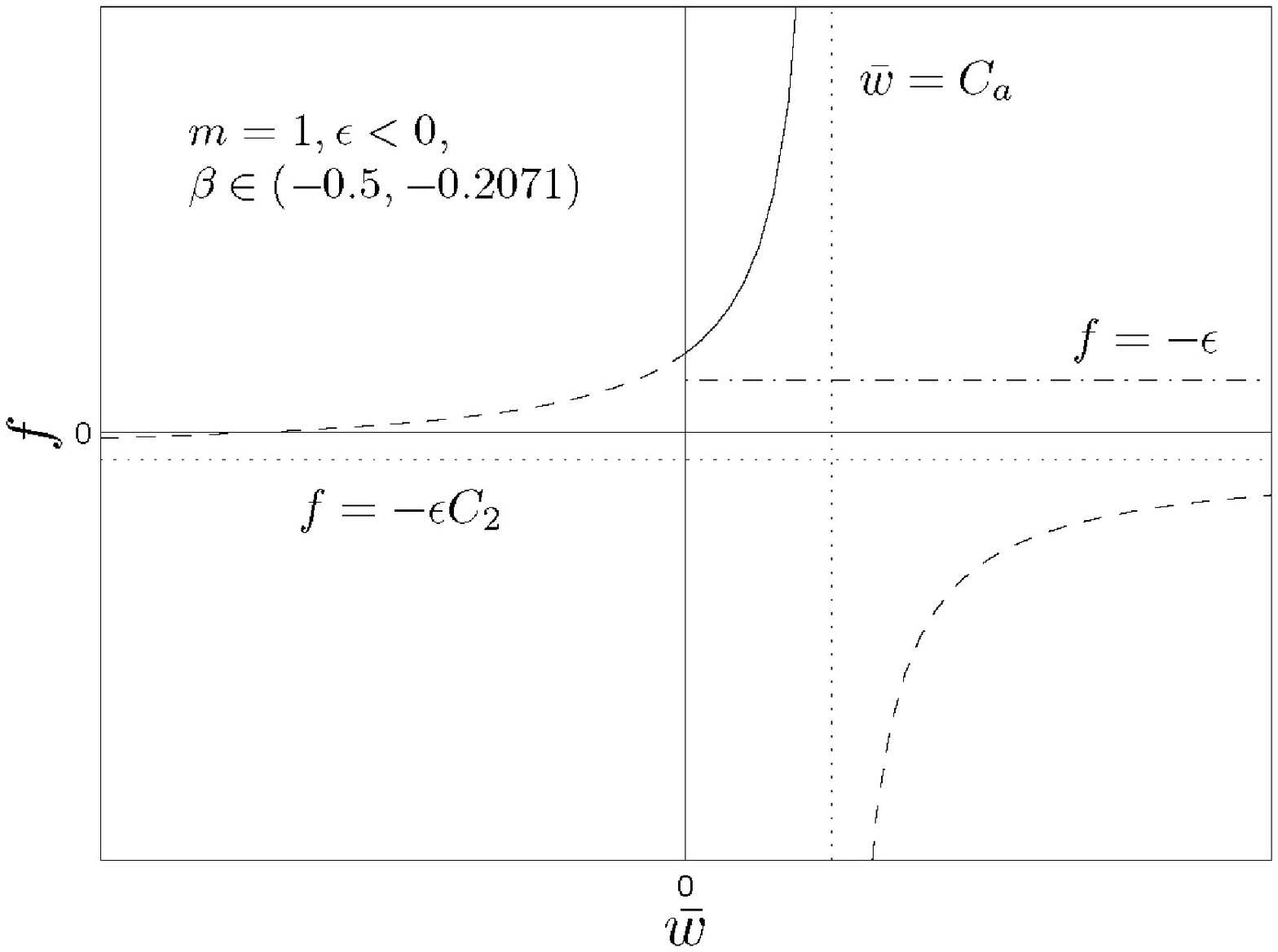}}
\hfill \subfigure[]{ \label{fig6c}
\includegraphics[angle=0,scale=0.45]{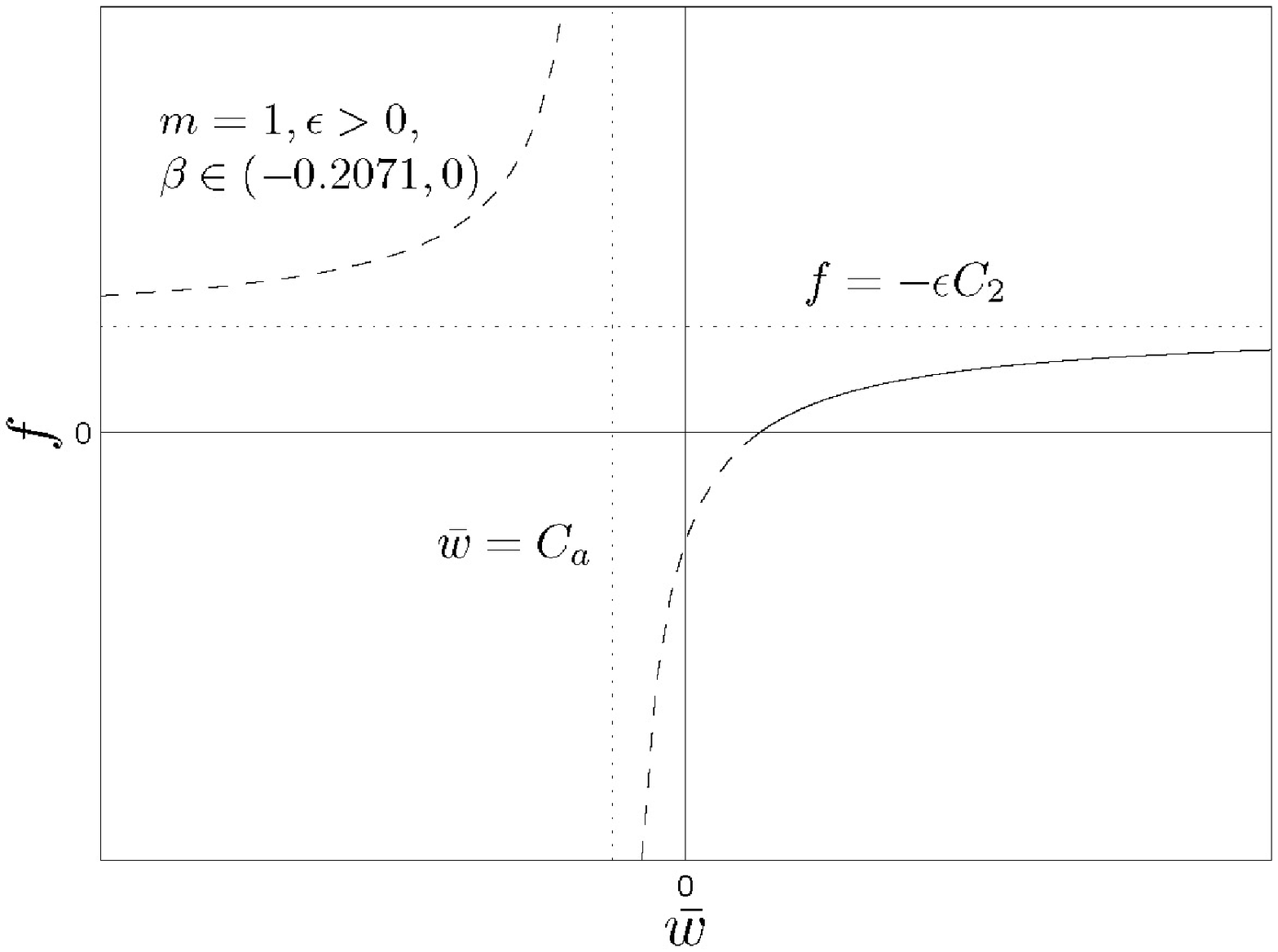}}
\hfill \subfigure[]{ \label{fig6d}
\includegraphics[angle=0,scale=0.45]{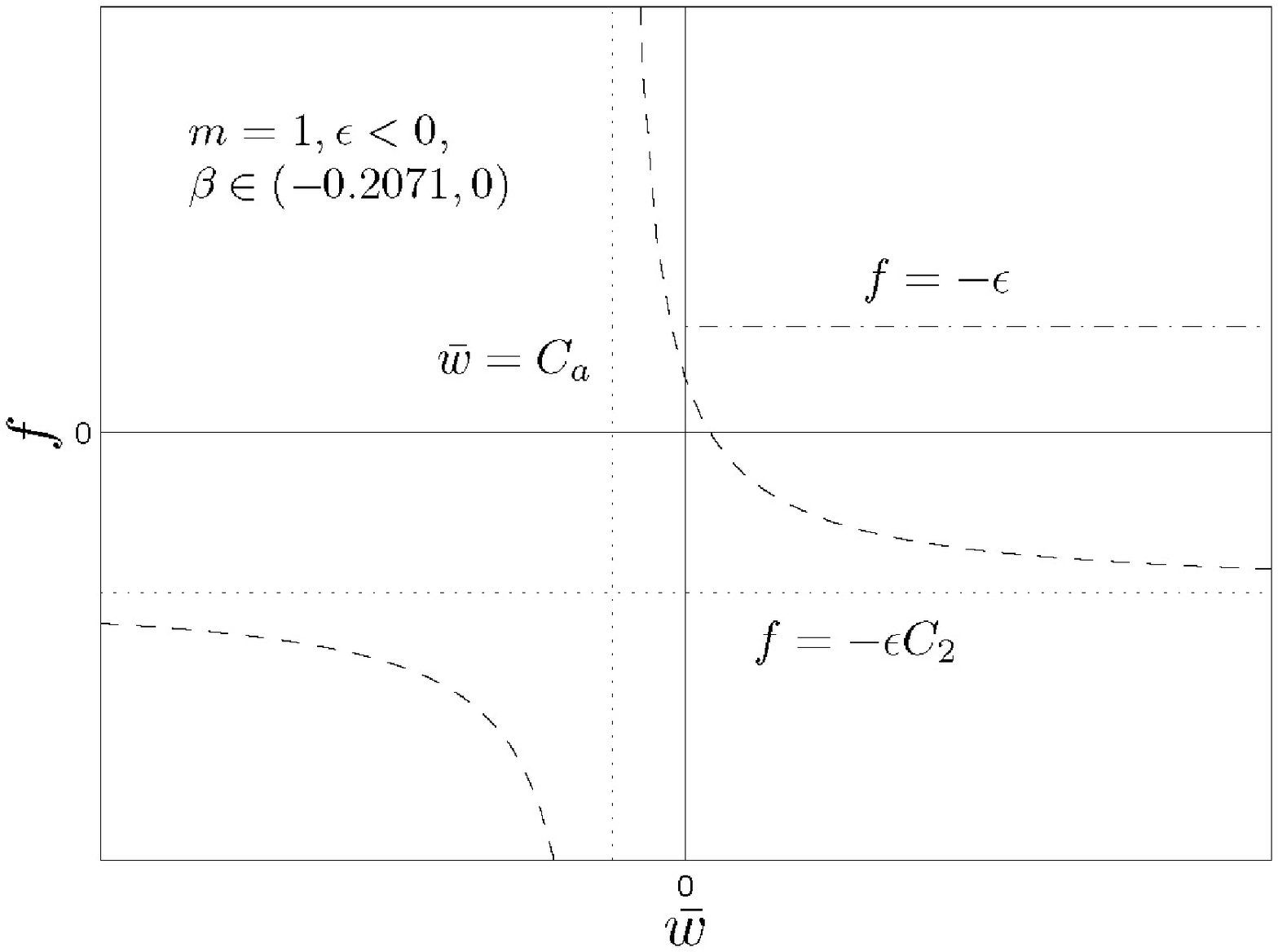}}
\hfill \subfigure[]{ \label{fig6e}
\includegraphics[angle=0,scale=0.45]{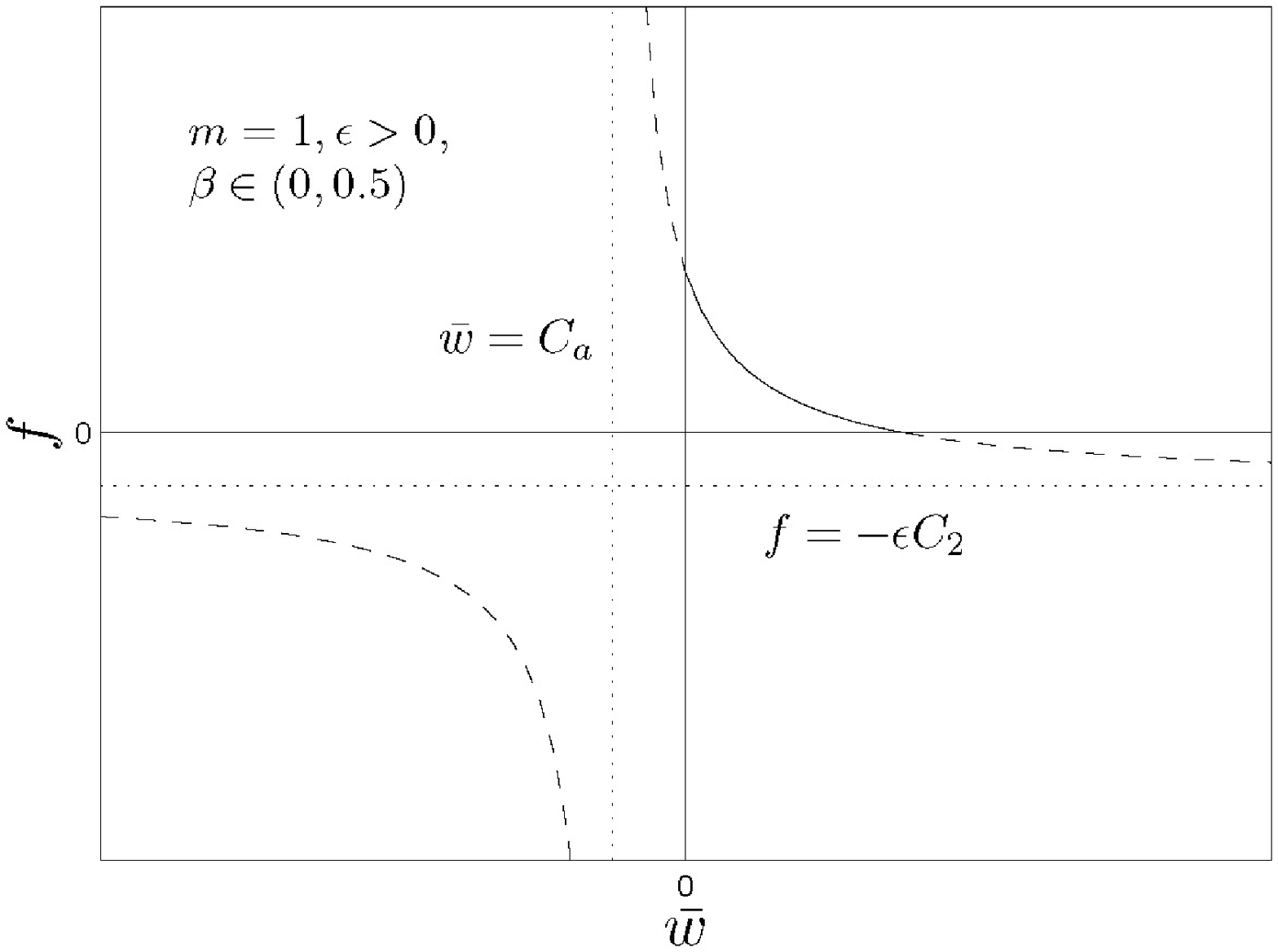}}
\hfill \subfigure[]{ \label{fig6f}
\includegraphics[angle=0,scale=0.45]{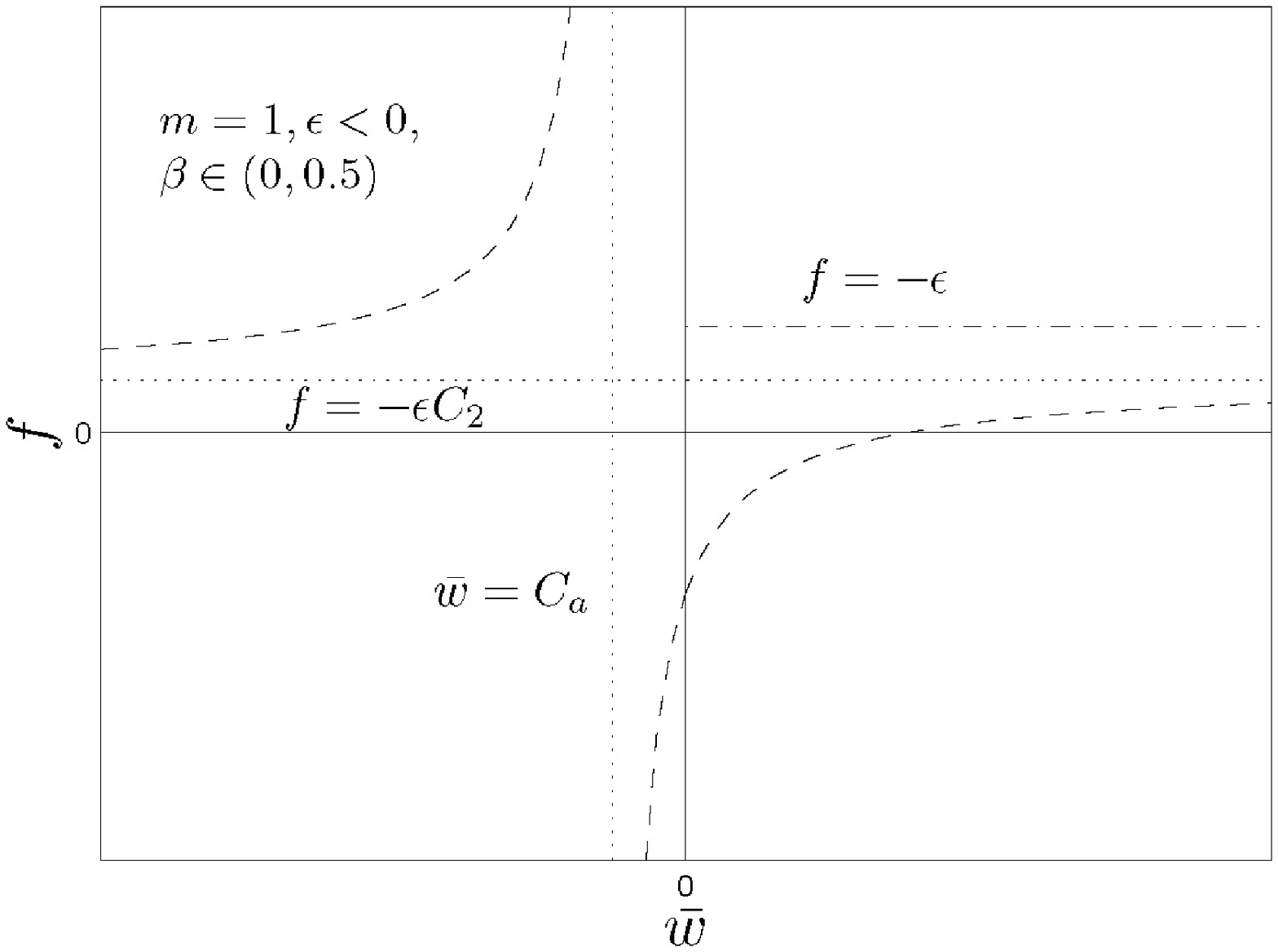}}
\caption{
Combinations of $\bar w$ (a measure for the disc temperature and
magnetic pressure together) and $f$ (a measure for the background
dark matter halo potential strength) for $m=2$ global stationary
MHD perturbation configurations with $\beta\in(-1/2,\ 1/2)$.
Panels (a), (c), (e) are the cases of $\epsilon>0$ (the disc
self-gravity overtakes the magnetic tension force), and panels
(b), (d), (f) are the cases of $\epsilon<0$ (the magnetic tension
force overtakes the disc self-gravity). The range of $\beta$ is
separated into three different segments: panels (a) and (b) are
for $\beta\in(-1/2,\ -0.2071)$; panels (c) and (d) are for
$\beta\in(-0.2071,\ 0)$; panels (e) and (f) are for $\beta\in(0,\
1/2)$. The physically meaningful parts are plotted in solid curves
and the unphysical parts are shown in dashed lines. The
dash-dotted lines indicate the constraint on $f$ with
$\epsilon<0$, due to $f+\epsilon\geq0$. Several asymptotes are
shown in dotted lines, $f=-\epsilon C_2$ and $\bar w=C_a$.
} \label{fig6}
\end{center}
\end{figure}

\subsection{Cases of $m\geq2$}

For cases of $m\geq2$, our numerical calculations show that
$1/2\leq C_a<1$, $C_1>0$ and $1/3<C_2<1$. The case $C_a=1/2$
is achieved for $m=2$ and $\beta=0$.

When $0<\epsilon\leq1$, the constraints are $f\geq0$ and
$\bar w\geq0$ [see panel (a) of Fig. C1]. We can further infer
\begin{equation}
f\geq0\ ,\qquad \bar w\geq 0\qquad
\Rightarrow\quad\frac{C_1}{(\bar w-C_a)}\geq C_2
\quad\Rightarrow\qquad
\bar w\in \left(C_a\ ,\ C_a+\frac{C_1}{C_2}\right]\ .
\end{equation}

When $\epsilon<0$, the constraints are $f+\epsilon\geq0$
and $\bar w\geq0$ [see panel (b) of Fig. C1]. We can
further infer
\begin{equation}
f+\epsilon\geq0\ ,\qquad \bar w\geq0 \qquad
\Rightarrow\qquad \mbox{max}\left\{C_a+\frac{C_1}{(C_2-1)}\ , \
0\right\}\leq\bar w<C_a\qquad\Rightarrow\qquad
\bar w\in [0,\ C_a)\ .
\end{equation}

\subsection{Cases of $m=1$}

When $m=1$, both coefficients $C_a$ and $C_1$ may become singular
at $\beta=(1-\sqrt{2})/2=-0.2071$. It follows that $C_a>0$ for
$\beta\in(-1/2,\ -0.2071)$ and $C_a<0$ for $\beta\in(-0.2071,\ 1/2)$,
respectively. We also have $C_1>0$ for $\beta\in(-1/2,\ -0.2071)
\cup(0,\ -1/2)$ and $C_1<0$ for $\beta\in(-0.2071,\ 0)$,
respectively. Finally, $C_2<0$ for $\beta\in(-1/2,\ 0)$
and $C_2>0$ for $\beta\in(0,\ 1/2)$, respectively. We now
consider several subcases separately.

\subsubsection{The subcase of $\beta\in(-1/2,\ -0.2071)$
with $C_a>0\ ,\ C_1>0\ ,\ C_2<0$\ . }

When $0<\epsilon\leq1$,
\begin{equation}
f\geq0\ ,\qquad\bar w\geq0 \qquad\Rightarrow 
\qquad\bar w\in(C_a,\ +\infty)\ ,
\end{equation}
and the range of $\bar w$ can be identified in panel (a) of Fig. C2.

When $\epsilon<0$,
\begin{equation}
f+\epsilon\geq0\ ,\qquad \bar w\geq0 \qquad
\Rightarrow\quad \mbox{max}\left\{C_a+\frac{C_1}{C_2-1}\ ,
\ 0\right\}\leq\bar w\leq C_a
\qquad \Rightarrow\qquad
\bar w\in[\ 0,\ C_a\ )\ ,
\end{equation}
and the range of $\bar w$ can be 
identified in panel (b) of Fig. C2.

\subsubsection{The subcase of $\beta\in(-0.2071,\ 0)$
with $C_a<0\ ,\ C_1<0\ ,\ C_2<0$\ . }

When $0<\epsilon\leq 1$, we infer
\begin{equation}
f\geq0\ ,\quad \bar w\geq0\quad \Rightarrow \quad
\bar w\geq\mbox{max}\left\{C_a+\frac{C_1}{C_2}\ ,\ 0\right\}
\quad\Rightarrow\quad
w\in\left[C_a+\frac{C_1}{C_2}\ ,\ +\infty\right)\ ,
\end{equation}
and the range of $\bar w$ can be
identified from panel (c) of Fig. C2.

When $\epsilon<0$, we infer
\begin{equation}
f+\epsilon\geq0\ ,\quad \bar w\geq0 \quad
\Rightarrow\quad C_a\leq\bar w\leq C_a+\frac{C_1}{C_2-1}
=\frac{-1}{1+2\beta}\ ,
\quad \bar w\geq0 \quad
\Rightarrow\quad \mbox{Such a }\bar w\mbox{ cannot exist}\ ,
\end{equation}
and no curve is within the physically allowed 
region in panel (d) of Fig. C2.

\subsubsection{The subcase of $\beta\in(0,\ 1/2)$
with $C_a<0\ ,\ C_1>0\ ,\ C_2>0$\ .}

When $0<\epsilon\leq1$, we infer
\begin{equation}
f\geq0\ ,\quad\bar w\geq0
\quad\Rightarrow\quad
\max\{C_a\ ,\ 0\}\leq\bar w\leq
C_a+\frac{C_1}{C_2}
\quad\Rightarrow\quad \bar w\in\left[0,\ C_a+\frac{C_1}{C_2}\right] \ ,
\end{equation}
and the range of $\bar w$ can be seen from panel (e) of Figure C2.

When $\epsilon<0$, we infer
\begin{equation}
f+\epsilon\geq0\ ,\quad \bar w\geq0
\quad\Rightarrow\quad \bar w< C_a\ ,\quad \bar w\geq0
\quad\Rightarrow\quad\mbox{Such a }\bar w\mbox{ cannot exist}\ ,
\end{equation}
and no curve is within the physically permitted
region in panel (f) of Figure C2.

\section[]{Several Properties of $\beta_1=1/4$ Discs}

For a 
disc system with $\beta_1=1/4$, the global stationary
dispersion relationship for MHD perturbations can be transformed into
\begin{equation}\label{b1}
f=\epsilon\left(\frac{D_1}{\bar w-D_0}-D_2\right)
\qquad\qquad\mbox{or}\qquad\qquad
(f+\epsilon D_2)(\bar w-D_0)=\epsilon D_1\ ,
\end{equation}
where
\begin{eqnarray}
D_0&\equiv &\frac{[m^2-2(1-\beta)]}{(m^2+2\beta+1/4+\nu^2)}\ ,
\\
D_1&\equiv &\frac{\pi}{2\beta Y_0(\beta)}
\left|\frac{\Gamma(m/2+1/4+\mathrm{i}\nu/2)}
{\Gamma(m/2+3/4+\mathrm{i}\nu/2)}\right|^2
\frac{[2m^2(1+\beta)+4\beta^2-7/4+\nu^2]}
{(m^2+2\beta+1/4+\nu^2)}\ ,
\\
D_2&\equiv &1-\frac{\pi(1+2\beta)}{2\beta Y_0(\beta)}
\left|\frac{\Gamma(m/2+1/4+\mathrm{i}\nu/2)}
{\Gamma(m/2+3/4+\mathrm{i}\nu/2)}\right|^2\ .
\end{eqnarray}
Numerical computations show the following results: $-8<D_0<0$
for $m=1$ and $0<D_0<1$ for $m\geq 2$, respectively; in all
circumstances, $D_1\geq 0$
and $D_1=0$ occurs when $m=1$, $\beta=-1/4$ and $\nu=0$;
in all circumstances, $0\leq D_2<1$
and $D_2=0$ occurs when $m=1$, $\beta=-1/4$ and $\nu=0$.
Note that $\beta\in (-1/2,\ 1/2)$.

Given $m$, $\nu$, $\beta$ and $\epsilon$ parameters, equation
(\ref{b1}) represents a hyperbolic curve in the $(\bar w,\ f)$
diagram. Two physical constraints are $\bar w\geq0$ and
$f\geq\max\{0,-\epsilon\}$, enclosing the physically meaningful
parts of the hyperbolic curve expressed as an area of $\bar w$.

The case of $\epsilon=0$ is special, in which $f$ becomes
arbitrary and $\bar w=D_0$. As constrained by $\bar w\geq0$,
neutral $m=1$ MHD perturbation mode cannot exist in
$\epsilon=0$ discs.

\subsection{Cases with $m\geq2$}

In these cases, we have $0<D_0<1$, $D_1>0$ and $0<D_2<1$ for $m\geq2$.
When $0<\epsilon\leq1$, the two physical constraints are $f\geq0$ and
$\bar w\geq0$. We therefore infer
\begin{equation}
f\geq0\ ,\qquad \bar w\geq 0
\qquad\Rightarrow\qquad
\frac{D_1}{\bar w-D_0}\geq D_2
\qquad\Rightarrow\qquad
\bar w\in \left(D_0\ ,\ D_0+\frac{D_1}{D_2}\right]\ .
\end{equation}
When $\epsilon<0$, the two physical constraints are
$f+\epsilon\geq0$ and $\bar w\geq0$. We then infer
\begin{equation}
f+\epsilon\geq0\ ,\quad\bar w\geq0
\quad\Rightarrow\quad
\mbox{max}\left\{D_0+\frac{D_1}{D_2-1}\ ,\ 0\right\}\leq\bar w<D_0
\quad\Rightarrow\quad
\bar w\in [\ 0,\ D_0)\ .
\end{equation}
The schemata of the above two cases ($\epsilon>0$ and $\epsilon<0$)
are the same as panels (a) and (b) of Fig. C1, respectively.

\subsection{Cases with $m=1$}

In these cases, we have $D_0<0$, $D_1>0$ and $0<D_2<1$ for
$m=1$; the $D_1=0$ and $D_2=0$ cases will be discussed 
separately below.
When $0<\epsilon\leq1$, we infer
\begin{equation}
f\geq0\ ,\quad\bar w\geq 0
\quad\Rightarrow\quad
\frac{D_1}{\bar w-D_0}\geq D_2
\quad\Rightarrow\quad
\bar w\in \left[0\ ,\ D_0+\frac{D_1}{D_2}\right]\ .
\end{equation}
When $\epsilon<0$, we infer
\begin{equation}
f+\epsilon\geq0\ ,\quad\bar w\geq0
\quad\Rightarrow\quad\bar w<D_0\ ,
\quad\bar w\geq0
\quad\Rightarrow\quad
\mbox{Such a }\bar w\mbox{ cannot exist}\ .
\end{equation}
The schemata of the above two cases ($\epsilon>0$ and $\epsilon<0$)
are the same as in panels (e) and (f) of Fig. C2, respectively.

\section{Logarithmic Spiral Discs with
$\nu\neq0\ $ and $\beta_1\neq1/4$}

When $\nu\neq0$ and $\beta_1\neq1/4$, a point in the 
($\bar w,\ f$) diagram is given by the two values 
$\bar w$ and $f$ explicitly expressed as
\begin{eqnarray}
\bar w&=&\Re(C)-\frac{\Re(B)\Im(C)}{\Im(B)}\ ,\\
f&=&-\epsilon\left\{\left[1+(1+2\beta)
\left(\Re(C)-\frac{\Re(B)\Im(C)}{\Im(B)}\right)\right]
\frac{\Im(B)}{\Im(C)}+1\right\}\ ,
\end{eqnarray}
where $\Re(\cdots)$ and $\Im(\cdots)$ represent the
real and imaginary parts of the argument, respectively,
\begin{eqnarray}
\Re(B)&\equiv&\frac{\Re[Y_m(\beta')]}{2\beta Y_0(\beta)}\ ,
\qquad\qquad
\Im(B)\equiv\frac{\Im[Y_m(\beta')]}{2\beta Y_0(\beta)}\ ,\\ \label{CRe}
\Re(C)&\equiv &\frac{[m^2-2(1-\beta)](m^2+2\beta+2\beta_1-4\beta_1^2+\nu^2)}
{(m^2+2\beta+2\beta_1-4\beta_1^2+\nu^2)^2+\nu^2(1-4\beta_1)^2}\ ,
\\ \label{CIm}
\Im(C)&\equiv &\frac{\nu(1-4\beta_1)[m^2-2(1-\beta)]}
{(m^2+2\beta+2\beta_1-4\beta_1^2+\nu^2)^2+\nu^2(1-4\beta_1)^2}\ .
\end{eqnarray}
In the limit of $\nu\rightarrow+\infty$, we have the following
asymptotic expressions to the leading orders according to
equations (\ref{Approx}), (\ref{CRe}) and (\ref{CIm}), namely
\begin{eqnarray}
\Re[Y_m(\beta')]&=&\frac{2\pi}{|\nu|}[1+O(\nu^{-2})]\ ,
\\
\Im[Y_m(\beta')]&=&\frac{\pi(1-4\beta_1)|\nu|}{\nu^3}[1+O(\nu^{-1})]\ ,
\\
\Re(C)&=&\frac{m^2-2(1-\beta)}{\nu^2}[1+O(\nu^{-2})]\ ,
\\
\Im(C)&=&\frac{(1-4\beta_1)[m^2-2(1-\beta)]}{\nu^3}[1+O(\nu^{-2})]\ .
\end{eqnarray}
To the leading order of large $\nu$, we
then have $\bar w$ and $f$ expressed as
\begin{eqnarray}\label{wvlimit}
\bar w
&=&-\frac{m^2-2(1-\beta)}{\nu^2}[1+O(\nu^{-1})]\ ,
\\
f \label{fvlimit}
&=&-\frac{\epsilon\pi|\nu|}{2\beta
Y_0(\beta)[m^2-2(1-\beta)]}[1+O(\nu^{-1})]\ .
\end{eqnarray}
According to equations (\ref{wvlimit}) and (\ref{fvlimit}) with
$\nu\rightarrow+\infty$, $\bar w<0$ when $m\geq2$ and $\bar w>0$
when $m=1$, respectively. For $m=1$, $f>0$ when $\epsilon>0$ and
$f<0$ when $\epsilon<0$, respectively.

The above analytical analysis based on the $\nu\rightarrow+\infty$
approximation gives the variation trends of $\bar w$ and $f$.
But, this approximation becomes invalid when $\nu\approx1$ or 
smaller. We resort to numerical experiments to probe solution 
properties. According to extensive numerical tests, we find 
that $\Re[Y_m(\beta')]>0$ and $\nu(1-4\beta_1)\Im[Y_m(\beta')]>0$ 
when $m\geq1$.

\subsection{Cases with $m\geq2$}

When $m\geq2\ ,$ we have $\Re(C)>0\ .$
Meanwhile, $\bar w$ can be written as
\begin{eqnarray}
\bar w=\Re(C)
\left\{1-\frac{\nu(1-4\beta_1)}
{(m^2+2\beta+2\beta_1-4\beta_1^2+\nu^2)}
\frac{\Re[Y_m(\beta')]}{\Im[Y_m(\beta')]}\right\}\ .
\end{eqnarray}
Extensive numerical results show that
\begin{eqnarray*}
\frac{\nu(1-4\beta_1)}{(m^2+2\beta+2\beta_1-4\beta_1^2+\nu^2)}
\frac{\Re[Y_m(\beta')]}{\Im[Y_m(\beta')]}>
\frac{\nu(1-4\beta_1)}{(m^2+1+2\beta_1-4\beta_1^2+\nu^2)}
\frac{\Re[Y_m(\beta')]}{\Im[Y_m(\beta')]}>1\ .
\end{eqnarray*}
Therefore when $m\geq2$, we have $\bar w<0$ by extensive
numerical explorations.

\subsection{Cases with $m=1$}

In these cases, we have
\begin{eqnarray}\nonumber
\bar w=\frac{-(1-2\beta)}
{[(1+2\beta+2\beta_1-4\beta_1^2+\nu^2)^2+\nu^2(1-4\beta_1)^2]}
\left\{1+2\beta+2\beta_1-4\beta_1^2+\nu^2-\nu(1-4\beta_1)
\frac{\Re[Y_1(\beta')]}{\Im[Y_1(\beta')]}\right\}\ .
\end{eqnarray}
Extensive numerical explorations
show the following inequalities
\begin{eqnarray*}
1+2\beta+2\beta_1-4\beta_1^2+\nu^2-\nu(1-4\beta_1)
\frac{\Re[Y_1(\beta')]}{\Im[Y_1(\beta')]}<
2+2\beta_1-4\beta_1^2+\nu^2-\nu(1-4\beta_1)
\frac{\Re[Y_m(\beta')]}{\Im[Y_m(\beta')]}<0\ ,
\end{eqnarray*}
and therefore we have $\bar w>0\ .$

The $f$ parameter can be written as
\begin{eqnarray}
f=\epsilon\left\{\frac{[1+(1+2\beta)\bar w]\Re(B)}
{[\bar w-\Re(C)]}-1\right\}\ .
\end{eqnarray}
We can also derive the following relationship
\begin{eqnarray*}
\bar
w-\Re(C)=-\frac{\Re(B)\Im(C)}{\Im(B)}=\frac{\nu(1-4\beta_1)(1-2\beta)}
{(1+2\beta+2\beta_1-4\beta_1^2+\nu^2)^2+\nu^2(1-4\beta_1)^2}
\frac{\Re[Y_1(\beta')]}{\Im[Y_1(\beta')]}>0\ .
\end{eqnarray*}

For $\epsilon<0$, we require $f+\epsilon\geq0$
on the ground of physics. However,
\begin{eqnarray*}
f+\epsilon=\epsilon\left\{
\frac{[1+(1+2\beta)\bar w]\Re(B)}
{[\bar w-\Re(C)]}\right\}<0\ ,
\end{eqnarray*}
leading to an obvious contradiction. Therefore for $\epsilon<0$,
stationary $m=1$ MHD perturbation modes do not exist.

When $\epsilon>0$, the physical constraint becomes
$f\geq0$, giving the following inequalities
\begin{eqnarray*}
\qquad
\frac{[1+(1+2\beta)\bar w]\Re(B)}{[\bar w-\Re(C)]}\geq1
\quad\Rightarrow\quad
(1+2\beta)\Re(C)+1-\frac{\Im(C)}{\Im(B)}[(1+2\beta)\Re(B)-1]\geq0\ .
\end{eqnarray*}
Extensive numerical examples show that the above inequalities always
hold true. We therefore infer numerically that when $\epsilon>0$ and
$f>0$, stationary $m=1$ MHD perturbation neutral modes do exist.

\section{Axisymmetric cases}

\begin{figure}
\begin{center}
\includegraphics[angle=0,scale=0.49]{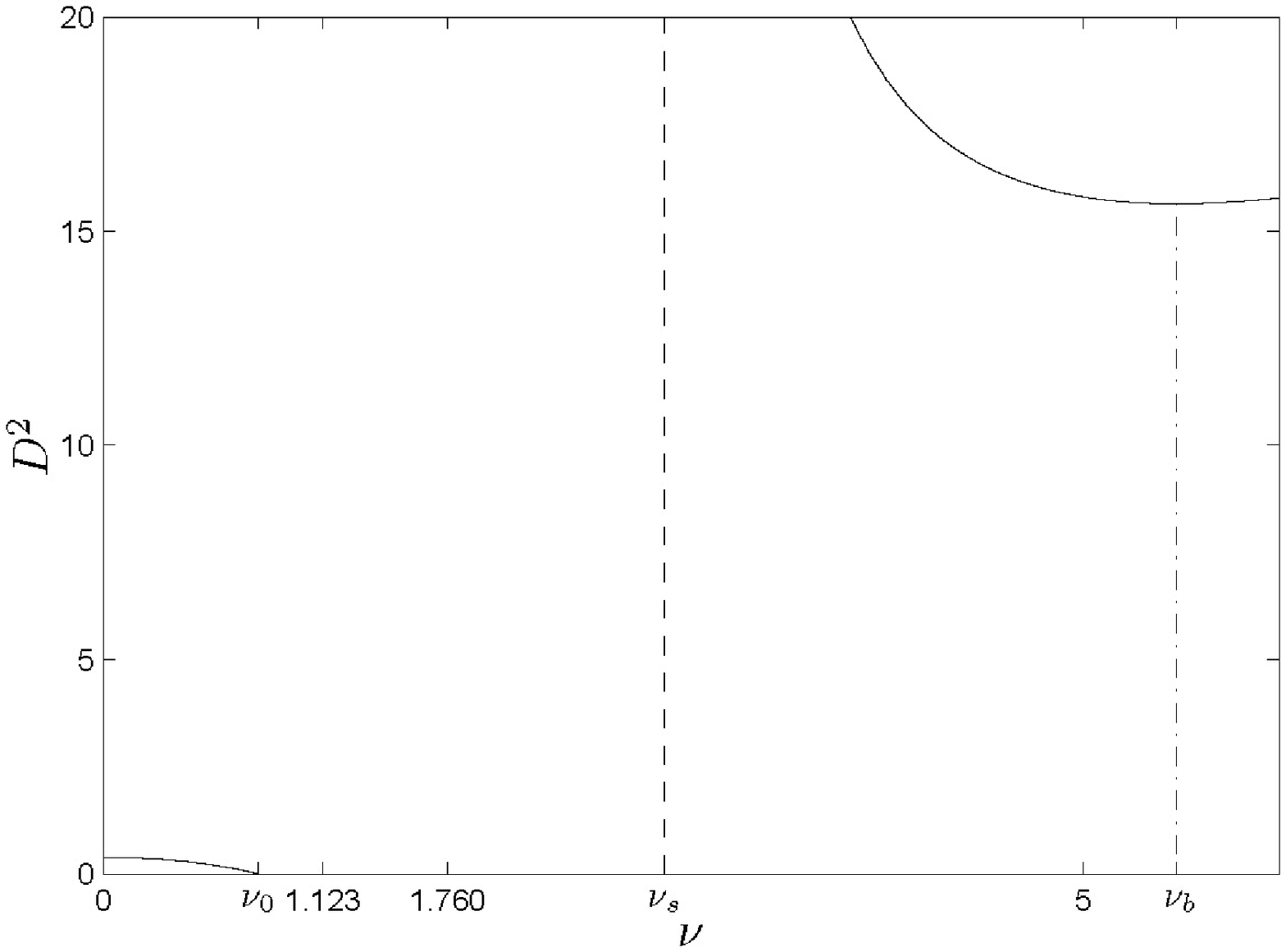}
\caption{The axisymmetric ($m=0$) marginal instability curves
(upper-right and lower-left solid curves) of $D^2\ $ ($1/\bar w$)
(the reciprocal of the disc temperature and magnetic pressure
together) versus $\nu$ (dimensionless radial wavenumber for
perturbations) for specified values of $f/\epsilon=0.5$, the
scale-free index $\beta_1=1/4$ for perturbations and the
background scale-free index $\beta=0$. Here, $\epsilon$ is taken
to be positive. Five specific points, $\nu_0$, $1.123$, $1.760$,
$\nu_s$ and $\nu_b$, can be found along the $\nu$-axis.
The dashed line of $\nu_s$ gives the asymptote and the
dash-dotted line marks the inflection point $\nu_b$.}
\end{center}
\end{figure}

When $m=0$, $\beta_1=1/4$ and $\beta=0$, we have
\begin{equation}\label{F1}
\frac{1}{\bar w}\equiv D^2=\frac{\mathcal{E}-{\mathcal N}_0(\nu)}
{{\mathcal N}_0(\nu)-8\mathcal{E}/(1+4\nu^2)}\ ,
\end{equation}
where the Kalnajs function ${\mathcal N}_0(\nu)$ is given by 
definition (\ref{Kalnajs}) and $\mathcal{E}$ is defined as
\begin{equation}\label{F2}
\mathcal{E}=1+{f}/{\epsilon}\ .
\end{equation}
Since we require inequality $f+\epsilon>0$ to hold, it follows that
$\mathcal{E}=1+f/\epsilon<0$ if $\epsilon<0$. When $\mathcal{E}<0$,
$\bar w$ remains always negative and thus unphysical. For this
reason, we only focus on the situation of $\epsilon>0$ and thus
$\mathcal{E}\geq1$.

When $\mathcal{E}-{\mathcal N}_0(\nu)=0$, one has $D^2=0$ by
equation (\ref{F1}). Numerical explorations show that ${\mathcal
N}_0(\nu)$ decreases with increasing $\nu$. When
$\mathcal{E}>{\mathcal N}_0(0) =4.377$, a positive
$\mathcal{E}-{\mathcal N}_0(\nu)$ can be always guaranteed.
We adopt the notation $\nu_0$ for the specific $\nu$ making
$\mathcal{E}-{\mathcal N}_0(\nu_0)=0$. Hence ${\mathcal
N}_0(\nu_0)=\mathcal{E}\geq 1$ corresponds to
$\nu_0\leq1.123$ because ${\mathcal N}_0(1.123)=1$. When
$\mathcal{E}>4.377$, there is no such $\nu_0$ root.

The singular point $\nu_s$ is to make
${\mathcal N}_0(\nu_s)-8\mathcal{E}/(1+4\nu_s^2)=0$ in equation
(\ref{F1}). Extensive numerical explorations show that $(1+4\nu^2)
{\mathcal N}_0(\nu)$ increases with increasing $\nu$. Therefore,
the singular point $\nu_s$ should be larger than 1.760 which is
the root of equation ${\mathcal N}_0(\nu)-8/(1+4\nu^2)=0$.

From equation (\ref{F1}), we readily obtain
\begin{eqnarray*}
\frac{\partial D^2}{\partial \mathcal{E}}=
\frac{(1+4\nu^2){\mathcal N}_0(\nu)(4\nu^2-7)}
{[(1+4\nu^2){\mathcal N}_0(\nu)-8\mathcal{E}]^2}\ .
\end{eqnarray*}
Thus as $\mathcal{E}=1+f/\epsilon$ increases, $D^2$
grows for $\nu>\sqrt{7}/2=1.323$ and decreases for
$0<\nu<\sqrt{7}/2=1.323$, respectively.

Meanwhile, we can also readily derive
\begin{eqnarray*}
\frac{\partial D^2}{\partial\nu}
=\frac{64\nu\mathcal{E}} {[(1+4\nu^2){\mathcal
N}_0(\nu)-8\mathcal{E}]^2}\left[{\mathcal
N}_0(\nu)-\frac{{\mathcal
N}'_0(\nu)(1+4\nu^2)(4\nu^2-7)}{64\nu}-\mathcal{E}\right]\ .
\end{eqnarray*}
We define a special value $\nu_b$ which is
the solution $\nu$ of the following equation
\begin{eqnarray}\label{F3}
{\mathcal N}_0(\nu)-\frac{{\mathcal
N}'_0(\nu)(1+4\nu^2)(4\nu^2-7)}{64\nu}-\mathcal{E}=0\ .
\end{eqnarray}
Numerical explorations clearly indicate that $\nu_b\geq 3.058$
which is the root of equation (\ref{F3}) with $\mathcal{E}=1$.
With the increase of $\nu$, $D^2$ decreases when $\nu<\nu_b$
and grows when $\nu>\nu_b$, respectively.
Moreover, numerical results show $\nu_b>\nu_s$.

Finally, we have an order of inequalities
$\nu_0\leq 1.123<1.323<1.760\leq\nu_s<\nu_b$ (see Fig. F1). When
$\nu\in[0,\ \nu_0]\bigcup(\nu_s,\ +\infty)$, we have $D^2\geq0$.
More specifically, $D^2$ is a decreasing function of $\nu$ when
$\nu\in[0,\nu_0]\bigcup(\nu_s,\ \nu_b)$ and an increasing function
of $\nu$ when $\nu\in(\nu_b,+\infty)$. In addition, $D^2$ is a
decreasing function of $\mathcal E$ when $\nu\in[0,\ \nu_0]$ and an
increasing function of $\mathcal E$ when $\nu\in(\nu_s,\ +\infty)$.
When ${\mathcal E}>4.377$, both $\nu_0$ and the range $[0,\ \nu_0]$
disappear. Since ${\mathcal E}=1+f/\epsilon$, the variation trend
of $D^2$ as $f$ and/or $\epsilon$ change can be readily determined.

\section{Angular Momentum Flux Transport}

In this appendix, we provide several proofs and discussions for
the angular momentum flux transport associated with global
coplanar MHD perturbations in a scale-free disc system with an
isopedic magnetic field geometry. In general, such angular
momentum flux transport contains three separate contributions
(Lynden-Bell \& Kalnajs 1972; Fan \& Lou 1999; Shen, Liu 
\& Lou 2005), namely, the flow advection transport 
$\Lambda^{\mathrm{A}}$ defined by
\begin{equation}\label{AdvectionAM}
\Lambda^{\mathrm{A}}\equiv R^2\Sigma_0\int_0^{2\pi}
\mathrm{d}\theta\Re(v_{R1})\Re(v_{\theta1})\ ,
\end{equation}
the gravity torque flux transport $\Lambda^{\mathrm{G}}$ defined by
\begin{equation}\label{GravityAM}
\Lambda^\mathrm{G}\equiv\frac{1}{4\pi G}
\int_0^{2\pi}\mathrm{d}\theta\int^{\infty}_{-\infty}\mathrm{d}z
R\left[\frac{\partial \Re(\phi_1)}{\partial \theta}\right]
\left[\frac{\partial \Re(\phi_1)}{\partial R}\right]\ ,
\end{equation}
and the magnetic torque flux transport $\Lambda^\mathrm{B}$
defined by
\begin{equation}\label{MagneticAM}
\Lambda^\mathrm{B}\equiv -\frac{R^2}{4\pi}
\int_0^{2\pi}\mathrm{d}\theta\int^{\infty}_{-\infty}\mathrm{d}z
\Re(b_R)\Re(b_\theta)\ .
\end{equation}
The three-dimensional gravitational potential perturbation
$\phi_1(R,\theta,z)$ is associated with a Fourier harmonic
component of a coplanar logarithmic spiral perturbation in 
the surface mass density
$\Sigma_1=c_1R^{-2\beta'-1}\exp(\mathrm{i}m\theta)=
c_1R^{-2\beta-1+\mathrm{i}\nu}\exp(\mathrm{i}m\theta)
=c_1R^{-2\beta-1}\exp[\mathrm{i}(m\theta+\nu\ln R)]$ where $c_1$
is a small constant amplitude coefficient. By the convention of
our notations, $\phi_1(R,\theta,0)\equiv\Phi_1$ with $\Phi_1$ 
being introduced in equation (\ref{Phi1}). Here, $b_R$ and 
$b_\theta$ represent the two components of the magnetic field 
perturbation in the isopedic magnetic field. From equation 
(2.14) of Shu \& Li (1997), we have 
\begin{equation}\label{brbtheta}
b_R=\frac{1}{2\pi G\Lambda}
\frac{\partial\phi_1}{\partial R}\ ,
\qquad\qquad
b_\theta=\frac{1}{2\pi G\Lambda R}\frac{\partial\phi_1}
{\partial\theta}\ .
\end{equation}
By definition $2\pi G\Lambda\equiv\lambda {G}^{1/2}$
and $1/\lambda^2=1-\epsilon$, we then have
\begin{equation}\label{magtogravity}
\Lambda^\mathrm{B}=-\frac{R^2}{4\pi}
\int_0^{2\pi}\mathrm{d}\theta\int^{\infty}_{-\infty}\mathrm{d}z
\Re(b_R)\Re(b_\theta) =-\frac{1}{4\pi G\lambda^2}
\int_0^{2\pi}\mathrm{d}\theta\int^{\infty}_{-\infty}\mathrm{d}z R
\left[\frac{\partial \Re(\phi_1)}{\partial \theta}\right]
\left[\frac{\partial \Re(\phi_1)}{\partial
R}\right]=(\epsilon-1)\Lambda^\mathrm{G}\ .
\end{equation}
Corresponding to $\Sigma_1=c_1R^{-2\beta-1}
\exp[\mathrm{i}(m\theta+\nu\ln R)]$, the gravitational
potential perturbation $\phi_1$ can be written as (e.g.,
Binney \& Tremaine 1987)
\begin{equation}\label{potentialPerturb}
\phi_1(R,\theta,z)=-2\pi G c_1\exp(\mathrm{i}m\theta)\int_0^\infty
\mathrm{d}k\exp(-k|z|)J_m(kR)\int_0^\infty \mathrm{d}R'
R'^{-2\beta_1+\mathrm{i}\nu} J_m(kR')\ ,
\end{equation}
where $J_m(x)$ is the cylindrical Bessel function
of order $m$ with an argument $x$. We then have
\begin{equation}\label{Rephi}
\Re(\phi_1)= -2\pi G c_1 \int_0^\infty \mathrm{d}k
\exp(-k|z|)J_m(kR)\int_0^\infty \mathrm{d}R'
R'^{-2\beta_1}\cos(m\theta+\nu\ln R') J_m(kR')\ .
\end{equation}
where $c_1$ is taken to be real.

It follows immediately that
\begin{equation}\label{Rephipartialtheta}
\frac{\partial \Re(\Phi_1)}{\partial\theta}=2\pi G mc_1
\int_0^\infty\mathrm{d}k_a\exp(-k_a|z|)J_m(k_aR)\int_0^\infty
\mathrm{d}R_aR_a^{-2\beta_1}\sin(m\theta+\nu\ln R_a)J_m(k_aR_a)\ ,
\end{equation}
and
\begin{equation}\label{RephipartialR}
\frac{\partial \Re(\Phi_1)}{\partial R}
=-2\pi G c_1\int_0^\infty
\mathrm{d}k_b \exp(-k_b|z|)
\left[-k_bJ_{m+1}(k_bR)+\frac{m}{R}J_m(k_bR)\right]\int_0^\infty
\mathrm{d}R_b R_b^{-2\beta_1}\cos(m\theta+\nu\ln R_b)J_m(k_bR_b)\ .
\end{equation}
Using the following relations
\begin{equation}\label{intZ}
\int^{\infty}_{-\infty}\mathrm{d}z\exp[-(k_a+k_b)|z|]
=2\int^{\infty}_0\mathrm{d}z\exp[-(k_a+k_b)z]
=\frac{2}{(k_a+k_b)}
\end{equation}
and
\begin{equation}\label{inttheta}
\begin{split}
&\int_0^{2\pi}\mathrm{d}\theta\cos(m\theta+\nu\ln
R_b)\sin(m\theta+\nu\ln R_a)\\
\qquad\qquad &\qquad
=\frac{1}{2}\int_0^{2\pi}\mathrm{d}\theta\{\sin[\nu\ln(R_a/R_b)]
+\sin[2m\theta+\nu\ln (R_aR_b)]\}
=\pi\sin\left[\nu\ln\left(\frac{R_a}{R_b}\right)\right]\ ,
\end{split}
\end{equation}
we obtain the gravity torque flux transport
\begin{equation}\label{LambdaG1}
\begin{split}
\Lambda^\mathrm{G}=-2m\pi^2Gc_1^2R\int^{\infty}_0
\mathrm{d}k_a\mathrm{d}k_b\mathrm{d}R_a \mathrm{d}R_b
R_a^{-2\beta_1}R_b^{-2\beta_1}J_m(k_aR_a)J_m(k_bR_b)\\
\times\sin\left[\nu\ln\left(\frac{R_a}{R_b}\right)\right]
\frac{J_m(k_aR)}{(k_a+k_b)}
\left[\frac{m}{R}J_m(k_bR)-k_bJ_{m+1}(k_bR)\right]\ .
\end{split}
\end{equation}
By taking the following integral transformations
\begin{equation}\label{integraltrans}
x_a=k_aR\ ,\quad x_b=k_bR\ ,\quad y_a=\frac{R_a}{R}\ ,\quad
y_b=\frac{R_b}{R}\quad\Rightarrow\quad k_a=\frac{x_a}{R}\ ,\quad
k_b=\frac{x_b}{R}\ ,\quad R_a=y_aR\ ,\quad R_b=y_bR\ ,
\end{equation}
we arrive at
\begin{equation}\label{LambdaG3}
\Lambda^\mathrm{G}=-2m\pi^2Gc_1^2R^{1-4\beta_1}
\mathcal{M}(\beta_1, m, \nu)\ ,
\end{equation}
where the function $\mathcal{M}(\beta_1, m, \nu)$ is defined by
\begin{equation}\label{MgravityAMF}
\mathcal{M}(\beta_1, m, \nu)\equiv\int^{\infty}_0
\mathrm{d}x_a\mathrm{d}x_b\mathrm{d}y_a \mathrm{d}y_b
(y_ay_b)^{-2\beta_1}J_m(x_ay_a)J_m(x_by_b)
\sin\left[\nu\ln\left(\frac{y_a}{y_b}\right)\right]
\frac{J_m(x_a)}{x_a+x_b}\left[mJ_m(x_b)-x_bJ_{m+1}(x_b)\right]\ .
\end{equation}
We correct here a few errors in Appendix C of Shen et al. (2005):
$J_m(k'r)$ is redundant in their equations (C1), (C2) and (C4);
$J_m(x')$ is redundant in equation (C6) and in the expression of
$\Xi^\mathrm{G}$ below equation (C8), and the heavy solid curve
of $\Xi^\mathrm{G}$ in their Figure C1 is thus invalid; an extra
numerical factor 2 should be multiplied in their equations (C4),
(C6) and (C8), respectively.

According to equations (\ref{vr1stationary})
and (\ref{vtheta1stationary}), we have
\begin{equation}
v_{R1}
=2mc_1R^{-2\beta_1+\beta}\mathcal{K}_R\mathrm{i}
\exp[{\mathrm{i}(m\theta+\nu\ln R)}]\ , \qquad v_{\theta1}
=-c_1R^{-2\beta_1+\beta}\mathcal{K}_\theta
\exp[{\mathrm{i}(m\theta+\nu\ln R)}]\ ,
\end{equation}
where
\begin{eqnarray*}
\quad \mathcal{K}_R\equiv\frac{\left[w\Theta b_0^2 -\epsilon c_0
GY_m(\beta')\right](1-\beta')}{c_0b_0\left[m^2-2(1-\beta)\right]}\ ,
\qquad\qquad
\mathcal{K}_\theta\equiv\frac{\left[w\Theta b_0^2 -\epsilon c_0
GY_m(\beta')\right]\left[m^2-2\beta'(1-\beta)\right]}
{c_0b_0\left[m^2-2(1-\beta)\right]}\ .
\end{eqnarray*}
We therefore come to
\begin{equation}\label{Revr}
\Re(v_{R1})=-2mc_1R^{-2\beta_1+\beta}\left[\Re(\mathcal{K}_R)
\sin(m\theta+\nu\ln R)
+\Im(\mathcal{K}_R)\cos(m\theta+\nu\ln R)\right]\ ,
\end{equation}
and
\begin{equation}\label{Revtheta}
\Re(v_{\theta1})=-c_1R^{-2\beta_1+\beta}\left[\Re(\mathcal{K}_\theta)
\cos(m\theta+\nu\ln R)
-\Im(\mathcal{K}_\theta)\sin(m\theta+\nu\ln R)\right]\ .
\end{equation}
Since
\begin{eqnarray*}
\int^{2\pi}_0\mathrm{d}\theta\left[\Re(\mathcal{K}_R)
\sin(m\theta+\nu\ln R)+\Im(\mathcal{K}_R)\cos(m\theta+\nu\ln R)\right]
\left[\Re(\mathcal{K}_\theta)\cos(m\theta+\nu\ln R)
-\Im(\mathcal{K}_\theta)\sin(m\theta+\nu\ln R)\right]
\\
=\frac{1}{2}\int^{2\pi}_0\mathrm{d}\theta\left\{-\Re(\mathcal{K}_R)
\Im(\mathcal{K}_\theta)\left[1-\cos(2m\theta+2\nu\ln R)\right]
+\Re(\mathcal{K}_\theta)\Im(\mathcal{K}_R)\left[1+\cos(2m\theta+2\nu\ln
R)\right]\right\}
\\
=\pi\left[\Re(\mathcal{K}_\theta)\Im(\mathcal{K}_R)-\Re(\mathcal{K}_R)
\Im(\mathcal{K}_\theta)\right]\ ,
\end{eqnarray*}
we derive in a straightforward manner that
\begin{equation}\label{LambdaA}
\Lambda^{\mathrm{A}}=R^2\Sigma_0\int_0^{2\pi}\mathrm{d}\theta
\Re(v_{R1})\Re(v_{\theta1})=2m\pi
c_1^2c_0R^{1-4\beta_1}\left[\Re(\mathcal{K}_\theta)\Im(\mathcal{K}_R)-\Re(\mathcal{K}_R)
\Im(\mathcal{K}_\theta)\right]\ .
\end{equation}
Finally, we write the total angular momentum
flux $\Lambda_\mathrm{total}$ as
\begin{equation}\label{totalLambda}
\Lambda_\mathrm{total}=\Lambda^{\mathrm{A}}
+\Lambda^{\mathrm{G}}+\Lambda^{\mathrm{B}}
=\Lambda^{\mathrm{A}}+\epsilon\Lambda^{\mathrm{G}}
=2mR^{1-4\beta_1}c_1^2\bigg\{\pi c_0
\left[\Re(\mathcal{K}_\theta)\Im(\mathcal{K}_R)
-\Re(\mathcal{K}_R)\Im(\mathcal{K}_\theta)\right]
-\epsilon\pi^2G\mathcal{M}(\beta_1,\ m,\ \nu)\bigg\}\ .
\end{equation}
As expected when $m=0$, we have $\Lambda_\mathrm{total}=0$. When
$\beta_1=1/4$, $\Lambda_\mathrm{total}$ is independent of $R$.
We note that all these foregoing analyses
are performed under the stationary assumption of $\omega=0$.

\end{appendix}

\bsp

\label{lastpage}

\begin{thebibliography}{99}
\bibitem{bal03}Balbus S. A., 2003, ARA\&A, 41, 555
\bibitem{bal98}Balbus S. A., Hawley J. F., 1998, Rev. Mod. Phys., 70, 1
\bibitem{bas94}Basu S., Mouschovias T. C., 1994, ApJ, 432, 720
\bibitem{becketal96}Beck R., Brandenburg A., Moss D.,
Shukurov A., Sokoloff D., 1996, ARA\&A, 34, 155
\bibitem{BertinLin}Bertin G., Lin C. C., 1996,
Spiral Structure in Galaxies. MIT Press, Cambridge
\bibitem{bin1987}Binney J., Tremaine S., 1987,
Galactic Dynamics. Princeton University Press, Princeton
\bibitem{dol2002}Dolag K., Schindler S.,
Govoni F., Feretti, L. 2001, A\&A, 378, 777
\bibitem{eva98a}Evans N. W., Read J. C. A., 1998a, MNRAS, 300, 83
\bibitem{eva98b}Evans N. W., Read J. C. A., 1998b, MNRAS, 300, 106
\bibitem{FanLou96}Fan Z. H., Lou Y.-Q., 1996, Nat., 383, 800
\bibitem{fan97}Fan Z. H., Lou Y.-Q., 1997, MNRAS, 291, 91
\bibitem{fan99}Fan Z. H., Lou Y.-Q., 1999, MNRAS, 307, 645
\bibitem{fuj05}Fujita Y., Kato T. N., 2005, MNRAS, 364, 247
\bibitem{gol79}Goldreich P., Tremaine S., 1979, ApJ, 233, 857
\bibitem{goo99}Goodman J., Evans N. W., 1999, MNRAS, 309, 599
\bibitem{hol71}Hohl F., 1971, ApJ, 168, 343
\bibitem{hu2004}Hu J., Lou Y.-Q., 2004, ApJ, 606, L1
\bibitem{hun97}Hungerford T. W., 1997, Algebra, 8th ed. Springer-Verlag, New York
\bibitem{kab2000}Kaburaki O., 2000, ApJ, 531, 210
\bibitem{kab2001}Kaburaki O., 2001, ApJ, 563, 505
\bibitem{kal71}Kalnajs A. J., 1971, ApJ, 166, 275
\bibitem{ken86}Kent S. M., 1986, AJ, 91, 130
\bibitem{ken87}Kent S. M., 1987, AJ, 93, 816
\bibitem{ken88}Kent S. M., 1988, AJ, 96, 514
\bibitem{kud96}Kudoh T., Kaburaki O., 1996, ApJ, 460, 199
\bibitem{lem91}Lemos J. P. S., Kalnajs A. J.,
Lynden-Bell D., 1991, ApJ, 375, 484
\bibitem{lin64}Lin C. C., Shu F. H., 1964, ApJ, 140, 646
\bibitem{lin66}Lin C. C., Shu F. H., 1966, Proc. Nat. Acad. Sci., 55, 229
\bibitem{lin87}Lin C. C., 1987, Selected Papers
of C. C. Lin. World Scientific, Singapore
\bibitem{liz89}Lizano S., Shu F. H., 1989, ApJ, 342, 834
\bibitem{lou02}Lou Y.-Q., 2002, MNRAS, 337, 225
\bibitem{loufan02}Lou Y.-Q., Fan Z. H., 2002, MNRAS, 329, L62
\bibitem{lou03}Lou Y.-Q., Shen Y., 2003, MNRAS, 343, 750
(astro-ph/0304270)
\bibitem{louwu04}Lou Y.-Q., Wu Y., 2005, MNRAS, 364, 475
(astro-ph/0508601)
\bibitem{lou04a}Lou Y.-Q., Zou Y., 2004, MNRAS, 350, 1220
(astro-ph/0312082)
\bibitem{lou06}Lou Y.-Q., Zou Y., 2006, MNRAS, 366, 1037
(astro-ph/0511348)
\bibitem{louBai06}Lou Y.-Q., Bai X. N., 2006, MNRAS, in press
\bibitem{lyn72}Lynden-Bell D., Kalnajs A. J., 1972, MNRAS, 157, 1
\bibitem{lyn93}Lynden-Bell D., Lemos J. P. S.,
1999 (astro-ph/9907093)
\bibitem{lyn67}Lynden-Bell D., Ostriker J. P., 1967, MNRAS, 136, 293
\bibitem{mak97}Makino N., 1997, ApJ, 490, 642
\bibitem{mes63}Mestel L., 1963, MNRAS, 157, 1
\bibitem{Mil70}Miller R. H., Prendergast K. H.,
Quirk W. J., 1970, ApJ, 161, 903
\bibitem{Mil78}Miller R. H., 1978, ApJ, 224, 32
\bibitem{Moffatt2000}Moffatt K., 2000, Dynamo Theory,
Encyclopedia of Astronomy and Astrophysics, ed. Paul Murdin
(Bristol: Institute of Physics Publishing 2001)
\bibitem{nak79}Nakano T., 1979, PASJ, 31, 697
\bibitem{ost73}Ostriker J. P., Peebles P. J. E., 1973, ApJ, 186, 467
\bibitem{qian92}Qian E., 1992, MNRAS, 257, 581
\bibitem{rub}Rubin V. C., Thonnard N. T., Ford W. K. Jr., 1982,
AJ, 87, 477
\bibitem{SheLL05}Shen Y., Liu X., Lou Y.-Q., 2005, MNRAS, 356, 1333
\bibitem{she03}Shen Y., Lou Y.-Q., 2003, MNRAS, 345, 1340 (astro-ph/0308063)
\bibitem{she04a}Shen Y., Lou Y.-Q., 2004a, MNRAS, 353, 249 (astro-ph/0405444)
\bibitem{she04b}Shen Y., Lou Y.-Q., 2004b, ChJAA, 4, 541 (astro-ph/0404190)
\bibitem{shu97}Shu F. H., Li Z.-Y., 1997, ApJ, 475, 251
\bibitem{shu00}Shu F. H., Laughlin G.,
Lizano S., Galli D., 2000, ApJ, 535, 190
\bibitem{shutremaine90}Shu F. H., Tremaine S.,
Adams F. C., Ruden S. P., 1990, ApJ, 358, 495
\bibitem{sof86}Sofue Y., Fujimoto M.,
Wielebinski R., 1986, ARA\&A, 24, 459
\bibitem{sye96}Syer D., Tremaine S., 1996, MNRAS, 281, 925
\bibitem{tay06}Taylor G. B., Gugliucci N. E., Fabian A. C., Sanders J. S., Gentile G., Allen S.
W., 2006, MNRAS, 368, 1500
\bibitem[]{toom77}Toomre A., 1977, ARA\&A, 15, 437
\bibitem{vallee04}Vall\'ee J. P., 2004,
New Astronomy Reviews, 48, 763
\bibitem{zan76}Zang T. A., 1976, PhD thesis, MIT, Cambridge MA
\end{thebibliography}
\end{document}